\documentclass[Jornal, comsoc]{IEEEtran}

\def\doublecolumn{1}

\usepackage[T1]{fontenc}
\usepackage{amsmath}
\usepackage[cmintegrals]{newtxmath}
\usepackage{bm} %need to manually load newtxmath before bm

\usepackage{enumitem}
\usepackage{tikz}
\usetikzlibrary{positioning} % below=of
\usepackage[space,noadjust]{cite}
\usepackage{subcaption}
\usepackage{etoolbox}
\makeatletter
\patchcmd{\@begintheorem}{\textit}{\textbf}{}{}
\patchcmd{\@begintheorem}{\itshape}{\bfseries}{}{}
\makeatother

\usepackage{enumitem}

\newcommand{\tail}[1]{{\small\mathtt{tail}}(#1)}
\newcommand{\head}[1]{{\small\mathtt{head}}(#1)}
\newcommand{\outset}[1]{{\small\mathtt{out}}({#1})}
\newcommand{\inset}[1]{{\small\mathtt{in}}({#1})}
\newcommand{\hatcal}[1]{\hat{\mathcal{#1}}}
\newcommand{\equalbydef}{\stackrel{\text{\tiny def}}{=}}

\newtheorem{theorem}{Theorem}
\newtheorem{corollary}{Corollary}[theorem]
\newtheorem{lemma}{Lemma}
\newtheorem{proposition}{Proposition}
\newtheorem{definition}{Definition}

\newtheorem{remark}{Remark}

\graphicspath{{.}{../pictures/}}

\begin{document}

%---------- Title ----------
\title{A Code Equivalence between\\ Secure Network and Index Coding}
\author{Lawrence Ong, J\"{o}rg Kliewer, Badri N. Vellambi, and Phee Lep Yeoh
\thanks{This work is supported by ARC grants FT140100219 and DE140100420, and US NSF grants CNS-1526547 and CCF-1439465.}
%\vspace{-1.7ex}
}
%\vspace*{-3ex}
\maketitle
%\vspace*{-4ex}

%\IEEEoverridecommandlockouts
%\vspace*{-3ex}
\maketitle
%\vspace*{-4ex}

\begin{abstract}

  A code equivalence between index coding and network coding was established, which shows that any index-coding instance can be mapped to a network-coding instance, for which any index code can be translated to a network code with the same decoding-error performance, and vice versa. Also, any network-coding instance can be mapped to an index-coding instance with a similar code translation. In this paper, we extend the equivalence to secure index coding and secure network coding, where eavesdroppers are present in the networks, and any code construction needs to guarantee security constraints in addition to decoding-error performance.
\end{abstract}

%%%%
%%%%
%%%%
%%%%
%\vspace{-3ex}
\section{Introduction}

%\subsection{Background}

Recently, equivalence results in information theory and network coding have been of significant interest to the community. Such reduction results uniquely map one communication problem to another equivalent problem that is potentially easier to study than the original problem. Some of the equivalence results already established include those between instances of multiple-unicast network coding and those of \textit{(i)} multiple-multicast network coding~\cite{DoughertyZeger06}, \textit{(ii)} secure network coding~\cite{Huangetal13}, and \textit{(iii)} index coding~\cite{effrosrouayheblangberg15,rouayhebsprintsongeorghiades10}.

%In particular, the latter results showed that a network-coding instance~\cite{ahlswedecai00} can be mapped to an equivalent index-coding instance~\cite{baryossefbirk11}. 
% , for which a code for one instance can be translated to the other, and vice versa. Similarly, an index-coding instance can be mapped to an equivalent network-coding instance, with a suitable code translation. We note that the equivalence maps between network and index coding in~\cite{effrosrouayheblangberg15,rouayhebsprintsongeorghiades10} focused on the non-secure communications scenario without eavesdroppers.

This paper focuses on the equivalence between index coding and network coding.
Index coding~\cite{baryossefbirk11} considers a one-hop network where a sender conveys multiple messages to multiple receivers through a noiseless broadcast medium, where each receiver wants some messages from the sender, but already knows some other messages. On the other hand, network coding~\cite{ahlswedecai00} considers a network of interconnected links with fixed capacities, where multiple senders send multiple messages to multiple receivers through these links.

Although these two problems appear different prima facie, the following equivalence between them has been demonstrated~\cite{rouayhebsprintsongeorghiades10,effrosrouayheblangberg15}: for any index-coding instance (specified by what each receiver has and wants), one can construct an equivalence network-coding instance (specified by how the links are connected, their capacities, and all sender and receiver locations), such that any index code (specified by the encoding function of the sender and the decoding functions of all the receivers) for the index-coding instance can be mapped to a network code for the same message sizes (specified by the encoding function of all nodes in the network, and the decoding functions of all receivers) for the network-coding instance, and vice versa. Similarly, for any network-coding instance, we can construct an equivalent index-coding instance with code mapping in both directions.

The equivalence was first shown for linear codes~\cite{rouayhebsprintsongeorghiades10} and then for non-linear codes (which include linear codes as a special case)~\cite{effrosrouayheblangberg15}. Furthermore, the equivalence has been shown for any (zero and non-zero) decoding error probability, that is, if the probability of decoding error for the network code is bounded above by a given value, the mapped index code also has this property, and vice versa. %We note that the equivalence focuses on decoding reliability.

In this paper, we further investigate if the equivalence holds if we impose another constraint besides decodability: security. Separately, the secure version of index coding and that of network coding have been studied, in which additional parties, \textit{eavesdroppers}, are present, and they attempt to obtain some information on the messages being communicated. More specifically, the secure version of index coding~\cite{dauskachekchee12} includes a number of eavesdroppers each of whom \textit{(i)} knows a subset of messages; \textit{(ii)} listens to the sender's broadcast; and \textit{(iii)} attempts to decode some messages. The secure version of network coding~\cite{caiyeung11} includes a number of eavesdropper each of whom \textit{(i)} can listen to a subset of links; and \textit{(ii)} attempts to decode some messages. A secure index code or a secure network code must prevent eavesdroppers from knowing the messages (where knowing is quantified by the information-theoretic security measure~\cite[Ch~22]{elgamalkim2001}), in addition to guaranteeing that all receivers can obtain their requested messages (by bounding the probability of decoding error). %An equivalence mapping between secure index coding and secure network coding has not been addressed to the best of our knowledge.

The non-secure equivalence results~\cite{effrosrouayheblangberg15,rouayhebsprintsongeorghiades10} do not trivially apply to the secure version of the problems. In particular, we pointed out~\cite{ongvellambiyeohklieweryuan2016} that equating the eavesdropper settings in secure network coding and secure index coding is not straightforward, as the eavesdroppers in the two problems have different characteristics (as described in the previous paragraph). Also, the non-secure equivalence was proven for deterministic code mapping. But randomised encoding is inevitable in some secure network-coding instances~\cite{caiyeung11}, and we have shown~\cite{ongvellambiyeohklieweryuan2016} that the non-secure equivalence breaks down for randomised encoding.

\subsection{Main Contributions} \label{section:contribution}

In this paper, we extend the code equivalence between index and network coding to the secured version.
Informally, in Theorem~\ref{theorem:index-to-network}, we show that any secure index-coding instance~$\mathbb{I}_1$ can be mapped to a secure network-coding instance ~$\mathbb{N}_1$, such that
  % \begin{enumerate}
  % \item
    any code for $\mathbb{I}_1$ can be translated to a code for $\mathbb{N}_1$ (and vice versa) with the same error decoding and security criteria. % criterion~$\eta$, we can devise a network code for $\mathbb{N}$ with the same decoding criterion~$\epsilon$ and security criterion~$\eta$, and vice versa.
  % \item for any network code for $\mathbb{N}$ that meets some error decoding criterion~$\epsilon$ and a security criterion~$\eta$, we can devise a index code for $\mathbb{I}$ with the same decoding criterion~$\epsilon$ and security criterion~$\eta$.
  % \end{enumerate}

In Theorem~\ref{theorem:network-to-index} and Corollary~\ref{cor:linear-codes}, we show that any secure network-coding instance  $\mathbb{N}_2$ can be mapped to a secure index-coding instance $\mathbb{I}_2$ such that
  \begin{enumerate}
\item any code for $\mathbb{N}_2$ can be translated to a code for $\mathbb{I}_2$ with the same error decoding and security criteria; % that meets some certain error decoding criterion~$\epsilon$ and a security criterion~$\eta$, we can devise a index code for $\mathbb{I}$ with the same decoding criterion~$\epsilon$ and security criterion~$\eta$; and
\item any code for $\mathbb{I}_2$
  \begin{enumerate}
  \item that has zero decoding error can be translated to code for $\mathbb{N}_2$ with the same error decoding and security criteria,
    \item \label{case} that has non-zero decoding error and is linear can be translated to a linear code for $\mathbb{N}_2$ with a security criterion that grows linearly in the codelength, and a decoding criterion that does not grow with the codelength. This implies that that strongly-secure index codes map to weakly-secure network codes. % is a function of only $\epsilon$, $\eta$, and the number of eavesdroppers.
  \end{enumerate}
\end{enumerate}

For all cases except \ref{case}, we establish an equivalence that preserves both the decodability and security criteria. %; for case~\ref{case}, we show that strongly-secure index codes map to weakly-secure network codes.

\subsection{Approaches}

To obtain the aforementioned results, we utilised the following ingredients:
\begin{enumerate}[label=\textsf{I.\arabic*}]
\item\label{item-1} a mapping between secure index-coding configurations and secure network-coding configurations, which specifies what each user and eavesdropper has access to and attempt to decode;
  \item\label{item-2} a mapping between index codes and network codes; and 
  \item\label{item-3} analysis of the performance of the mapped index code, in terms of decoding error and security criteria, given the performance of the original network code; and vice versa.
\end{enumerate}

For \ref{item-1}, extending the configuration mapping proposed by Effros et al.~\cite{effrosrouayheblangberg15}, we propose a mapping for the eavesdroppers. Briefly, for each eavesdropper in a index-coding instance, who knows a subset of messages, the corresponding eavesdropper in the network-coding instance will have access to a particular link as well as all the outgoing links from the source nodes of the corresponding messages. In the other direction, for each eavesdropper in a network-coding instance, who has access to a subset of links, the corresponding eavesdropper in the index-coding instance will have the messages corresponding to the links as side information. %The key novelty here lies in the mapping from index coding to network coding, where each eavesdropper in the latter is constructed to have access to \textit{all} outgoing links from the relevant source nodes.

Note that unlike the mapping of the users (the number of users always increase when we map one instance to the other) and the messages (the number of messages always increase from when we map a network-coding instance to an index-coding instance), the number of eavesdroppers in both instances remains the same, and there is a one-to-one correspondence among the eavesdroppers in both problems.

For \ref{item-2}, we build on the code mapping proposed by Effros et al.~\cite{effrosrouayheblangberg15}. At first sight, this mapping fails when we map a randomised network code to a randomised index code. To rectify this issue, we introduce the concept of an augmented secure network-coding instance to capture the randomness in the encoding. This increases the number of messages in the network-coding instance, but converts all randomised encoding functions to deterministic encoding functions.

For \ref{item-3}, difficulties arise in obtaining an equivalence for non-zero error and leakage due to the fact that the eavesdroppers in both instances observe different signals: messages for index coding and functions of messages transmitted on links for network coding. If decoding error at the receivers is allowed, these two types of messages do not necessarily match, making it difficult to guarantee the same amount of leakage.

This problem is even more severe for case~\ref{case} mentioned in Section~\ref{section:contribution}, in which we need to select certain parameters for the index code to obtain the required network code, and the parameters must simultaneously satisfy both error and leakage criteria.
To obtain the above equivalence result, 
we use the hypothesis that decoding is correct ($1-\epsilon$) fraction of the time for $\mathbb{I}_2$ to bound the distance between the probability mass functions (pmf) of the messages in both instances.%, which is exploited to bound the leakage of the network code.

%To obtain the above equivalence result, we bound the difference between the probability mass functions (pmf) of the messages in both instances. When decoding is correct (($1-\epsilon$) fraction of the time), the pmfs are ``close'', which is exploited to bound the leakage of the network code.

\section{Problem Definition and Notation}

For a strictly ordered set $\mathcal{S} = \{s_1, s_2, \dotsc s_{|\mathcal{S}|}\}$, with a binary relation $<$ where $s_1 < s_2 \dotsm < s_{|\mathcal{S}|}$, let $\bm{X}_{\mathcal{S}}  \equalbydef ( X_{s_1}, X_{s_2}, \dotsc, X_{s_{|\mathcal{S}}|})$.
Consider a directed graph $G=(\mathcal{V},\mathcal{E})$ with node set $\mathcal{V}$ and edge set $\mathcal{E}$. For an edge $e= (u \rightarrow v) \in \mathcal{E}$, $u,v \in \mathcal{V}$, its tail is $\tail{e} \equalbydef u$, and its head is $\head{e} \equalbydef v$. For any node $v \in \mathcal{V}$,  the set of incoming edges  is denoted by $\inset{v} \equalbydef \{ e \in \mathcal{E}: \head{e}=v\}$, and the set of outgoing edges by $\outset{v} \equalbydef \{ e \in \mathcal{E}: \tail{e}=v\}$.
For any positive integer $a \in \mathbb{Z}^+$, denote $[a] \equalbydef \{1,2,\dotsc,a\}$. For two ordered sets of discrete random variables $\bm{X}_{\mathcal{S}_1}$ and $\bm{Y}_{\mathcal{S}_2}$, $\bm{X}_{\mathcal{S}_1} \stackrel{\text{d}}{=} \bm{Y}_{\mathcal{S}_2}$ means that they have the same probability mass functions (pmf), and all corresponding pairs of random variables (one with index from $\mathcal{S}_1$ and another one from $\mathcal{S}_2$) have the same range/alphabet.

\subsection{Secure network coding}

\subsubsection{Network-coding instances} \label{sec:network-coding}
We follow Chan and Grant's secure network-coding definition~\cite{changrant08}. It includes Bhattad and Narayanan's weakly secure network-coding definition~\cite{bhattadnarayanan05} and Cai and Yeung's strongly secure network coding definition~\cite{caiyeung11} as special cases.
A secure network-coding instance, denoted by $\mathbb{N} = ( G, C, W)$, is defined as follows:
\begin{itemize}
\item $G = (\mathcal{V}, \mathcal{E})$ is an acyclic directed graph with vertex set $\mathcal{V}$ and edge set $\mathcal{E}$.\footnote{Here, $\mathcal{E}$ is a strictly ordered set, with natural ordering by the head and tail vertices.} Each edge $e \in \mathcal{E}$ is a directed noiseless communication link with a \textit{capacity} of $c_e \in \mathbb{R}^+_0 \equalbydef [0,\infty)$ bits per use. This means that if the link is used $n \in \mathbb{Z}^+$ times, vertex $\tail{e}$ can send a message $X \in [2^{\lfloor c_e n \rfloor}]$ to vertex $\head{e}$ with no error.
\item $C = (\mathcal{S}, O, \mathcal{D})$ is the connection requirement. The strictly ordered set $\mathcal{S}$ is the collection of source-message indices, where the messages are denoted by $\{X_s: s \in \mathcal{S}\}$. %As we will see later, $n$ is the number of times we use each link, and is the same for all messages. Here, $R_s$ denotes the rate of the message $X_s$. % , i.e., number of bits per use of the links
The source-location mapping $O: \mathcal{S} \rightarrow \mathcal{V}$ specifies the unique originating node $O(s)$ for source message $X_s$. The destination-location mapping $\mathcal{D}: \mathcal{S} \rightarrow 2^{\mathcal{V}}$ specifies the set of nodes $\mathcal{D}(s)$ that requires message $X_s$. Note that multiple source messages can originate from a node, multiple destination nodes can demand a particular source message, and a destination node can demand multiple source messages.
  \item $W = ((\mathcal{A}_r,\mathcal{B}_r):r \in \mathcal{R})$ defines the eavesdropping pattern a set of eavesdroppers indexed by $\mathcal{R}$. Each eavesdropper $r \in \mathcal{R}$ observes the set of links $\mathcal{B}_r \subseteq \mathcal{E}$ and tries to reconstruct a subset of source messages indexed by $\mathcal{A}_r \subseteq \mathcal{S}$, i.e., $\bm{X}_{\mathcal{A}_r}$.% We assume that $\bm{X}_{\mathcal{A}_r}$ are each requested by some node. Otherwise, we need not send that message, and it will be protected from all eavesdroppers.
  \end{itemize}

 We assume that vertices with no incoming links are originating nodes for some source messages, and  vertices with no outgoing links are destinations for some source messages. Otherwise, they can be deleted without any consequence.

 \subsubsection{Deterministic network codes}
 Given $(G,C)$, let the source messages $\{X_s: s \in \mathcal{S}\}$ be mutually independent, and each message $X_s$ be distributed over a finite alphabet $\mathcal{X}_s$ according to some pmf $p_{X_s}$. 
 
 A deterministic network code $(\mathsf{E},\mathsf{D})$ consists of a collection of deterministic encoding functions $\mathsf{E} = \{\mathsf{e}_e:e \in \mathcal{E}\}$ for the edges, and deterministic decoding functions $\mathsf{D} = \{\mathsf{d}_u: u \in \mathcal{V}\}$ for the vertices satisfying the following: Consider $n \in \mathbb{Z}^+$ network uses, meaning that each link is used $n$ times.
 \begin{itemize}
   \item  The local encoding function $\mathsf{e}_e$ for edge $e$  takes in all incoming messages $\inset{\tail{e}}$ to node~$\tail{e}$ and source messages $\bm{X}_{O^{-1}(\tail{e})}$ originating at node $\tail{e}$, and  outputs a random variable associated with link~$e$, denoted by $X_e \in [2^{\lfloor c_e n \rfloor}]$. %$\{\mathsf{e}_e\}$ are commonly called the local encoding function.

Given that $G$ is acyclic, each edge message $X_e$ can be written as a function of source messages originating from its predecessors, denoted by $\mathsf{g}_e$. This is known as the global encoding function, and it can be recursively calculated (following the topology of the graph) using \textit{(i)} $\mathsf{g}_e = \mathsf{e}_e(\bm{X}_{O^{-1}(\tail{e})})$ if $\tail{e}$ has no incoming links, and \textit{(ii)} $\mathsf{g}_e = \mathsf{e}_e(\mathsf{g}_{\inset{\tail{e}}},\bm{X}_{O^{-1}(\tail{e})})$. So, in general, we write $\mathsf{g}_e(\bm{X}_\mathcal{S})$ for all $e \in \mathcal{E}$. %, where $\{e_1, e_2, \dotsc, e_n\} = \inset{\tail{e}}$.

\item The decoding function $\mathsf{d}_u$ for a node~$u \in \mathcal{V}$ takes in random variables associated with links $\inset{u}$ and source messages originating at node $u$, and outputs an estimate of $\bm{X}_{\{s \in \mathcal{S}: u \in \mathcal{D}(s)\}}$, denoted by $\bm{X}_{\{s \in \mathcal{S}: u \in \mathcal{D}(s)\}}^{(u)}$. 
% In this paper, we only consider \textit{zero-error} decoding.
\end{itemize}

Let the probability of the event that one or more destination nodes make a decoding error be denoted as

\ifx\doublecolumn\undefined
% ==== put single-column equations here
\begin{equation}
  P_\text{e} = \Pr \{ \bm{X}_{\{s \in \mathcal{S}: u \in \mathcal{D}(s)\}}^{(u)} \neq \bm{X}_{\{s \in \mathcal{S}: u \in \mathcal{D}(s)\}} \text{ for at least one destination node } u\}. \label{eq:network-coding-decoding-prob}
\end{equation}
\else
% ==== put double-column equations here
\begin{align}
  P_\text{e} &= \Pr \{ \bm{X}_{\{s \in \mathcal{S}: u \in \mathcal{D}(s)\}}^{(u)} \neq \bm{X}_{\{s \in \mathcal{S}: u \in \mathcal{D}(s)\}} \text{ for at least} \nonumber\\
  &\quad\quad\; \text{one destination node } u\}. \label{eq:network-coding-decoding-prob}
\end{align}
\fi

For some $\epsilon  \in \mathbb{R}^+_0$, a network code $(\mathsf{E},\mathsf{D})$ said to have at most $\epsilon$ error if and only if $P_\text{e} \leq \epsilon$.

When $\eta=0$, we say that the code allows perfect decoding. %In this paper, we consider both both perfect decoding and non-zero-error decoding.

%In the conference version of this work [ref], we considered only zero error, that is, $\epsilon=0$.
 
\subsubsection{Randomised network codes}
A network code is said to be randomised if there exists an edge function $\mathsf{e}_e$ that is not a deterministic function of the random variables associated with $\inset{\tail{e}}$ and source messages originating at node $\tail{e}$.

Any randomised network code can be implemented by an equivalent deterministic network code by generating an independent random variable $Z_u$ at each node $u \in \mathcal{V}$, and defining a deterministic map from $\bm{X}_{\inset{\tail{e}}}$, $\bm{X}_{O^{-1}(\tail{e})}$, and $Z_{\tail{e}}$ to $X_e$ for each edge $e \in \mathcal{E}$~\cite{changrant08}. %  where $Z_u$ takes values in a finite alphabet % with size $\prod_{e \in \outset{u}} 2^{\lceil c_e n \rceil }$, where $\outset{u}$ is defined as the set of all outgoing edges from node $u$
% [Chan and Grant].
These random variables $\{Z_u: u \in \mathcal{V}\}$ are assumed to be mutually independent, and are often referred to as random keys.

A randomised network code $(\mathsf{E}',\mathsf{D})$ is similar to a deterministic network code $(\mathsf{E},\mathsf{D})$, except that each edge encoding function $\mathsf{e}_e'$ is a deterministic function of (i) random variables associated with $\inset{\tail{e}}$, (ii) source messages originating at node $\tail{e}$, and (iii) the random key $Z_{\tail{e}}$.

\subsubsection{Secure network codes}
A deterministic or randomised network code $(\mathsf{E},\mathsf{D})$ for $(G,M)$ is said to be secure against an eavesdropping pattern $W$ if each eavesdropper~$r$ gains not more than a specific amount of information about  $\bm{X}_{\mathcal{A}_r}$ that it attempts to reconstruct after observing $\bm{X}_{\mathcal{B}_r}$ on the links it has access to. Formally, the information leakage to eavesdropper~$r$ is calculated as $I(\bm{X}_{\mathcal{A}_r}; \bm{X}_{\mathcal{B}_r})$.
% \ifx\doublecolumn\undefined
% % ==== put single-column equations here
% \begin{alignat}{3}
%  \text{Perfect security:} & \quad\quad I(\bm{X}_{\mathcal{A}_r}; \bm{X}_{\mathcal{B}_r}) = 0, &\quad\quad  \text{ for all }r\in \mathcal{R}. \label{eq:secure-nc-condition-1}\\
%  \text{$\eta$-leakage:} & \quad\quad I(\bm{X}_{\mathcal{A}_r}; \bm{X}_{\mathcal{B}_r}) \leq \eta, &\quad\quad \text{ for all } r\in \mathcal{R}, \text{ for some } \eta  \in \mathbb{R}^+_0. \label{eq:secure-nc-condition-2}
% \end{alignat}
% \else
% % ==== put double-column equations here
% \begin{align}
%  \text{Perfect security:}  \quad\quad &I(\bm{X}_{\mathcal{A}_r}; \bm{X}_{\mathcal{B}_r}) = 0,\nonumber \\ \quad\quad  &\text{for all }r\in \mathcal{R}. \label{eq:secure-nc-condition-1}\\
%  \text{$\eta$-leakage:} \quad\quad &I(\bm{X}_{\mathcal{A}_r}; \bm{X}_{\mathcal{B}_r}) \leq \eta, \nonumber \\ \quad\quad &\text{for all } r\in \mathcal{R}, \text{ for some } \eta  \in \mathbb{R}^+_0. \label{eq:secure-nc-condition-2}
% \end{align}
% \fi

For some $\eta  \in \mathbb{R}^+_0$, a network code is said to be have at most $\eta$ leakage if and only if
\begin{equation}
  I(\bm{X}_{\mathcal{A}_r}; \bm{X}_{\mathcal{B}_r}) \leq \eta, \quad\quad \text{for all } r\in \mathcal{R}. \label{eq:secure-nc-condition-2}
\end{equation}

When $\eta=0$, we say that the code is perfectly secure.
%In this paper, we consider both perfect security and non-zero leakage.

%In other words, $(\mathsf{E},\mathsf{D})$ is a secure network code for the secure network-coding instance $\mathbb{N}$.

\subsubsection{Secure network-coding feasibility} \label{sec:nc-feasibility-definition}
A secure network-coding instance $\mathbb{N}$ is said to be $(\mathcal{S}^*, (p_{X_s}:s \in \mathcal{S}^*),\epsilon, \eta ,n)$-feasible if and only if there exists a joint pmf $p_{\bm{X}_{\mathcal{S} \setminus \mathcal{S}^*}} = \prod_{s \in \mathcal{S} \setminus \mathcal{S}^*} p_{X_s}$ for messages $\bm{X}_{\mathcal{S} \setminus \mathcal{S}^*}$ and a secure network code over $n$ network uses with at most $\epsilon$ error and $\eta$ leakage for the message joint pmf $p_{\bm{X}_\mathcal{S}}(\bm{x}_\mathcal{S}) = p_{\bm{X}_{\mathcal{S}*}}(\bm{x}_{\mathcal{S}*}) p_{\bm{X}_{\mathcal{S} \setminus \mathcal{S}^*}}(\bm{x}_{\mathcal{S} \setminus \mathcal{S}^*})$.

Note that our message setup is sufficiently general, which include the problem formulations:
\begin{enumerate}[label=\textsf{F.\arabic*}]
\item Given a joint message pmf $p_{\bm{X}_{\mathcal{S}}}$, we want to find the minimum number of network uses $n$ required to achieve certain decoding and leakage requirements.
\item\label{formulation-2} Let each message $X_s$ be uniformly distributed over $|\mathcal{X}_s|$, and define $R_s \equalbydef (\log_2 |\mathcal{X}_s|)/n$ as the average message rate per network use. Given a number of network uses $n$, we want to find rate tuples $\bm{R}_{\mathcal{S}}$ that satisfy certain decoding and leakage requirements.
\item\label{formulation-3} For each message index $s \in \mathcal{S}$, consider a pmf $p_s$ over $[L_s]$ for some $L_s \in \mathbb{Z}^+$. For every $s \in \mathcal{S}$, let $X_s = (X_{s_1}, X_{s_2}, \dotsc, X_{s_m})$ where each $X_{s_i}$ are independently distributed according to $p_s$. This means $|\mathcal{X}_s| = L_s^m$. We want to find the maximum source-channel rate $m/n$  that satisfies certain decoding and leakage requirements.
\end{enumerate}

For formulations~\ref{formulation-2} and \ref{formulation-3} above, one can fix the rate ($\bm{R}_{\mathcal{S}}$ for \ref{formulation-2} or $m/n$ for \ref{formulation-3}) and find a sequence of network codes with increasing $n$ to get following notions of security criteria:
\begin{align*}
\text{Strong security: }&  \lim_{n \rightarrow \infty} I(\bm{X}_{\mathcal{A}_r}; \bm{X}_{\mathcal{B}_r}) = 0, \quad \forall r\in \mathcal{R},\\
\text{Weak security: }& \lim_{n \rightarrow \infty} \frac{1}{\ell} I(\bm{X}_{\mathcal{A}_r}; \bm{X}_{\mathcal{B}_r}) = 0, \quad \forall r\in \mathcal{R}. %\label{eq:secure-nc-condition-3}
\end{align*}
where $\ell = n$ for \ref{formulation-2}, and $\ell = m$ for \ref{formulation-3}.

% \begin{remark}
%   We have used the message-connection setting by Chan and Grant~\cite{changrant08} instead that by Effros, El Rouayheb, and Langberg~\cite{effrosrouayheblangberg15}, although they are equivalent as far as classical network coding is concerned (i.e., without security constraints). In the definition by Effros et al., each source message is generated by one \textit{source node} without any incoming links. The definition by Chan and Grant, used in this paper, can be easily converted to that by Effros et al.\ as follows: For each node~$u$ that is (i) generating multiple source messages or (ii) generating source message(s) and having incoming link(s), we remove the source messages at $u$ and add source nodes each generating a source message and each with a link to $u$. %However, by doing so, we change the min-cut of the message flows in the network, thereby making the secure-network-coding results incompatible with that in Cai and Yeung~\cite{caiyeung11}.
% \end{remark}

\subsection{Secure index coding}

\subsubsection{Secure index-coding instances}
We follow Dau, Skachek, and Chee's secure index-coding definition~\cite{dauskachekchee12}. A secure index-coding instance, denoted by $\mathbb{I} = (\hatcal{{S}}, \hatcal{{T}}, \{(\hatcal{{W}}_t,\hatcal{{H}}_t): t \in \hatcal{T}\}, \hat{W})$, is defined as follows:
\begin{itemize}
\item $\hatcal{{S}}$ is a strictly ordered set of indices of source messages available at a sender.
\item $\hatcal{{T}}$ is an strictly ordered set of receiver indices.
\item $\hatcal{{W}}_t$ is the set of the indices of the messages required by receiver~$t \in \hatcal{{T}}$.
\item $\hatcal{{H}}_t$ is the set of indices of the messages known a priori to receiver~$t \in \hatcal{{T}}$.
  \item $\hat{W} = ((\hatcal{A}_r,\hatcal{B}_r): r \in \hatcal{{R}})$ is the eavesdropping pattern. % \footnote{The notation used by Dau, Skachek, and Chee is reversed, i.e., $\mathcal{A}$ for the indices of messages the eavesdropper can access and $\mathcal{B}$ for the indices of messages the eavesdropper tries to reconstruct. Here, we have used the notation that mirrors that for secure network coding.}
    Each eavesdropper~$r \in \hatcal{{R}}$ has access to the codeword broadcast by the sender and a subset of the messages $\bm{X}_{\hatcal{B}_r}$, and tries to reconstruct $\bm{X}_{\hatcal{A}_r}$, where $\hatcal{A}_r,\hatcal{B}_r \subseteq \hatcal{{S}}$, and $\hatcal{A}_r \cap \hatcal{B}_r = \emptyset$.
\end{itemize}

\subsubsection{Deterministic index codes}
Let the messages $\{\hat{X}_s: s \in \hatcal{{S}}\}$ be mutually independent, and for each $s \in \hat{ \mathcal{S}}$,  $\hat X_s$ be distributed over a finite alphabet $\hatcal{X}_s$ according to some pmf $p_{\hat{X}_s}$.
A deterministic index code $(\hat{\mathsf{e}},\hat{\mathsf{D}})$, where $\hat{\mathsf{D}} =\{\hat{\mathsf{d}}_t: t \in \hatcal{{T}} \})$, consists of
\begin{itemize}
  \item a deterministic encoding function $\hat{\mathsf{e}}$ by the sender,  which takes
in random variables $\hat{\bm{X}}_{\hatcal{{S}}}$ and outputs a
random variable $\hat{X}_\text{b} = \hat{\mathsf{e}}(\hat{\bm{X}}_{\hatcal{{S}}}) \in [2^{\hat{n}}]$, for some $\hat{n} \in \mathbb{Z}^+$, and
\item a deterministic decoding function $\hat{\mathsf{d}}_{t}$ for each receiver~$t \in \hatcal{T}$,
which takes in the sender's codeword $\hat{X}_\text{b}$ and its prior
messages $\hat{\bm{X}}_{\hatcal{{H}}_t}$ and outputs an estimate of the
messages $\hat{\bm{X}}_{\hatcal{{W}}_t}$ it requires, denoted by $\hat{\bm{X}}_{\hatcal{{W}}_{t}}^{(t)}$.
\end{itemize}
%We see that the length of the codes (sent by the sender) is $n$ bits.

\begin{remark}
This index-code definition is consistent with the index-coding literature~\cite{blasiakkleinberglubertzky13, unalwagner16, arbabjolfaei13, shanmugamdimakislangberg13}, but is different from that by Effros et al., where the sender transmits $X_\text{b} \in [2^{\hat{c}_\text{b}n}]$, and $\hat{c}_\text{b}$ is then chosen to be a function of the link capacities of the equivalent network-coding instance. Our choice results in a scaling factor of the alphabet size for the index-coding messages, but avoids the issue of $2^{\hat{c}_\text{b} n}$ not being an integer.
\end{remark}

As with network coding, let the probability of the event that one or more destination nodes make a decoding error be denoted as
\begin{equation}
  \hat{P}_\text{e} \equalbydef \Pr \{ \hat{\bm{X}}_{\hatcal{{W}}_{t}}^{(t)} \neq \hat{\bm{X}}_{\hatcal{{W}}_{t}}  \text{ for at least one destination node } t \in \hatcal{T}\}.
\end{equation}
%In this paper, we only consider \textit{zero-error} decoding.

For some $\epsilon \in \mathbb{R}^+_0$, an index code $(\hat{\mathsf{e}},\hat{\mathsf{D}})$ said to have at most $\epsilon$ error if and only if $\hat{P}_\text{e} \leq \epsilon$. %We consider both perfect decoding ($\epsilon=0$) and non-zero-error decoding in this paper.
%In the conference version of this work [ref], we considered only zero error.

\subsubsection{Randomised index codes}
A randomised index code $(\hat{\mathsf{e}}',\hat{\mathsf{D}})$ is defined similar to the deterministic index codes except that the sender's encoding function takes in an independent random key $\hat{Z}$ in addition to $\hat{\bm{X}}_{\hatcal{{S}}}$. 
Unlike the model by Mojahedian, Aref, and Gohari~\cite{mojahediangohariaref15}, the randomness allowed in the encoding in our setting  is generated locally at the sender, and is not shared with the receivers or the eavesdroppers.

\subsubsection{Secure index codes}
A deterministic or randomised index code
$(\hat{\mathsf{e}},\hat{\mathsf{D}})$ is said to be secure against the
eavesdropping pattern $\hat{W}$ if each eavesdropper $r \in \hatcal{R}$ gains no information
about the message set $\hat{\bm{X}}_{\hatcal{A}_r}$ it tries to reconstruct by observing the
sender's codeword $\hat{X}_{\text{b}}$ and its side information $\hat{\bm{X}}_{\hatcal{B}_r}$. % As with network coding, we consider the following security constraints:
% \ifx\doublecolumn\undefined
% % ==== put single-column equations here
% \begin{alignat}{3}
%  \text{Perfect security:} & \quad\quad I(\hat{\bm{X}}_{\hatcal{A}_r} ; \hat{X}_{\text{b}}, \hat{\bm{X}}_{\hatcal{B}_r}) = 0, &\quad\quad  \text{ for all }r \in \hatcal{{R}}. \label{eq:secure-ic-condition-1}\\
%  \text{$\eta$-leakage:} & \quad\quad I(\hat{\bm{X}}_{\hatcal{A}_r} ; \hat{X}_{\text{b}}, \hat{\bm{X}}_{\hatcal{B}_r}) \leq \eta, &\quad\quad \text{ for all } r \in \hatcal{{R}}, \text{ for some } \eta  \in \mathbb{R}^+_0. \label{eq:secure-ic-condition-2}
% \end{alignat}
% \else
% % ==== put double-column equations here
% \begin{align}
%  \text{Perfect security:} \quad\quad &I(\hat{\bm{X}}_{\hatcal{A}_r} ; \hat{X}_{\text{b}}, \hat{\bm{X}}_{\hatcal{B}_r}) = 0,\nonumber \\ \quad\quad  &\text{ for all }r \in \hatcal{{R}}. \label{eq:secure-ic-condition-1}\\
%  \text{$\eta$-leakage:}  \quad\quad &I(\hat{\bm{X}}_{\hatcal{A}_r} ; \hat{X}_{\text{b}}, \hat{\bm{X}}_{\hatcal{B}_r}) \leq \eta, \nonumber \\ \quad\quad &\text{ for all } r \in \hatcal{{R}}, \text{ for some } \eta  \in \mathbb{R}^+_0. \label{eq:secure-ic-condition-2}
% \end{align}
% \fi
Similar to network coding, the leakage to eavesdropper~$r$ is calculated as $I(\hat{\bm{X}}_{\hatcal{A}_r} ; \hat{X}_{\text{b}}, \hat{\bm{X}}_{\hatcal{B}_r})$. For any $\eta  \in \mathbb{R}^+_0$, we say that an index code has at most $\eta$ leakage if and only if
\begin{equation}
  I(\hat{\bm{X}}_{\hatcal{A}_r} ; \hat{X}_{\text{b}}, \hat{\bm{X}}_{\hatcal{B}_r}) \leq \eta, \quad\quad \text{ for all } r \in \hatcal{{R}}. \label{eq:secure-nc-condition-2}
\end{equation}
Also, when $\eta=0$, we say that the index code is perfectly secure. %, and we consider both perfect security and non-zero leakage.

\subsubsection{Secure index-coding feasibility}
Similar to the feasibility notion for secure network coding, a secure index-coding instance $\mathbb{I}$ is said to be $(\hatcal{S}^*, (p_{\hat{X}_s}:s \in \hatcal{S}^*),\epsilon, \eta ,n)$-feasible if and only if there exists a joint pmf $p_{\hat{\bm{X}}_{\hatcal{S} \setminus \hatcal{S}^*}} = \prod_{s \in \hatcal{S} \setminus \hatcal{S}^*} p_{\hat{X}_s}$ for messages $\hat{\bm{X}}_{\hatcal{S} \setminus \hatcal{S}^*}$ and a secure network code of length $n$ with at most $\epsilon$ error and $\eta$ leakage for the message joint pmf $p_{\hat{\bm{X}}_{\hatcal{S}}}(\hat{\bm{x}}_{\hatcal{S}}) = p_{\hat{\bm{X}}_{\hatcal{S}*}}(\hat{\bm{x}}_{\hatcal{S}*}) p_{\hat{\bm{X}}_{\hatcal{S} \setminus \hatcal{S}^*}}(\hat{\bm{x}}_{\hatcal{S} \setminus \hatcal{S}^*})$. The general message definition here also allows us to define different index-coding problem formulations similar to those for network coding mentioned earlier.

%The secure index-coding instance is said to be $(\epsilon,\eta, \hat{\bm{R}}_{\hatcal{{S}}},n)$-feasible for the message joint probability mass function $p_{\hat{\bm{X}}_{\hatcal{S}}}$ if and only if there exists at least one secure index code with the associated source-message rates $\hat{\bm{R}}_{\hatcal{{S}}}$, of block size $n$, with at most $\epsilon$ error and $\eta$ leakage.

\section{Mapping from Secure Index Coding to Secure Network Coding}

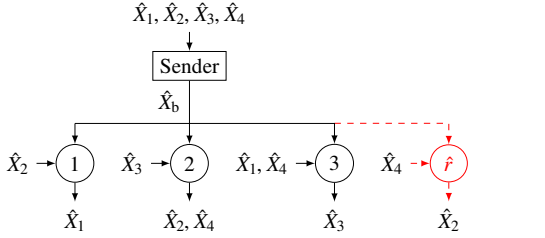
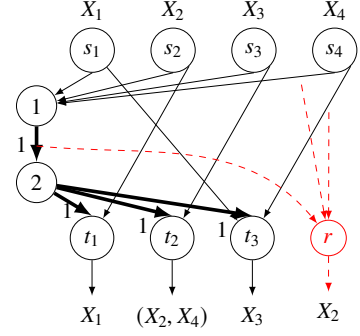
\begin{figure*}
  \centering
  \begin{subfigure}[b]{0.45\textwidth}
    \centering
        \resizebox{40ex}{!}{%
\begin{tikzpicture} [node distance=2ex,
receiver/.style={circle,draw},
point/.style={circle,inner sep=0pt}
]
\node (sender) [draw,rectangle] {Sender};
\node (message) [above=of sender] {$\hat{X}_1,\hat{X}_2,\hat{X}_3,\hat{X}_4$};
\node (joint2) [point,below=of sender,label=left:{$\hat{X}_{\text{b}}$}] {};
\node (joint) [point,below=of joint2] {};
\node (2) [receiver,below=of joint] {2};
\node (k2) [left=of 2] {$\hat{X}_3$};
\node (1) [receiver,left=of k2] {1};
\node (k1) [left=of 1] {$\hat{X}_2$};
\coordinate (joint-coor) at (joint);
\node (k3) [right=of 2] {$\hat{X}_1,\hat{X}_4$};
\node (3) [receiver,right=of k3] {3};
\node (k4) [right=of 3] {$\hat{X}_4$};
\node (4) [receiver,right=of k4,color=red] {\color{red}$\hat{r}$};
\node (t3) [point,above=of 3] {};
\coordinate (t3-coor) at (t3);
\node (t4) [point,above=of 4] {};
\coordinate (t4-coor) at (t4);
\node (t1) [point,above=of 1] {};
\coordinate (t1-coor) at (t1);
\node (b1) [below=of 1] {$\hat{X}_1$};
\node (b2) [below=of 2] {$\hat{X}_2,\hat{X}_4$};
\node (b3) [below=of 3] {$\hat{X}_3$};
\node (b4) [below=of 4] {$\hat{X}_2$};
%\node (a1) [right=of sender] {};
%\node (a2) [right=of a1, cloud,draw, aspect=4, cloud puffs=15, inner sep=0, fill=pink] {Eavesdropper};

\draw[->,>=latex] (message) -- (sender);
\draw[->,>=latex] (sender) -- (2);
\draw[->,>=latex] (k2) -- (2);
\draw[->,>=latex] (k1) -- (1);
\draw (joint-coor) -- (t1-coor);
\draw[->,>=latex]  (t1-coor) -- (1);
\draw[->,>=latex] (k3) -- (3);
\draw[->,>=latex,color=red,dashed] (k4) -- (4);
\draw[->,>=latex,color=red,dashed] (t3-coor) -| (4);
%\draw[->,>=latex] (t4-coor) -- (4);
\draw[->,>=latex] (joint-coor) -| (3);
\draw[->,>=latex] (1) -- (b1);
\draw[->,>=latex] (2) -- (b2);
\draw[->,>=latex] (3) -- (b3);
\draw[->,>=latex,color=red,dashed] (4) -- (b4);
\end{tikzpicture}
}
        \caption{A secure index-coding instance $\mathbb{I}$, where an eavesdropper~$\hat{r}$ has access to the broadcast message $\hat{X}_{\text{b}}$, side information $\hat{X}_4$, and tries to reconstruct $\hat{X}_2$}
        \label{fig:index-coding-1}
      \end{subfigure}
      \quad
    \begin{subfigure}[b]{0.45\textwidth}
      \centering
      \resizebox{30ex}{!}{%
\begin{tikzpicture} [node distance=3.5ex,
receiver/.style={circle,draw},
point/.style={circle,inner sep=0pt}
]
\node (1) [receiver] {1};
\node (2) [receiver,below=of 1] {2};
\draw[->,>=latex,ultra thick] (1) -- node[midway,left] (12) {$1$}  (2);

\node (t1) [receiver, below right=of 2] {$t_1$};
\node (t2) [receiver, right=of t1] {$t_2$};
\node (t3) [receiver, right=of t2] {$t_3$};
\node (e) [receiver, right=of t3, color=red] {\color{red}$r$};

\node (s1) [receiver, above right=of 1,label=above:{$X_1$}] {$s_1$};
\node (s2) [receiver, right=of s1,label=above:{$X_2$}] {$s_2$};
\node (s3) [receiver, right=of s2,label=above:{$X_3$}] {$s_3$};
\node (s4) [receiver, right=of s3, ,label=above:{$X_4$}] {$s_4$};

\node (w1) [point, below=of t1,label=below:{$X_1$}] {};
\node (w2) [point, below=of t2,label=below:{$(X_2,X_4)$}] {};
\node (w3) [point, below=of t3,label=below:{$X_3$}] {};
\node (we) [point, below=of e,label=below:{$X_2$}] {};

\draw[->,>=latex] (t1) -- (w1);
\draw[->,>=latex] (t2) -- (w2);
\draw[->,>=latex] (t3) -- (w3);
\draw[->,>=latex,color=red,dashed] (e) -- (we);

\draw[->,>=latex] (s1.south) -- (1);
\draw[->,>=latex] (s2.south) -- (1);
\draw[->,>=latex] (s3.south) -- (1);
\draw[->,>=latex] (s4.south) -- node[very near start] (41) {} (1);

\draw[->,>=latex,ultra thick] (2) -- node[near end,left] (1-2) {$1$}  (t1.north);
\draw[->,>=latex,ultra thick] (2) -- node[near end,below] (1-2) {$1$}  (t2.north);
\draw[->,>=latex,ultra thick] (2) -- node[very near end,below] (1-2) {$\!\!\!1$}  (t3.north);

\draw[->,>=latex] (s1) -- (t3);
\draw[->,>=latex] (s2.south east) -- (t1);
\draw[->,>=latex] (s3.south east) --  (t2);
\draw[->,>=latex] (s4.south east) -- node[near start] (43) {} (t3);

\draw[->,>=latex, dashed, color=red, bend right] (12) to [out=10,in=145] (e);
\draw[->,>=latex, dashed, color=red, bend right] (43) -- (e);
\draw[->,>=latex, dashed, color=red, bend right] (41) -- (e);
\end{tikzpicture}
}
        \caption{A secure network-coding instance $\mathbb{N}$, where an eavesdropper~$r$ has access to link $(1 \rightarrow 2)$,  all outgoing links from node $s_4$, and tries to reconstruct $X_2$. The capacity of all links given by thick arrows is 1~bit per channel use}
        \label{fig:network-coding}
      \end{subfigure}
      \caption{A secure index-coding instance $\mathbb{I}$ and its
        corresponding secure network-coding instance $\mathbb{N}$}
      %\vspace{-2.5ex}
      \label{fig:index-to-network-example}
    \end{figure*}

\subsection{Index-to-network coding configuration mapping}
\label{sec:i-2-n}

   Given a configuration $\mathbb{I} = (\hatcal{{S}}, \hatcal{{T}}, \{(\hatcal{{W}}_t,\hatcal{{H}}_t): t \in \hatcal{T}\}, \hat{W})$ of a secure index-coding instance. Let $\hatcal{{S}}= [k]$ and $\hatcal{{T}} = [\ell]$ for some positive integers $k$ and $\ell$.

We follow the mapping for $G$ and $C$ by Effros et al.~\cite{effrosrouayheblangberg15}:
\begin{itemize}
\item The graph $G = (\mathcal{V},\mathcal{E})$ consists of $k+\ell+2$ vertices labelled as $\mathcal{V}=\{s_1, s_2, \dotsc, s_{k}, t_1, t_2, \dotsc, t_{\ell}, 1, 2\}$. For each $i \in \hatcal{{S}}$, vertex $s_i$ has an outgoing link to vertex~1 and to each vertex in $\{t_j: i \in \hatcal{H}_j \}$. Each of these links from vertex $s_i$ are of sufficiently large capacity. Vertex~1 has a link of capacity 1~bit per use to vertex~2, and vertex~2 has a link of capacity 1~bit per use to each vertex in $\{t_i: i \in \hatcal{{T}}\}$.
\item The connection requirement $C$ consists of the following: $\mathcal{S} = \hatcal{{S}}$.
  For each message $X_i$, $i \in \mathcal{S}$, the source locations are $O(i) = s_i$, i.e., the message $X_i$ originates at vertex~$s_i$, and is destined for $\mathcal{D}(i) = \{t_j: i \in \hatcal{{W}}_j \}$.
\end{itemize}
  Note that by construction, for each $i \in \hatcal{{T}}$,
  \begin{itemize}
  \item $\hatcal{{W}}_i = \{j \in \mathcal{S}: t_i \in \mathcal{D}(j) \}$, that means, the requested messages are the same in both instances; and
  \item $\hatcal{{H}}_i = \{ j \in \mathcal{S}: (s_j \rightarrow t_i) \in \mathcal{E}\}$, that means, side information in $\mathbb{I}$ manifests itself in incoming links from corresponding source nodes in $\mathbb{N}$.
  \end{itemize}
Also, the vertices $\mathcal{V} \setminus \{t_1, \dotsc, t_{\ell} \}$ are not the destinations of any source message.

We propose the following mapping $W$ for the eavesdroppers:
\begin{itemize}
\item The eavesdropping pattern $W$ is defined as $\mathcal{R} = \hatcal{{R}}$, $\mathcal{B}_r = \{ (1 \rightarrow 2), \{\outset{s_i} : i \in \hatcal{B}_r \}\}$, and $\mathcal{A}_r = \hatcal{A}_r$, for each $r \in \hatcal{R}$.
\end{itemize}
Note that different from the mapping $C$, we propose that the side information of an eavesdropper in $\mathbb{I}$ be mapped to an eavesdropper in $\mathbb{N}$ having access to all outgoing links from the corresponding source nodes as well as the link $1 \rightarrow 2$.

Figure~\ref{fig:index-to-network-example} depicts an example of such a mapping.

% Let $H_\text{b}()$ denote the binary entropy function.

\subsection{Equivalence results}
\label{sec:i-t-n-result}

With the above conversion, we now state an equivalence between these two instances:
  \begin{theorem}\label{theorem:index-to-network}
    Let $\mathbb{I}$ be a secure index-coding instance, and $\mathbb{N}$ be the corresponding secure network-coding instance using the index-to-network coding mapping. For any $\epsilon, \eta \in \mathbb{R}^+_0$, and $n \in \mathbb{Z}^+$, the instance~$\mathbb{I}$ is $(\hatcal{S}, (p_{\hat{X}_s}:s \in \hatcal{S}),\epsilon, \eta ,n)$-feasible if and only if $\mathbb{N}$ is $(\mathcal{S}, (p_{X_s}:s \in \mathcal{S}),\epsilon, \eta ,n)$-feasible with deterministic coding functions for vertices $\{s_i:i \in \hatcal{S}\}$, where $\hat{\bm{X}}_{\hatcal{S}} \stackrel{\text{d}}{=} \bm{X}_\mathcal{S}$.
  \end{theorem}

   The theorem above preserves the message size, as well as the decodability and security criteria.
  We will prove Theorem~\ref{theorem:index-to-network} in the next two sections.

  \section{Proof of Theorem~\ref{theorem:index-to-network} -- the forward direction} \label{proof:1-f}
    We will now prove Theorem~\ref{theorem:index-to-network} for the forward direction, that is $\mathbb{I}$ is $(\hatcal{S}, (p_{\hat{X}_s}:s \in \hatcal{S}),\epsilon, \eta ,n)$-feasible $\Rightarrow$ $\mathbb{N}$ is $(\mathcal{S}, (p_{X_s}:s \in \mathcal{S}),\epsilon, \eta ,n)$-feasible.

  \subsection{Code construction}

  Let $(\hat{\mathsf{e}},\hat{\mathsf{D}})$ be a secure index code (which can be randomised) for $\mathbb{I}$ that is $(\hatcal{S}, (p_{\hat{X}_s}:s \in \hatcal{S}),\epsilon, \eta ,n)$-feasible. We now adapt the code mapping by El Rouayheb et al.\ and Effros et al. to obtain a network code for $\mathbb{N}$. The decoding fidelity for this code mapping has been proven for deterministic codes. Here, we will prove that this code mapping also satisfy
  \begin{itemize}
  \item the same decoding criterion for randomised index codes, and
  \item the security criteria.
  \end{itemize}

The secure network code $(\mathsf{E},\mathsf{D})$ is as follows:
    \begin{itemize}
    \item Set a deterministic edge function $\mathsf{e}_{s_i}(X_{O^{-1}(s_i)}) = \mathsf{e}_{s_i}(X_i) =X_i$ for all outgoing edges from each vertex in $\{s_i: i \in \mathcal{{S}}\}$. This is possible since vertex~$s_i$ is the originating vertex for the message~$X_i$, and the link capacity is sufficiently large.
    \item Set $X_{1 \rightarrow 2} = \mathsf{e}_{1 \rightarrow 2}(\cdot) = \hat{\mathsf{e}}(\bm{X}_{\mathcal{S}},Z_1) \in [2^n]$ for the edge $1 \rightarrow 2$. $Z_1$ (which is the random key used in the encoding function of vertex~1 in $\mathbb{N}$) is independent of all the source messages $\bm{X}_{\mathcal{S}}$ and has the same distribution as $\hat{Z}$ (which is the random key in the encoding function in $\mathbb{I}$). This means $(\bm{X}_{\mathcal{S}},Z) \stackrel{\text{d}}{=} (\hat{\bm{X}}_{\mathcal{S}},\hat{Z})$. Again this is possible as vertex~1 receives $\bm{X}_\mathcal{S}$ from the incoming links, and the link $(1 \rightarrow 2)$ and all outgoing links from vertex~2 have the capacity of $n$~bits per use. 
    \item Set $X_e = \mathsf{e}_e(\cdot) = X_{1 \rightarrow 2}$ for all $e \in \outset{2}$
    \item Set $\mathsf{d}_{t_i}(\cdot) = \hat{\mathsf{d}}_i(\cdot)$ for all $i \in \hatcal{{T}}$, and $\mathsf{d}_u=0$ for all other vertices.
    \end{itemize}

\subsection{Decoding criteria}

Note that, in the network-coding instance $\mathbb{N}$, only receivers $\{t_i$: $i \in \hatcal{T}\}$ need to decode messages, and each of them receives $\hat{\mathsf{e}}(\bm{X}_\mathcal{S},Z_1)$ and $\bm{X}_{\hatcal{H}_i}$ over its coming links. These are the same functions that each receiver $i \in \hatcal{T}$ receives in the index-coding instance $\mathbb{I}$. By using the same decoding functions for receivers $\{t_i$: $i \in \hatcal{T}\}$ in $\mathbb{N}$, if $\hat{P}_\text{e} \leq \epsilon$ for  $\mathbb{I}$, we also must have $P_\text{e} \leq \epsilon$ for  $\mathbb{N}$.

\subsection{Security criteria}

Each eavesdropper $r \in \mathcal{R}$ in $\mathbb{N}$ has access to messages $\bm{X}_{\mathcal{B}_r}$ on the link set $\mathcal{B}_r$ consisting of
    \begin{itemize}
    \item link $(1 \rightarrow 2)$, which carries $X_{{1\rightarrow 2}}=\hat{\mathsf{e}}(\bm{X}_{\mathcal{S}},Z)$, and
      \item links $\{ \outset{s_i}: i \in \hatcal{B}_r\}$, which carry messages $\bm{X}_{\hatcal{B}_r}$, because by construction, each outgoing link from node $s_i$ carries $X_i$. 
      \end{itemize}
      Now, we know that, for $\mathbb{I}$, $I(\hat{\bm{X}}_{\hatcal{A}_r} ; \hat{\mathsf{e}}(\hat{\bm{X}}_{\hatcal{S}},\hat{Z}), \hat{\bm{X}}_{\hatcal{B}_r}) \leq \eta$, for all  $r \in \hatcal{{R}}$. Since $(\bm{X}_{\mathcal{S}},Z) \stackrel{\text{d}}{=} (\hat{\bm{X}}_{\mathcal{S}},\hat{Z})$ by construction, we have
      \ifx\doublecolumn\undefined
%==== put single-column equations here 
\begin{equation}
        (\bm{X}_{\mathcal{A}_r}, \hat{\mathsf{e}}(\bm{X}_{\mathcal{S}},Z), \bm{X}_{\hatcal{B}_r}) \stackrel{\text{d}}{=} (\hat{\bm{X}}_{\mathcal{A}_r}, \hat{\mathsf{e}}(\hat{\bm{X}}_{\mathcal{S}},\hat{Z}), \hat{\bm{X}}_{\hatcal{B}_r}) = (\hat{\bm{X}}_{\hatcal{A}_r}, \hat{\mathsf{e}}(\hat{\bm{X}}_{\hatcal{S}},\hat{Z}), \hat{\bm{X}}_{\hatcal{B}_r}). \label{eq:change-of-var}
      \end{equation}
\else
%==== put double-column equations here
\begin{subequations}
\begin{align}
        (\bm{X}_{\mathcal{A}_r}, \hat{\mathsf{e}}(\bm{X}_{\mathcal{S}},Z), \bm{X}_{\hatcal{B}_r}) &\stackrel{\text{d}}{=} (\hat{\bm{X}}_{\mathcal{A}_r}, \hat{\mathsf{e}}(\hat{\bm{X}}_{\mathcal{S}},\hat{Z}), \hat{\bm{X}}_{\hatcal{B}_r})\\ 
&= (\hat{\bm{X}}_{\hatcal{A}_r}, \hat{\mathsf{e}}(\hat{\bm{X}}_{\hatcal{S}},\hat{Z}), \hat{\bm{X}}_{\hatcal{B}_r}). \label{eq:change-of-var}
\end{align}
\end{subequations}
\fi
      
      For $\mathbb{N}$, we now show that
      \begin{subequations}
      \begin{align}
                I(\bm{X}_{\mathcal{A}_r}; \bm{X}_{\mathcal{B}_r}) &= I(\bm{X}_{\mathcal{A}_r}; \hat{\mathsf{e}}(\bm{X}_{\mathcal{S}},Z),  \bm{X}_{\hatcal{B}_r}) \\
               &=  I(\hat{\bm{X}}_{\hatcal{A}_r} ; \hat{\mathsf{e}}(\hat{\bm{X}}_{\hatcal{S}},\hat{Z}), \hat{\bm{X}}_{\hatcal{B}_r}) \label{eq:follow}\\
        & \leq \eta.
      \end{align}
    \end{subequations}
    where \eqref{eq:follow} follows from \eqref{eq:change-of-var} with a change of variables (from non-hatted to hatted). This completes the security proof for $\mathbb{N}$.

      \section{Proof of Theorem~\ref{theorem:index-to-network} -- the backward direction} \label{proof:1-b}
    We will now prove Theorem~\ref{theorem:index-to-network} for the backward direction, that is, $\mathbb{N}$ is $(\mathcal{S}, (p_{X_s}:s \in \mathcal{S}),\epsilon, \eta ,n)$-feasible  $\Rightarrow$   $\mathbb{I}$ is $(\hatcal{S}, (p_{\hat{X}_s}:s \in \hatcal{S}),\epsilon, \eta ,n)$-feasible.

\subsection{Code construction}
Let $(\mathsf{E},\mathsf{D})$ be a secure network code for $\mathbb{N}$ that is $(\hatcal{S}, (p_{\hat{X}_s}:s \in \hatcal{S}),\epsilon, \eta ,n)$-feasible such that the outgoing links from the sources $\{s_i: i \in \mathcal{S}\}$ are each deterministic functions of the source messages $X_i$, i.e., for each $i \in \mathcal{S}$, we have 
\begin{equation}
X_e = \mathsf{e}_e(X_i), \quad \text{for each } e \in \outset{s_i}. 
\end{equation}
This means, for a given message realisation, the only randomness in the code is due to $Z_1$ and $Z_2$, which are the independent random keys injected by nodes 1 and 2 respectively (refer to the definition of network codes). This implies that a global encoding function $\mathsf{g}_{1 \rightarrow 2}(\bm{X}_{\mathcal{S}},Z_1)$ can be written for the link $1 \rightarrow 2$.

We see that as $Z_2$ is independent of all $\{\bm{X}_{\mathcal{S}}, Z_1\}$, we have the following Markov chain:
\ifx\doublecolumn\undefined
% ==== put single-column equations here
\begin{equation}
  \bm{X}_{\mathcal{W}_i} - (\mathsf{g}_{1 \rightarrow 2}(\bm{X}_{\mathcal{S}},Z_1), \{\mathsf{e}_{s_j \rightarrow t_i}(X_j):j \in \hatcal{H}_i\}) - (\mathsf{e}_{2 \rightarrow t_i}(\mathsf{g}_{1 \rightarrow 2}(\bm{X}_{\mathcal{S}},Z_1)
  ,Z_2), \{\mathsf{e}_{s_j \rightarrow t_i}(X_j):j \in \hatcal{H}_i\}),
\end{equation}
\else
% ==== put double-column equations here
\begin{multline}
  \bm{X}_{\mathcal{W}_i} - (\mathsf{g}_{1 \rightarrow 2}(\bm{X}_{\mathcal{S}},Z_1), \{\mathsf{e}_{s_j \rightarrow t_i}(X_j):j \in \hatcal{H}_i\})\\ - (\mathsf{e}_{2 \rightarrow t_i}(\mathsf{g}_{1 \rightarrow 2}(\bm{X}_{\mathcal{S}},Z_1)
  ,Z_2), \{\mathsf{e}_{s_j \rightarrow t_i}(X_j):j \in \hatcal{H}_i\}),
\end{multline}
\fi
for each $i \in \mathcal{S}$. Recall that $\mathsf{g}_e$ is the global encoding function of $\mathsf{e}_e$. 
By data-processing inequality, the probability of decoding error $P_\text{e}$ cannot increase if we replace $X_{2 \rightarrow t_i}$ with $X_{1 \rightarrow 2}$ in each receiver $t_i$'s observations. Also, by definition, none of the links $\{2 \rightarrow t_i: i \in \hatcal{T}\}$ can be accessed by any eavesdropper. Consequently, for any network code $\mathbb{N}$ (mapped from an index code $\mathbb{I}$) that is $(\hatcal{S}, (p_{\hat{X}_s}:s \in \hatcal{S}),\epsilon, \eta ,n)$-feasible, setting
\begin{equation}
X_{2 \rightarrow t_i} = X_{1 \rightarrow 2},\quad \text{ for all } i \in \hatcal{T}, \label{eq:set-equal-from-1}
\end{equation}
will result in another $(\hatcal{S}, (p_{\hat{X}_s}:s \in \hatcal{S}),\epsilon, \eta ,n)$-feasible network code for $\mathbb{N}$. This is because doing so can only improve decodability, and will not affect security. Without loss of rate performance, for the remaining of this section, we will consider network codes only of the form \eqref{eq:set-equal-from-1}.

With this, we now construct the required secure index code $(\hat{\mathsf{e}}, \hat{\mathsf{D}})$.
The construction is the same by El Rouayheb et al.\ and Effros et al., except for a modification that allows the network code to be randomised, where the randomness is restricted to node 1 (manifested in $Z_1$). For this code construction, an equivalence under the decoding criterion has been proven for deterministic codes. Here, we will prove an equivalence under both decoding and security criteria for randomised codes.

The index code is chosen as follows:
\begin{itemize}
\item Select  $\hat{Z}$, such that $(\hat{\bm{X}}_{\hatcal{S}},\hat{Z}) \stackrel{\text{d}}{=} (\bm{X}_\mathcal{S},Z_1)$.
%  \item Define $\mathcal{U}' = \{s_i: i \in \mathcal{S}\}$ to be the source vertices in $\mathbb{N}$, and generate $(\hat{\bm{Z}}_\mathcal{U}', \hat{Z}_1)$ at the sender.
\item Set the sender's transmitted code to be $\hat{\mathsf{e}}(\cdot) = \mathsf{g}_{1 \rightarrow 2}(\hat{\bm{X}}_{\hatcal{S}}, \hat{Z}) \in [2^n]$.
  \item Set the decoding function of receiver $i \in \hatcal{T}$ to be  $\hat{\mathsf{d}}_i = \mathsf{d}_{t_i} (\hat{X}_{1 \rightarrow 2},(\mathsf{e}_{s_j \rightarrow t_i}(\hat{X}_j): j \in \hatcal{H}_i))$. This is feasible since receiver~$i$ observes $\hat{\mathsf{e}}(\cdot) = \hat{X}_{1 \rightarrow 2}$ from the sender and has side information $\bm{\hat{X}}_{\hatcal{H}_i}$.
\end{itemize}

\subsection{Decoding criteria}

For the network-coding instance $\mathbb{N}$, where each receiver $t_i$ tries to decode $\bm{X}_{\hatcal{W}_i}$ from $\mathsf{g}_{1 \rightarrow 2}(\bm{X}_{\mathcal{S}},Z_1)$ and $\{\mathsf{e}_{s_j \rightarrow t_i}(X_j):j \in \hatcal{H}_i\}$, we have $P_{\text{e}} \leq \epsilon$. For the index-coding instance $\mathbb{I}$, since each receiver~$i$ tries to decode $\hat{\bm{X}}_{\hatcal{W}_i}$ from $\hat{X}_\text{b} = \mathsf{g}_{1 \rightarrow 2}(\hat{\bm{X}}_{\hatcal{S}},\hat{Z})$ and $\{\mathsf{e}_{s_j \rightarrow t_i}(\hat{X}_j):j \in \hatcal{H}_i\}$, and $(\hat{\bm{X}}_{\hatcal{S}},\hat{Z}) \stackrel{\text{d}}{=} (\bm{X}_{\mathcal{S}},Z_1)$, we must have $\hat{P}_\text{e} \leq \epsilon$.

\subsection{Security criteria}

From the security condition of $\mathbb{N}$, we have $I(\bm{X}_{\mathcal{A}_r}; \bm{X}_{\mathcal{B}_r}) < \eta$, where $\mathcal{B}_r = \{ (1 \rightarrow 2), \{\outset{s_i} : i \in \hatcal{B}_r \}\}$ are the indices of all outgoing links from sources nodes $\{ s_i: i \in \hatcal{B}_r \}$ plus the link $1 \rightarrow 2$, which are observed by the eavesdropper~$r$, . $\mathcal{A}_r = \hatcal{A}_r$ are the indices of the messages that eavesdropper~$r$ wants to obtain.

Showing that the index code also satisfy a similar security condition is not trivial, as the eavesdroppers in $\mathbb{I}$ can access the messages themselves, instead of just functions of the messages as in $\mathbb{N}$. Note that these functions may not necessarily allow one to recover the messages, as we allow non-zero error decoding probability.   So, it seems that the eavesdroppers in $\mathbb{I}$ have ``better'' observations, which may lead to a larger leakage in the code.

We will show that this is not the case. First, note the following:
\textit{(a)} $\{\bm{X}_\mathcal{S},Z_1\}$ are mutually independent;
\textit{(b)} $\bm{X}_{\outset{s_i}}$, for each $i \in \mathcal{S}$, are each a deterministic function of $X_i$;
\textit{(c)} $\hatcal{B}_r \cap \mathcal{A}_r = \emptyset$.
With these, we have the following Markov chain for every $r$:
\begin{equation}
  \bm{X}_{\hatcal{B}_r} - \bm{X}_{\{\outset{s_i} : i \in \hatcal{B}_r \}} - (Z_1, \bm{X}_{\mathcal{A}_r}, \bm{X}_{\mathcal{S} \setminus (\mathcal{A}_r \cup \hatcal{B}_r) }),
  \end{equation}
  which is equivalent to
\ifx\doublecolumn\undefined
  % ==== put single-column equations here
  \begin{subequations}
    \begin{align}
      0 &= I(\bm{X}_{\hatcal{B}_r} ; Z_1, \bm{X}_{\mathcal{A}_r}, \bm{X}_{\mathcal{S} \setminus (\mathcal{A}_r \cup \hatcal{B}_r)} | \bm{X}_{\{\outset{s_i} : i \in \hatcal{B}_r \}}) \\
        &= I(\bm{X}_{\hatcal{B}_r} ; Z_1, \bm{X}_{\mathcal{A}_r}, \bm{X}_{\mathcal{S} \setminus (\mathcal{A}_r \cup \hatcal{B}_r)},\bm{X}_{\{\outset{s_i} : i \in \hatcal{B}_r \}} | \bm{X}_{\{\outset{s_i} : i \in \hatcal{B}_r \}})\\
        &= I(\bm{X}_{\hatcal{B}_r} ; Z_1, \bm{X}_{\mathcal{S} \setminus \hatcal{B}_r}, \bm{X}_{\{\outset{s_i} : i \in \hatcal{B}_r \}}, X_{\{ s_i \rightarrow 1: i \in \mathcal{S}\}} | \bm{X}_{\{\outset{s_i} : i \in \hatcal{B}_r \}})\\
        &= I(\bm{X}_{\hatcal{B}_r} ; Z_1,  \bm{X}_{\mathcal{S} \setminus \hatcal{B}_r}, \bm{X}_{\{\outset{s_i} : i \in \hatcal{B}_r \}}, X_{\{ s_i \rightarrow 1: i \in \mathcal{S}\}}, X_{1 \rightarrow 2} | \bm{X}_{\{\outset{s_i} : i \in \hatcal{B}_r \}})\\
      %  & \geq I(\bm{X}_{\hatcal{B}_r} ; Z_1, \bm{X}_{\mathcal{A}_r}, \bm{X}_{\mathcal{S} \setminus (\mathcal{A}_r \cup \hatcal{B}_r)}, \bm{X}_{\{\outset{s_i} : i \in \hatcal{B}_r \}}, X_{\{ s_i \rightarrow 1: i \in \mathcal{S}\}} | \bm{X}_{\{\outset{s_i} : i \in \hatcal{B}_r \}}, X_{1 \rightarrow 2})\\
      & \geq I(\bm{X}_{\hatcal{B}_r} ; \bm{X}_{\mathcal{A}_r},  X_{1 \rightarrow 2} | \bm{X}_{\{\outset{s_i} : i \in \hatcal{B}_r \}})\\
        & \geq I(\bm{X}_{\hatcal{B}_r} ;  \bm{X}_{\mathcal{A}_r} | \bm{X}_{\{\outset{s_i} : i \in \hatcal{B}_r \}}, X_{1 \rightarrow 2})\\
      &= I(\bm{X}_{\hatcal{B}_r};\bm{X}_{\mathcal{A}_r}| \bm{X}_{\mathcal{B}_r})   \geq 0.
    \end{align}
  \end{subequations}
\else
% ==== put double-column equations here
\begin{subequations}
    \begin{align}
      0 &= I(\bm{X}_{\hatcal{B}_r} ; Z_1, \bm{X}_{\mathcal{A}_r}, \bm{X}_{\mathcal{S} \setminus (\mathcal{A}_r \cup \hatcal{B}_r)} | \bm{X}_{\{\outset{s_i} : i \in \hatcal{B}_r \}}) \\
        &= I(\bm{X}_{\hatcal{B}_r} ; Z_1, \bm{X}_{\mathcal{A}_r}, \bm{X}_{\mathcal{S} \setminus (\mathcal{A}_r \cup \hatcal{B}_r)},\bm{X}_{\{\outset{s_i} : i \in \hatcal{B}_r \}} \nonumber \\ &\quad \quad | \bm{X}_{\{\outset{s_i} : i \in \hatcal{B}_r \}})\\
        &= I(\bm{X}_{\hatcal{B}_r} ; Z_1, \bm{X}_{\mathcal{S} \setminus \hatcal{B}_r}, \bm{X}_{\{\outset{s_i} : i \in \hatcal{B}_r \}}, X_{\{ s_i \rightarrow 1: i \in \mathcal{S}\}} \nonumber \\ &\quad \quad | \bm{X}_{\{\outset{s_i} : i \in \hatcal{B}_r \}})\\
        &= I(\bm{X}_{\hatcal{B}_r} ; Z_1,  \bm{X}_{\mathcal{S} \setminus \hatcal{B}_r}, \bm{X}_{\{\outset{s_i} : i \in \hatcal{B}_r \}}, X_{\{ s_i \rightarrow 1: i \in \mathcal{S}\}}, X_{1 \rightarrow 2} \nonumber \\ &\quad \quad | \bm{X}_{\{\outset{s_i} : i \in \hatcal{B}_r \}})\\
      %  & \geq I(\bm{X}_{\hatcal{B}_r} ; Z_1, \bm{X}_{\mathcal{A}_r}, \bm{X}_{\mathcal{S} \setminus (\mathcal{A}_r \cup \hatcal{B}_r)}, \bm{X}_{\{\outset{s_i} : i \in \hatcal{B}_r \}}, X_{\{ s_i \rightarrow 1: i \in \mathcal{S}\}} | \bm{X}_{\{\outset{s_i} : i \in \hatcal{B}_r \}}, X_{1 \rightarrow 2})\\
      & \geq I(\bm{X}_{\hatcal{B}_r} ; \bm{X}_{\mathcal{A}_r},  X_{1 \rightarrow 2} | \bm{X}_{\{\outset{s_i} : i \in \hatcal{B}_r \}})\\
        & \geq I(\bm{X}_{\hatcal{B}_r} ;  \bm{X}_{\mathcal{A}_r} | \bm{X}_{\{\outset{s_i} : i \in \hatcal{B}_r \}}, X_{1 \rightarrow 2})\\
      &= I(\bm{X}_{\hatcal{B}_r};\bm{X}_{\mathcal{A}_r}| \bm{X}_{\mathcal{B}_r})   \geq 0.
    \end{align}
  \end{subequations}
\fi
This means that eavesdropper~$r$, having observed the links $\bm{X}_{\mathcal{B}_r}$, does not gain any more information about $\bm{X}_{\hatcal{A}_r}$ even if it can also observe the sources messages $\bm{X}_{\hatcal{B}_r}$. Now, we show that the eavesdropper cannot do better if we replace its observation of the outgoing links from the sources with the source messages:
\begin{subequations}
  \begin{align}
    &I(\bm{X}_{\hatcal{B}_r},X_{1 \rightarrow 2};\bm{X}_{\mathcal{A}_r}) \nonumber\\
&= I(\bm{X}_{\hatcal{B}_r},X_{1 \rightarrow 2},\bm{X}_{\{\outset{s_i} : i \in \hatcal{B}_r \}};\bm{X}_{\mathcal{A}_r})\\
    &= I(\bm{X}_{\hatcal{B}_r},\bm{X}_{\mathcal{B}_r};\bm{X}_{\mathcal{A}_r})\\
                                                    &= I(\bm{X}_{\mathcal{B}_r};\bm{X}_{\mathcal{A}_r}) + I(\bm{X}_{\hatcal{B}_r};\bm{X}_{\mathcal{A}_r}| \bm{X}_{\mathcal{B}_r}) \\
                                                    &= I(\bm{X}_{\mathcal{B}_r};\bm{X}_{\mathcal{A}_r})\\
    &\leq \eta.                                                      
  \end{align}
\end{subequations}

Since we set $\hat{X}_\text{b} = \mathsf{g}_{1 \rightarrow 2}(\hat{\bm{X}}_{\hatcal{S}}, \hat{Z})$, we have $(\hat{\bm{X}}_{\hatcal{S}}, \hat{Z},\hat{X}_\text{b}) \stackrel{\text{d}}{=} (\bm{X}_{\mathcal{S}},Z_1,X_{1 \rightarrow 2})$. Also, by definition, $\hatcal{A}_r = \mathcal{A}_r$. So,  $I(\hat{\bm{X}}_{\hatcal{B}_r}, \hat{X}_\text{b};\hat{\bm{X}}_{\hatcal{A}_r}) \leq \eta$ for $\mathbb{I}$.
This shows that the index code $(\hat{\mathsf{e}},\hat{\mathsf{D}})$ is $(\hatcal{S}, (p_{\hat{X}_s}:s \in \hatcal{S}),\epsilon, \eta ,n)$-feasible. \hfill
$\blacksquare$
% \end{IEEEproof}

\section{Mapping from Secure Network Coding to Secure Index Coding}

\subsection{Network-to-index coding mapping}
\label{sec:n-2-i}
In the other direction, consider a secure network-coding instance $\mathbb{N} = ( G, C, W)$. Let $\mathcal{S} = [S]$ and $\mathcal{V} = [V]$. Without loss of generality, we assume that each message is requested by at least one destination. Otherwise, it can be removed from the system without affecting decodability and security.

To map $\mathbb{N}$ to an index-coding instance $\mathbb{I} = (\hatcal{{S}}, \hatcal{{T}}, \{(\hatcal{{W}}_t,\hatcal{{H}}_t): t \in \hatcal{T}\}, \hat{W})$, we perform the following steps:
\begin{itemize}
\item We first construct an augmented secure network-coding instance $\mathbb{N}'$ from any (possibly randomised) secure network-coding instance $\mathbb{N}$.\footnote{We will see later that this step is required for the code mapping.}
\item We then following the mapping by Effros et al.\ to obtain $\hatcal{{S}}, \hatcal{{T}}, \{(\hatcal{{W}}_t,\hatcal{{H}}_t): t \in \hatcal{T} \}$ from $\mathbb{N}'$, except that we omitting one receiver in $\hatcal{T}$. We will show that omitting this receiver will not affect the result.
  \item We will propose a mapping for the eavesdroppers to get $\hat{W}$.
\end{itemize}

For $\mathbb{I}$, we set
\begin{equation}
  \hat{n} = \sum_{e \in \mathcal{E}} \lfloor c_e n \rfloor \label{eq:normalisation-nc}
\end{equation}
This means the number of bit that the sender can transmit in $\mathbb{I}$ equals the total number of bits that can be transmitted on all the edges in $\mathbb{N}$.

% Given a secure network-coding instance $\mathbb{N} = ( G, C, W)$, we first construct an augmented secure network-coding instance with deterministic encoding, and then construct an equivalent  secure index-coding instance $\mathbb{I} = (\hatcal{{S}}, \hatcal{{T}}, \{\hatcal{{W}}_{\hat{t}}\}, \{ \hatcal{{H}}_{\hat{t}} \}, \hat{W} )$.
% For any choice of $n$ for  $\mathbb{N}$, we define%we normalise the edge capacities in $\mathbb{N}$---meaning that we scale each $c_e$ by the same factor---such that

% \begin{equation}
%   \hat{n} = \sum_{e \in \mathcal{E}} \lfloor c_e n \rfloor \label{eq:normalisation-nc}
% \end{equation}
% for $\mathbb{I}$.
% %Results for the general $\bm{c}_{\mathcal{E}}$ can then be obtained by scaling the the results in this paper by adjusting $n$ accordingly.
% This means the number of bit that the sender can transmit in $\mathbb{I}$ equals the total number of bits that can be transmitted on all the edges in $\mathbb{N}$.

Now, we describe the configuration mapping in detail:
\subsubsection{Augmented secure network coding}
We construct an \emph{augmented} secure network-coding instance $\mathbb{N}' = ( G', C', W')$ as follows:
\begin{itemize}
\item $G'= (\mathcal{V}',\mathcal{E}')=(\mathcal{V},\mathcal{E})=G$, and $c'_e = c_e$ for all $e \in \mathcal{E}'$. The vertices, the edges, and the edge capacities remain the same.
\item  The
  connection requirement is augmented as follows: $\mathcal{S}' = \mathcal{S} \cup \{S+1, S+2, \dotsc, S+V\}$, where we introduce an additional independent source $X'_{S+v}$ originating at each vertex $v \in [V]$ that takes the role of and has the same distribution as the random key $Z_v$ used in the randomised encoding at vertex~$v$ in $\mathbb{N}$. So, $O'(S+v) = v$ and $\mathcal{D}'(S+v) = \emptyset$, meaning that $X'_{S+v}$ originates at vertex~$v$, and is not requested by any vertex. Also, for any vertex $v \in [V]$ that has no outgoing edge, there is no encoding function associated with it, and we set $X'_{S+v} = \alpha$.
  For $s \in \mathcal{S}$, $O'(s) = O(s)$, and $\mathcal{D}'(s) = \mathcal{D}(s)$.  %Note that $R'_s = R_s$ for all $s \in [S]$, and $R'_{S+v} = k_v$ for all $v \in [V]$.
  
  \item $W'=W$, which means $\mathcal{R}' = \mathcal{R}$, $\mathcal{B}'_r = \mathcal{B}_r$, and $\mathcal{A}'_r = \mathcal{A}_r$. The adversarial setting remains the same. Thus, the random keys $\{X'_{S+v}: v \in [V]\}$ are neither known to the adversaries nor required to be protected.
  %\item Note that the messages $\{X_s: s \in \{S+1, \dotsc, S+|\mathcal{V}'|\}\}$ are mutually independent and are independent of all other messages, but they need not be uniformly distributed.
  \end{itemize}

  By choosing $\bm{X}'_{\mathcal{S}'} \stackrel{\text{d}}{=} (\bm{X}_{\mathcal{S}},\bm{Z}_{\mathcal{V}})$, any deterministic or randomised secure network code for $\mathbb{N}$ is equivalent to a
  deterministic secure network code for $\mathbb{N}'$, where each node $v$ is assigned an
  additional source~$X'_{S+v}$ that is not required to be decoded by any node.
Note that for vertices $v \in [V]$ that has no outgoing edge, we set $Z_v = \alpha$.
  
%  Since $\{Z_v\}$ and $\{X_{S+v}\}$ play the role of random keys for security in secure network coding, we assume that they are each uniformly distributed over an alphbet of size $2^{k_v n}$ respecively, where $2^{k_v n} \leq \prod_{e \in \outset{v}} 2^{c_en}$. It follows that $R'_s = R_s$ for $s \in [S]$ and $R'_{S+v} = k_v$ for $v \in |\mathcal{V}'|]$.

  Denote the set of vertices in $\mathbb{N}'$ that are destinations for some source messages by $\mathcal{U}' = \{j \in \mathcal{V}': j \in \mathcal{D}'(i) \text{ for some } i \in \mathcal{S}'\}$. Note that $O'(\cdot)$ can map different source indices to one vertex, and so $O'^{-1}(j)$ returns a set of indices of messages originating at vertex~$j$.

\begin{figure*}
  \centering
  \begin{subfigure}[b]{0.19\textwidth}
    \centering
    \resizebox{\textwidth}{!}{%
\begin{tikzpicture} [node distance=5ex,
receiver/.style={circle,draw, minimum size=5ex},
point/.style={circle,inner sep=0pt}
]
\node (1) [receiver,label=above:{$X_1$}] {1};
\node (m) [point, below=of 1] {};
\node (2) [receiver,below=of m] {2};
\draw[->,>=latex, bend left] (1) to node[near start,right] (es2) {$\mathsf{e}_{e_2}(X_1,Z_1)$} node[midway] (em2) {}  (2);
\draw[->,>=latex, bend right] (1) to node[near start,left] (es1) {$\mathsf{e}_{e_1}(X_1,Z_1)$} node[midway] (em1) {}  (2);

\node (v1) [receiver, left=of 2, color=red] {\color{red}$r_1$};
\node (v2) [receiver, right=of 2,color=red] {\color{red}$r_2$};

\node (bv1) [point, below=of v1] {$X_1$};
\node (b2) [point, below=of 2] {$X_1$};
\node (bv2) [point, below=of v2] {$X_1$};

\draw[->,>=latex, color=red, dotted] (v1) to (bv1);
\draw[->,>=latex] (2) to (b2);
\draw[->,>=latex, color=red, dashed] (v2) to (bv2);
\draw[->,>=latex, color=red, dashed, bend right] (em1) to (v1);
\draw[->,>=latex, color=red, dashed, bend left] (em2) to (v2);
\end{tikzpicture}
}%
        \caption{$\mathbb{N}$ with randomised encoding}
        \label{fig:network-coding-1}
      \end{subfigure}
      \quad \quad
  \begin{subfigure}[b]{0.19\textwidth}
    \centering
    \resizebox{\textwidth}{!}{%
\begin{tikzpicture} [node distance=5ex,
receiver/.style={circle,draw, minimum size=5ex},
point/.style={circle,inner sep=0pt}
]
\node (1) [receiver,label=above:{$X'_1,X'_2$}] {1};
\node (m) [point, below=of 1] {};
\node (2) [receiver, label=left:{$X'_3\!$}, below=of m] {2};
\draw[->,>=latex, bend left] (1) to node[near start,right] (es2) {$\mathsf{e}'_{e_2}(X'_1,X'_2)$} node[midway] (em2) {}  (2);
\draw[->,>=latex, bend right] (1) to node[near start,left] (es1) {$\mathsf{e}'_{e_1}(X'_1,X'_2)$} node[midway] (em1) {}  (2);

\node (v1) [receiver, left=of 2, color=red] {\color{red}$r'_1$};
\node (v2) [receiver, right=of 2,color=red] {\color{red}$r'_2$};

\node (bv1) [point, below=of v1] {$X'_1$};
\node (b2) [point, below=of 2] {$X'_1$};
\node (bv2) [point, below=of v2] {$X'_1$};

\draw[->,>=latex, color=red, dashed] (v1) to (bv1);
\draw[->,>=latex] (2) to (b2);
\draw[->,>=latex, color=red, dashed] (v2) to (bv2);
\draw[->,>=latex, color=red, dashed, bend right] (em1) to (v1);
\draw[->,>=latex, color=red, dashed, bend left] (em2) to (v2);
\end{tikzpicture}
}
        \caption{$\mathbb{N}'$ with deterministic encoding, where $X'_3=\alpha$}
        \label{fig:network-coding-2}
      \end{subfigure}
      \quad
    \begin{subfigure}[b]{0.52\textwidth}
\centering
\resizebox{\textwidth}{!}{%
\begin{tikzpicture} [node distance=2.5ex,
receiver/.style={circle,draw},
point/.style={circle,inner sep=0pt}
]
\node (sender) [draw,rectangle] {Sender};
\node (message) [above=of sender] {$\hat{X}_1,\hat{X}_2,\hat{X}_{e_1},\hat{X}_{e_2}, (\hat{X}_3 = \alpha)$};
\node (joint2) [point,below=of sender,label=left:{$\hat{X}_{\text{b}}$}] {};
\node (joint) [point,below=of joint2] {};
\node (2) [receiver,below=of joint] {$\hat{t}_2$};
\node (k2) [left=of 2] {$\hat{X}_{e_1},\hat{X}_{e_2}$};
\node (1) [receiver,left=of k2] {$\hat{t}_{e_2}$};
\node (k1) [left=of 1] {$\hat{X}_1,\hat{X}_2$};
\coordinate (joint-coor) at (joint);
\node (k3) [right=of 2] {$\hat{X}_{e_1}$};
\node (3) [receiver,right=of k3,color=red] {\color{red}$\hat{r}_1$};
\node (k4) [right=of 3] {$\hat{X}_{e_2}$};
\node (4) [receiver,right=of k4,color=red] {\color{red}$\hat{r}_2$};
\node (t3) [point,above=of 3] {};
\coordinate (t3-coor) at (t3);
\node (t4) [point,above=of 4] {};
\coordinate (t4-coor) at (t4);
\node (t1) [point,above=of 1] {};
\coordinate (t1-coor) at (t1);
\node (b1) [below=of 1] {$\hat{X}_{e_2}$};
\node (b2) [below=of 2] {$\hat{X}_1$};
\node (b3) [below=of 3] {$\hat{X}_1$};
\node (b4) [below=of 4] {$\hat{X}_1$};
\node (0) [receiver,left=of k1] {$\hat{t}_{e_1}$};
\node (k0) [left=of 0] {$\hat{X}_1,\hat{X}_2$};
\node (b0) [point, below=of 0] {$\hat{X}_{e_1}$};
\node (t0) [point,above=of 0] {};
\coordinate (t0-coor) at (t0);
%\node (a1) [right=of sender] {};
%\node (a2) [right=of a1, cloud,draw, aspect=4, cloud puffs=15, inner sep=0, fill=pink] {Eavesdropper};

\draw[->,>=latex] (message) -- (sender);
\draw[->,>=latex] (sender) -- (2);
\draw[->,>=latex] (k2) -- (2);
\draw[->,>=latex] (k1) -- (1);
\draw[->,>=latex] (k0) -- (0);
\draw (joint-coor) -- (t0-coor);
\draw[->,>=latex]  (t1-coor) -- (1);
\draw[->,>=latex]  (t0-coor) -- (0);
\draw[->,>=latex,color=red,dashed] (k3) -- (3);
\draw[->,>=latex,color=red,dashed] (k4) -- (4);
\draw[->,>=latex,color=red,dashed] (t3-coor) -| (4);
%\draw[->,>=latex] (t4-coor) -- (4);
\draw[->,>=latex,color=red,dashed] (joint-coor) -| (3);
\draw[->,>=latex] (1) -- (b1);
\draw[->,>=latex] (2) -- (b2);
\draw[->,>=latex,color=red,dashed] (3) -- (b3);
\draw[->,>=latex,color=red,dashed] (4) -- (b4);
\draw[->,>=latex] (0) -- (b0);
\end{tikzpicture}
}%
        \caption{$\mathbb{I}$ with deterministic encoding}
        \label{fig:index-coding-2}
      \end{subfigure}
      \quad
 
      \caption{A secure network-coding instance ${I}$, its augmented
        version $\mathbb{N}'$, and the corresponding secure index-coding instance
        $\mathbb{I}$, where $r_1, r_2, r'_1, r'_2, \hat{r}_1, \hat{r}_2$ are
        eavesdroppers}
      %\vspace{-2ex}
      \label{fig:network-to-index-example}
\end{figure*}
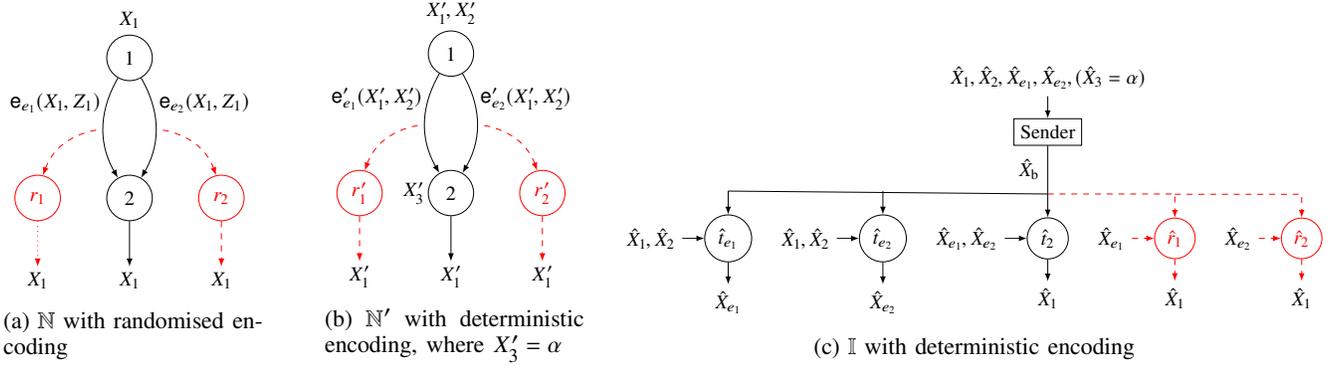

\subsubsection{Network-to-index coding mapping}
Now, we map $\mathbb{N}'$ to a secure index-coding instance $\mathbb{I}$.
\begin{itemize}
\item $\hatcal{{S}} = \mathcal{S}' \cup \mathcal{E}'$. It consists of one message $\hat{X}_s$ for each  $s \in \mathcal{S}'$ in $\mathbb{N}'$, and one $\hat{X}_e$ for each $e \in \mathcal{E}'$ in $\mathbb{N}'$. 
\item $\hatcal{{T}} = \{\hat{t}_i\}_{i \in \mathcal{U}'} \cup \{\hat{t}_e\}_{e \in \mathcal{E}'}$. This means $\mathbb{I}$ has $|\mathcal{U}'|+|\mathcal{E}'|$ receivers: one corresponds to each destination node in $\mathbb{N}'$, and one corresponds to each edge in $\mathbb{N}'$.
\item For each $\hat{t}_e \in \hatcal{{T}}$ where $e \in \mathcal{E}'$, we set $\hatcal{{H}}_{\hat{t}_e} = \inset{\tail{e}} \cup O'^{-1}(\tail{e})$, and $\hatcal{{W}}_{\hat{t}_e} = \{ e \}$.
\item For each $\hat{t}_i \in \hatcal{{T}}$ where $i \in \mathcal{U}'$, we set  $\hatcal{{H}}_{\hat{t}_i} = \inset{i} \cup  O'^{-1}(i)$, and $\hatcal{{W}}_{\hat{t}_i} = \{s \in [S]: i \in \mathcal{D}'(s) \}$.
\item The eavesdropper setting $W'$: $\hatcal{{R}} = \mathcal{R}'$. For each $\hat{r} \in \hatcal{{R}}$, $\hatcal{B}_{\hat{r}} = \mathcal{B}'_{\hat{r}}$, and $\hatcal{A}_{\hat{r}} = \mathcal{A}'_{\hat{r}}$.
%  \item We set the broadcast rate as $\hat{c}_{\text{b}} = \sum_{e \in \mathcal{E}'}c'_e$.
  \end{itemize}

Figure~\ref{fig:network-to-index-example} depicts an example of such a mapping.

  \begin{remark}
    This network-to-index coding mapping is slightly different from that of Effros et al.~\cite{effrosrouayheblangberg15} because we do not require the use of an additional receiver $\hat{t}_\text{all}$ in $\mathbb{I}$. Briefly, receiver $\hat{t}_\text{all}$ has $\hatcal{H}_{\hat{t}_\text{all}} = \hat{\bm{X}}_{\mathcal{S}'}$ and wants $\hatcal{W}_{\hat{t}_\text{all}} = \hat{\bm{X}}_{\mathcal{E}'}$. This additional receiver was added to guarantee the following useful property: For any broadcast message $\hat{x}_\text{b} \in [2^{\hat{n}}]$ and any realisation $\hat{\bm{x}}_{\mathcal{S}'}$, if we know that all receivers can decode their requested messages correctly, then there can be only one unique realisation $\hat{\bm{x}}_{\mathcal{E}'}$ which has led to the broadcast message $\hat{x}_\text{b}$. We will show that this required property remains true even without receiver $\hat{t}_\text{all}$.
\end{remark}

Part of the result for the network-to-index coding mapping will be expressed in term of the total variation distance of probability measures. Let $p$ and $q$ be two pmfs on an finite discrete alphabet $\Omega$. The total variation distance\footnote{For two probability measures $P$ and $Q$ on a measurable space $(X,\Sigma)$, the total variation distance is defined as $\delta(P,Q) \equalbydef \sup_{A \in \Sigma} |P(A) - Q(A)|$.} between $p$ and $q$ can be expressed in $\text{L}^1$ norms as $\delta(p,q) = \frac{1}{2} \lVert p - q \rVert_1 = \frac{1}{2} \sum_{\sigma \in \Omega} | p(\sigma) - q(\sigma)|$. Also, denote the uniform distribution on a finite set $\Omega$ by $\mathtt{unif}(\Omega)$.
  
\subsection{Equivalence results}
  With the above-mentioned conversion, we now state an equivalence between $\mathbb{N}$ and $\mathbb{I}$ through $\mathbb{N}'$:
  \begin{theorem}\label{theorem:network-to-index}
    Let $\mathbb{N}$ be a secure network-coding instance, $\mathbb{N}'$ be its augmented instance, and $\mathbb{I}$ be the corresponding secure index-coding instance obtained using the network-to-index coding mapping from $\mathbb{N}'$. For any $\eta \in \mathbb{R}^+_0 $, $\epsilon \in [0,0.5]$, and $n \in \mathbb{Z}^+$, we have the following:
    \begin{enumerate}
    \item If $\mathbb{N}$, in which all messages $\bm{X}_{\mathcal{S}}$ are independent and uniformly distributed, is $(\mathcal{S}, (p_{X_s}:s \in \mathcal{S}),\epsilon, \eta ,n)$-feasible, then $\mathbb{I}$ is $(\mathcal{S}, (p_{\hat{X}_s}:s \in \mathcal{S}),\epsilon, \eta ,\hat{n})$-feasible, where $\hat{\bm{X}}_{\mathcal{S}} \stackrel{\text{d}}{=} \bm{X}_{\mathcal{S}}$.
    \item If $\mathbb{I}$, in which all messages $\hat{\bm{X}}_{\mathcal{S}' \cup \mathcal{E}'}$ are independent and uniformly distributed, is $(\mathcal{S}, (p_{\hat{X}_s}:s \in \mathcal{S}),\epsilon, \eta ,\hat{n})$-feasible, where $\hat{X}_e \in [2 ^{\lfloor c_e n \rfloor}]$  then
      \begin{enumerate}
        \item For $\epsilon=0$, $\mathbb{N}$ is $(\mathcal{S}, (p_{X_s}:s \in \mathcal{S}),\epsilon, \eta ,n)$-feasible; and
        \item Otherwise (for $0 < \epsilon \leq 0.5$), $\mathbb{N}$ is $(\mathcal{S}, (p_{X_s}:s \in \mathcal{S}), |\mathcal{R}| \eta + \zeta, \gamma ,n)$-feasible,
        \end{enumerate}
      \end{enumerate}
      where $\bm{X}_\mathcal{S} \stackrel{\text{d}}{=} \hat{\bm{X}}_\mathcal{S}$, $\zeta$ is a function of $(\epsilon,n)$, and $\gamma$ is a function of $(\zeta,\epsilon,\eta,n)$, defined as follows:
      %Here, $\hat{n} = \sum_{e \in \mathcal{E}} \lfloor c_e n \rfloor$, and\footnote{{\color{red} Note that we have improve this result since the ISIT submission.}}
      \ifx\doublecolumn\undefined
      % ==== put single-column equations here
\begin{align*}
        \gamma &\equalbydef \min \Big\{ (|\mathcal{R}| \eta + \zeta) \left( \frac{1}{1 - \epsilon} + \frac{\log e  + \hat{n}}{1-(|\mathcal{R}| \eta + \zeta)}+ \log |\mathcal{X}_{\mathcal{S}'}|  \right) \nonumber \\
      &\quad + \frac{1}{1 - \epsilon} |\mathcal{R}| H_\text{b}(\epsilon) - \log \left(1 - (|\mathcal{R}| \eta + \zeta)\right),\,\, \hat{n} \Big\}\\
  \zeta &\equalbydef \min \Big\{ \epsilon [1 + 2 \delta(p_{\hat{X}_\text{b}},\mathtt{unif}([2^{n}])) ] , \,\, \epsilon [ 1 + \epsilon 2^{\hat{n}}],\,\, 1 \Big\}.
  % \begin{cases}
  %   (1 + 2\alpha) \epsilon, & \text{ if $\delta(p_{\hat{X}_\text{b}},\mathtt{unif}([2^{n}])) \leq \alpha \epsilon$}\\
  %   \min \{\epsilon ( 1 + \epsilon 2^{\hat{n}}), 1 \}, & \text{ in general}.
  % \end{cases}
\end{align*}
\else
% ==== put double-column equations here
\begin{align*}
        \gamma &\equalbydef \min \Big\{ (|\mathcal{R}| \eta + \zeta) \left( \frac{1}{1 - \epsilon} + \frac{\log e  + \hat{n}}{1-(|\mathcal{R}| \eta + \zeta)}+ \log |\mathcal{X}_{\mathcal{S}'}|  \right) \nonumber \\
      &\quad + \frac{1}{1 - \epsilon} |\mathcal{R}| H_\text{b}(\epsilon) - \log \left(1 - (|\mathcal{R}| \eta + \zeta)\right),\,\, \hat{n} \Big\},\\
  \zeta &\equalbydef \min \Big\{ \epsilon [1 + 2 \delta(p_{\hat{X}_\text{b}},\mathtt{unif}([2^{n}])) ] , \,\, \epsilon [ 1 + \epsilon 2^{\hat{n}}],\,\, 1 \Big\}.
  % \begin{cases}
  %   (1 + 2\alpha) \epsilon, & \text{ if $\delta(p_{\hat{X}_\text{b}},\mathtt{unif}([2^{n}])) \leq \alpha \epsilon$}\\
  %   \min \{\epsilon ( 1 + \epsilon 2^{\hat{n}}), 1 \}, & \text{ in general}.
  % \end{cases}
\end{align*}
\fi
\end{theorem}

\begin{IEEEproof}
  See Sections~\ref{proof:2-1} and \ref{proof:2-2}.
\end{IEEEproof}

  Part 1 of the above theorem is proven by setting the pmfs of the rest of the messages in $\mathbb{I}$ (which are $\hat{\bm{X}}_{\hatcal{S} \setminus \mathcal{S}}$) as follows:
As mentioned above, we choose $\bm{X}'_{\mathcal{S}'} \stackrel{\text{d}}{=} (\bm{X}_{\mathcal{S}},\bm{Z}_{\mathcal{V}})$ for $\mathbb{N}'$ to get an equivalent network-coding instance. For $\mathbb{I}$, we choose $\hat{\bm{X}}_{\mathcal{S}'} \stackrel{\text{d}}{=} \bm{X}'_{\mathcal{S}'} \stackrel{\text{d}}{=} (\bm{X}_{\mathcal{S}},\bm{Z}_{\mathcal{V}})$, and each $\hat{X}_e$, $e \in \mathcal{E}'$, to be uniformly distributed over $[2^{\lfloor c'_e n \rfloor}]$. We will see that using uniformly distributed $\hat{X}_e$ is the key to ensuring security.
Note that unlike the index-to-network mapping, here $\hat{\bm{X}}_{\mathcal{E}'}$ and $\bm{X}'_{\mathcal{E}'}$ have different distributions. $\{\hat{X}_i: i \in \mathcal{S}' \cup \mathcal{E}'\}$ in $\mathbb{I}$ are mutually independent, while $\{X'_i: i \in \mathcal{E}'\}$ in $\mathbb{N}'$ are functions of $\bm{X}'_{\mathcal{S}'}$ and may be correlated.\footnote{This property is also true in the mapping of Effros et al.}

In Part 2b of Theorem~\ref{theorem:network-to-index}, the upper bounds on decoding error and leakage increase exponentially with $n$. We can tighten the bounds for linear codes:

\begin{corollary} \label{cor:linear-codes}
  Let $\mathbb{N}$ be a secure network-coding instance and $\mathbb{I}$  be the corresponding secure index-coding instance obtained using the network-to-index coding mapping. For any $\eta \in \mathbb{R}^+_0 $, $\epsilon \in (0,0.5]$, and $n \in \mathbb{Z}^+$, we have the following: If $\mathbb{I}$ is $(\mathcal{S}, (p_{\hat{X}_s}:s \in \mathcal{S}),\epsilon, \eta ,\hat{n})$-feasible using a \textbf{linear} index code with cardinality $2^{\hat{n}}$, where $\hat{\bm{X}}_{\mathcal{S}' \cup \mathcal{E}'}$ are independent and uniformly distributed, then $\mathbb{N}$ is $(\mathcal{S}, (p_{X_s}:s \in \mathcal{S}), |\mathcal{R}| \eta + \epsilon, \gamma' ,n)$-feasible, where % $\hat{n} = \sum_{e \in \mathcal{E}} \lfloor c_e n \rfloor$, and
  \ifx\doublecolumn\undefined
  % ==== put single-column equations here
\begin{align*}
        \gamma' &\equalbydef \min \Big\{ (|\mathcal{R}| \eta + \epsilon) \left( \frac{1}{1 - \epsilon} + \frac{\log e  + \hat{n}}{1-(|\mathcal{R}| \eta + \epsilon)}+ \log |\mathcal{X}_{\mathcal{S}'}|  \right) \nonumber \\
      &\quad + \frac{1}{1 - \epsilon} |\mathcal{R}| H_\text{b}(\epsilon) - \log \left(1 - (|\mathcal{R}| \eta + \epsilon)\right), \,\, \hat{n} \Big\}.
\end{align*}
\else
% ==== put double-column equations here
\begin{align*}
        \gamma' &\equalbydef \min \Big\{ (|\mathcal{R}| \eta + \epsilon) \left( \frac{1}{1 - \epsilon} + \frac{\log e  + \hat{n}}{1-(|\mathcal{R}| \eta + \epsilon)}+ \log |\mathcal{X}_{\mathcal{S}'}|  \right) \nonumber \\
      &\quad + \frac{1}{1 - \epsilon} |\mathcal{R}| H_\text{b}(\epsilon) - \log \left(1 - (|\mathcal{R}| \eta + \epsilon)\right), \,\, \hat{n} \Big\}.
\end{align*}
\fi
\end{corollary}

Note here that, for linear codes, the error probability for $\mathbb{N}$ is independent of $n$, and is solely a function of $\epsilon$, $\eta$, and the number of eavesdroppers $|\mathcal{R}|$; the leakage for $\mathbb{N}$ is a linear function of $n$, and the coefficient of $n$ can be made arbitrarily small by choosing arbitrarily small $\eta$ and $\epsilon$. This means a sequence of strongly-secure index codes for $\mathbb{I}$ translates to a sequence of weakly-secure network codes for $\mathbb{N}$ (with appropriate rate scaling).

\begin{IEEEproof}[Proof of Corollary~\ref{cor:linear-codes}]
  Using linear codes for $\mathbb{I}$, if the messages are uniformly distributed, then the codeword $\hat{X}_\text{b}$ is uniformly distributed over its support. So, $\delta(p_{\hat{X}_\text{b}},\mathtt{unif}([2^{n}]))=0$, which implies $\zeta = \epsilon$,  and Corollary~\ref{cor:linear-codes} follows directly from Part~2b of Theorem~\ref{theorem:network-to-index}.
\end{IEEEproof}

  \section{Proof of Theorem~\ref{theorem:network-to-index} -- Part 1 (the forward direction)} \label{proof:2-1}
    We will now prove Part 1 in Theorem~\ref{theorem:network-to-index}, that is $\mathbb{N}$ is $(\mathcal{S}, (p_{X_s}:s \in \mathcal{S}),\epsilon, \eta ,n$-feasible  $\Rightarrow$ $\mathbb{I}$ is   $(\mathcal{S}, (p_{\hat{X}_s}:s \in \mathcal{S}),\epsilon, \eta ,\hat{n})$-feasible.

\subsection{Code construction}

   First, note that $\mathbb{N}$ is $(\mathcal{S}, (p_{X_s}:s \in \mathcal{S}),\epsilon, \eta ,n)$-feasible for $p_{\bm{X}_{\mathcal{S}}}$  if and only if $\mathbb{N}'$ is $(\mathcal{S}, (p_{X'_s}:s \in \mathcal{S}),\epsilon, \eta ,n)$--feasible with $\bm{X}'_{\mathcal{S}'} \stackrel{\text{d}}{=} (\bm{X}_{\mathcal{S}},\bm{Z}_{\mathcal{V}})$ for some $\bm{Z}_{\mathcal{V}}$ using \textit{deterministic} network encoding functions $\{\mathsf{e}'_e\}$ derived from $\{\mathsf{e}_e\}$ for $\mathbb{N}$, where all the randomness $\{Z_v: v \in \mathcal{V}\}$ in the network code for $\mathbb{N}$ is realised using $\{X'_{S+v}: v \in \mathcal{V}\}$ in $\mathbb{N}'$.

   Since the network code for $\mathbb{N}'$ is deterministic, we use the same code mapping as that proposed by Effros et al.~\cite{effrosrouayheblangberg15}: The sender broadcasts $\hat{X}_{\text{b}} = [\hat{X}_{\text{b},e}: e \in \mathcal{E}']  $, where
   \begin{equation}
     \hat{X}_{\text{b}, e} = \hat{X}_e + \mathsf{g}'_e(\hat{\bm{X}}_{\mathcal{S}'}) \mod 2^{\lfloor c_e n \rfloor}. \label{eq:edge-code}
   \end{equation}
   Note that each $\hat{X}_e, \mathsf{g}'_e \in [2^{\lfloor c'_e n \rfloor}] = [2^{\lfloor c_e n \rfloor}]$, and therefore $\hat{X}_\text{b} \in [ \prod_{e \in \mathcal{E}'} 2^{\lfloor c_e n \rfloor} ] = [ 2^{\sum_{e \in \mathcal{E}'}\lfloor c_e n \rfloor} ] = [2^{\hat{n}}]$. 

\subsection{Decoding criteria}   

In $\mathbb{N}$, according to definition~\eqref{eq:network-coding-decoding-prob}, with probability of at least $(1-\epsilon)$ (over the messages $p_{\bm{X}_\mathcal{S}}$),  every  vertex~$v \in \mathcal{U}'$ can decode all messages that it requires from the message on all incoming edges and messages originating at $v$. Since, only messages $\bm{X}_\mathcal{S}$ of all messages $\bm{X}_{\mathcal{S}'}$ in $\mathbb{N}'$ need to be decoded, it follows that, in $\mathbb{N}'$, with probability of at least $(1-\epsilon)$, every $v \in \mathcal{U}'$ satisfies the following:
\ifx\doublecolumn\undefined
% ==== put single-column equations here
\begin{align}
     \Pr \Big\{ \bm{X}'_{\{s \in \mathcal{S}': v \in \mathcal{D}'(s)\}} &= \bm{X}'_{\{s \in [S]: v \in \mathcal{D}'(s)\}} = \mathsf{d}'_v(\bm{X}'_{\inset{v} \cup O'^{-1}(v)}) \Big\} \geq 1 - \epsilon,  \label{eq:nc-ic-different-pmf} \\
     \intertext{or equivalently,}
     \Pr \Big\{ \bm{X}'_{\{s \in \mathcal{S}': v \in \mathcal{D}'(s)\}} &= \bm{X}'_{\{s \in [S]: v \in \mathcal{D}'(s)\}} = \mathsf{d}'_v( [\mathsf{g}'_e(\bm{X}'_{\mathcal{S}'})]_{e \in \inset{v})}, \bm{X}'_{O'^{-1}(v)}) \Big\} \geq 1 - \epsilon. \label{eq:nc-ic-different-pmf-2}
   \end{align}
\else
% ==== put double-column equations here
\begin{align}
     &\Pr \Big\{ \bm{X}'_{\{s \in \mathcal{S}': v \in \mathcal{D}'(s)\}} = \bm{X}'_{\{s \in [S]: v \in \mathcal{D}'(s)\}} = \mathsf{d}'_v(\bm{X}'_{\inset{v} \cup O'^{-1}(v)}) \Big\}\nonumber \\ & \quad \geq 1 - \epsilon,  \label{eq:nc-ic-different-pmf} \\
     \intertext{or equivalently,}
     &\Pr \Big\{ \bm{X}'_{\{s \in \mathcal{S}': v \in \mathcal{D}'(s)\}} = \bm{X}'_{\{s \in [S]: v \in \mathcal{D}'(s)\}} \nonumber \\ & \quad  = \mathsf{d}'_v( [\mathsf{g}'_e(\bm{X}'_{\mathcal{S}'})]_{e \in \inset{v})}, \bm{X}'_{O'^{-1}(v)}) \Big\} \geq 1 - \epsilon. \label{eq:nc-ic-different-pmf-2}
   \end{align}
\fi

We first consider receivers $\hat{t}_i \in \hatcal{{T}}$ where $i \in \mathcal{U}'$: 
 As mentioned above, while source messages $\bm{X}'_{O'^{-1}(v)}$ in $\mathbb{N}'$ and $\hat{\bm{X}}_{O'^{-1}(v)}$ in $\mathbb{I}$ have the same distribution, edge messages $\bm{X}'_{\inset{v}}$ in $\mathbb{N}'$ and $\hat{\bm{X}}_{\inset{v}}$  $\mathbb{I}$ may not. So, though a node $\hat{i}_i \in \hat{\mathcal{T}}$ in $\mathbb{I}$ has side information $(\hat{\bm{X}}_{\inset{i}}, \hat{\bm{X}}'_{O'^{-1}(i)})$, directly porting \eqref{eq:nc-ic-different-pmf} to $\mathbb{I}$ will not work, as the pmf $(\bm{X}'_{\mathcal{S}'},\bm{X}'_{\mathcal{E}'})$ and that of  $(\hat{\bm{X}}_{\mathcal{S}'}, \hat{\bm{X}}_{\mathcal{E}'}) = \hat{\bm{X}}_{\hatcal{S}}$ are different.
To deal with this issue, consider the broadcast message $\hat{X}_{\text{b}}$. From \eqref{eq:edge-code}, any receiver that knows $\hat{X}_e$ can obtain $\mathsf{g}'_e(\hat{\bm{X}}_{\mathcal{S}'})$ from the broadcast message $\hat{X}_\text{b}$, where $(\bm{X}'_{\mathcal{S}'},[\mathsf{g}'_e(\bm{X}'_{\mathcal{S}'})]_{e \in \mathcal{E}'})$ and  $(\hat{\bm{X}}_{\mathcal{S}'},[\mathsf{g}'_e(\hat{\bm{X}}_{\mathcal{S}'})]_{e \in \mathcal{E}'})$ have the same distribution.

In $\mathbb{I}$, as $\hatcal{{H}}_{\hat{t}_i} = \inset{i} \cup  O'^{-1}(i)$ by the mapping, receiver $\hat{t}_i$ knows $\hat{\bm{X}}_{O'^{-1}(i)}$ and can obtain $[\mathsf{g}'_e(\hat{\bm{X}}_{\mathcal{S}'})]_{ e \in \inset{i})}$ from $\hat{X}_\text{b}$ and $\hat{\bm{X}}_{\inset{i}}$ using \eqref{eq:edge-code}. % This is possible if $\hat{\bm{X}}_{\mathcal{S}'}$ and $\hat{\bm{X}}_{\mathcal{S}'}$ have the same distribution.
So, using \eqref{eq:nc-ic-different-pmf-2} with a change of variables (from non-hatted to hatted), receiver $\hat{t}_i \in \hatcal{{T}}$ can decode the messages it requires correctly with probability of at least $(1-\epsilon)$ because
\ifx\doublecolumn\undefined
% ==== put single-column equations here
\begin{equation*}
     \Pr \Big\{ \hat{\bm{X}}_{\hatcal{{W}}_{\hat{t}_i}} = \hat{\bm{X}}_{\{s \in [S]: i \in \mathcal{D}'(s) \}} = \mathsf{d}'_i( [\mathsf{g}'_e(\hat{\bm{X}}_{\mathcal{S}'})]_{ e \in \inset{i})}, \hat{\bm{X}}_{O'^{-1}(i)}) \Big\} \geq 1 - \epsilon,
   \end{equation*}
\else
% ==== put double-column equations here
\begin{align*}
     &\Pr \Big\{ \hat{\bm{X}}_{\hatcal{{W}}_{\hat{t}_i}} = \hat{\bm{X}}_{\{s \in [S]: i \in \mathcal{D}'(s) \}} = \mathsf{d}'_i( [\mathsf{g}'_e(\hat{\bm{X}}_{\mathcal{S}'})]_{ e \in \inset{i})}, \hat{\bm{X}}_{O'^{-1}(i)}) \Big\} \nonumber \\ &\quad \geq 1 - \epsilon,
   \end{align*}
\fi
   because $(\bm{X}'_{\mathcal{S}'},[\mathsf{g}'_e(\bm{X}'_{\mathcal{S}'})]_{e \in \mathcal{E}'}) \stackrel{\text{d}}{=} (\hat{\bm{X}}_{\mathcal{S}'},[\mathsf{g}'_e(\hat{\bm{X}}_{\mathcal{S}'})]_{e \in \mathcal{E}'})$.
 %  This means, with probability of at least $(1-\epsilon)$, all receivers  $\hat{t}_i \in \hatcal{{T}}$ where $i \in \mathcal{U}'$ can correctly decode the messages they want.

   Now, we consider receivers $\hat{t}_e \in \hatcal{{T}}$ where $e \in \mathcal{E}'$. Recall that $\hatcal{{H}}_{\hat{t}_e} = \inset{\tail{e}} \cup O'^{-1}(\tail{e})$, and $\hatcal{{W}}_{\hat{t}_e} = \{ e \}$. Receiver $\hat{t}_e$ performs the following steps:
   \begin{enumerate}[label=(\roman*)]
   \item  As it knows $\{\hat{X}_d: d \in \inset{\tail{e}}\}$, it can obtain $\{\mathsf{g}'_{d}(\hat{\bm{X}}_{\mathcal{S}'}):d \in \inset{\tail{e}}\}$ from \eqref{eq:edge-code}.
   \item Since it also knows $\hat{\bm{X}}_{O'^{-1}(\tail{e})}$ as side information, it then calculates\\ $\mathsf{e}'_e([\mathsf{g}'_{d}(\hat{\bm{X}}_{\mathcal{S}'}):d \in \inset{\tail{e}}],\hat{\bm{X}}_{O'^{-1}(\tail{e})})$, which equals $\mathsf{g}'_e(\hat{\bm{X}}_{\mathcal{S}'})$, where $\mathsf{e}'_e$ is the local encoding function of edge $e$ in $\mathbb{N}'$.
     \item With $\mathsf{g}'_e(\hat{\bm{X}}_{\mathcal{S}'})$ and the broadcast message $\hat{X}_{\text{b},e}$, it obtains the required $\hat{X}_e$ using \eqref{eq:edge-code}.
   \end{enumerate}
 So, receiver $\hat{t}_e$ for each $e \in \mathcal{E}'$ must be able to correctly decode the required $\hat{X}_e$ without error.

   Combining these two classes of receivers, we have shown that all receivers in $\mathbb{I}$ can correctly decode their required messages with probability of at least $(1-\epsilon)$.

% Receiver $\hat{t}_e \in \hatcal{{T}}$ uses \eqref{eq:edge-code} to obtain the required $\hat{X}_e$ from $\hat{X}_{\text{b}}(e) - \mathsf{g}'_e(\hat{\bm{X}}_{\mathcal{S}'})$, where the first term is the broadcast message available to the receiver $\hat{t}_e$. To obtain the second term, express the global encoding function as its local encoding function, $\mathsf{g}'_e(\hat{\bm{X}}_{\mathcal{S}'}) = \mathsf{e}'_e([\mathsf{g}'_{e'}(\hat{\bm{X}}_{\mathcal{S}'})]_{e' \in \inset{\tail{e}}},\hat{\bm{X}}_{O'^{-1}(\tail{e})})$, where $\hat{\bm{X}}_{O'^{-1}(\tail{e})}$ is available to receiver $\hat{t}_e$ as side information. From the broadcast message, receiver $\hat{t}_e$ can obtain
% %   \begin{equation}
%      $\mathsf{g}'_{e'}(\hat{\bm{X}}_{\mathcal{S}'}) = \hat{X}_{\text{b}}(e') - \hat{X}_{e'},$
% %     \end{equation}
%      as it has $\hat{X}_{e'}$, $e' \in \inset{\tail{e}},$ as side information. With this, we have shown that each $\hat{t} \in \hatcal{{T}}$ can decode the messages that it requires.

   \subsection{Security criteria}

Given $I(\bm{X}_{\mathcal{A}_r}; \bm{X}_{\mathcal{B}_r}) \leq \eta$ for $\mathbb{N}'$, we need to show $I(\hat{\bm{X}}_{\hatcal{A}_r} ; \hat{X}_{\text{b}}, \hat{\bm{X}}_{\hatcal{B}_r}) \leq \eta$ for $\mathbb{I}$.

     We now consider the security constraints. % Note that, from the mapping, $\hatcal{B}_{\hat{r}}$ contains only edge indices. Knowing only edges messages $\{\hat{X}_e: e \in \hatcal{B}_{\hat{r}}\}$ and the broadcast message $\hat{X}_{\text{b}}$, eavesdropper~$\hat{r}$ can only obtain $\{\mathsf{g}'_e(\hat{\bm{X}}_{\mathcal{S}'}):e \in \hatcal{B}_{\hat{r}}\}$, as $\{\mathsf{g}'_{e'}(\hat{\bm{X}}_{\mathcal{S}'}): e' \notin \hatcal{B}_{\hat{r}}\}$ have been randomised by independently and uniformly distributed $\hat{X}_{e'}$. So,
     For each $\hat{r} \in \hatcal{{R}}$,
     \begin{subequations}
       \begin{align}
         & H(\hat{\bm{X}}_{\hatcal{A}_{\hat{r}}} | \hat{X}_{\text{b}}, \hat{\bm{X}}_{\hatcal{B}_{\hat{r}}})\nonumber \\
         &= H(\hat{\bm{X}}_{\hatcal{A}_{\hat{r}}} | \{\hat{X}_{\text{b},e}:e \in \mathcal{E}'\}, \{ \hat{X}_{e'}: e' \in \hatcal{B}_{\hat{r}} \}) \\
         &= H(\hat{\bm{X}}_{\hatcal{A}_{\hat{r}}} | \{\hat{X}_{\text{b},e}:e \in \hatcal{B}_{\hat{r}}\}, \{ \hat{X}_{e'}: e' \in \hatcal{B}_{\hat{r}} \}) \label{eq:blocked} \\
         &= H(\hat{\bm{X}}_{\hatcal{A}_{\hat{r}}} | \{\hat{X}_{\text{b},e}, \hat{X}_{e}, \mathsf{g}'_e(\hat{\bm{X}}_{\mathcal{S}'}):e \in \hatcal{B}_{\hat{r}}\}) \label{eq:edge-function-2} \\
         &= H(\hat{\bm{X}}_{\hatcal{A}_{\hat{r}}} | \{\hat{X}_{e}, \mathsf{g}'_e(\hat{\bm{X}}_{\mathcal{S}'}):e \in \hatcal{B}_{\hat{r}}\}) \label{eq:edge-function-3} \\
         & = H(\hat{\bm{X}}_{\hatcal{A}_{\hat{r}}} | \{ \mathsf{g}'_e(\hat{\bm{X}}_{\mathcal{S}'}):e \in \hatcal{B}_{\hat{r}}\}) \label{eq:must-be-equality} \\
         &= H(\hat{\bm{X}}_{\mathcal{A}'_{\hat{r}}} | \{ \mathsf{g}'_e(\hat{\bm{X}}_{\mathcal{S}'}):e \in \mathcal{B}'_{\hat{r}}\})\\
         &= H(\bm{X}'_{\mathcal{A}'_{\hat{r}}} | \{ \mathsf{g}'_e(\bm{X}'_{\mathcal{S}'}):e \in \mathcal{B}'_{\hat{r}}\}) \label{eq:change-var-1}\\
         &= H(\bm{X}'_{\mathcal{A}'_{\hat{r}}}|\bm{X}'_{\mathcal{B}'_{\hat{r}}}) = H(\bm{X}_{\mathcal{A}_{\hat{r}}}|\bm{X}_{\mathcal{B}_{\hat{r}}}), \label{eq:change-var-2}
       \end{align}
     \end{subequations}
    where \eqref{eq:blocked} follows from the Markov chain
    $$\hat{\bm{X}}_{\hatcal{A}_{\hat{r}}} - \left(\{\hat{X}_{\text{b},e}:e
    \in \hatcal{B}_{\hat{r}}\}, \{ \hat{X}_{e'}: e' \in
    \hatcal{B}_{\hat{r}} \}\right) - (\{\hat{X}_{\text{b},e}:e \notin
  \hatcal{B}_{\hat{r}}\}),$$
  where  $\{\hat{X}_{\text{b}}(e):e \notin \hatcal{B}_{\hat{r}}\}$ are independent of $(\hat{\bm{X}}_{\hatcal{A}_{\hat{r}}}, \{\hat{X}_{\text{b},e}:e
    \in \hatcal{B}_{\hat{r}}\}, \{ \hat{X}_{e'}: e' \in
    \hatcal{B}_{\hat{r}} \})$, because the former has been randomised by independently and uniformly distributed $\{\hat{X}_e:e \notin \hatcal{B}_{\hat{r}}\}$ (which are independent of $(\hat{\bm{X}}_{\hatcal{A}_{\hat{r}}}, \hat{\bm{X}}_{\hatcal{B}_{\hat{r}}},\hat{\bm{X}}_{\mathcal{S}'})$, see \eqref{eq:edge-code});\\
  \eqref{eq:edge-function-2} follows from \eqref{eq:edge-code};\\
  \eqref{eq:edge-function-3} is derived because $\hat{X}_{\text{b},e}$ is a deterministic function of $(\hat{X}_{e}, \mathsf{g}'_e(\hat{\bm{X}}_{\mathcal{S}'} ))$;\\
  \eqref{eq:must-be-equality} follows from the Markov chain
$$\hat{\bm{X}}_{\hatcal{A}_{\hat{r}}} - \{ \mathsf{g}'_e(\hat{\bm{X}}_{\mathcal{S}'}):e \in \hatcal{B}_{\hat{r}}\} - \{\hat{X}_{e}:e \in \hatcal{B}_{\hat{r}}\},$$
which can be derived from noting that $\{\hat{X}_e: e \in \mathcal{E}'\}$ are independent of $(\hat{\bm{X}}_{\hatcal{A}_{\hat{r}}},\hat{\bm{X}}_{\mathcal{S}'})$;\\
\eqref{eq:change-var-1} follows from a change of variables (from hatted to non-hatted);\\
\eqref{eq:change-var-2} is obtained from noting that $\{ \mathsf{g}'_e(\bm{X}'_{\mathcal{S}'}):e \in \mathcal{B}'_{\hat{r}}\} = \bm{X}'_{\mathcal{B}'_{\hat{r}}}$

Now, for $\mathbb{N}$, if $I(\bm{X}_{\mathcal{A}_r}; \bm{X}_{\mathcal{B}_r}) \leq \eta$, then
\begin{subequations}
  \begin{align}
    I(\hat{\bm{X}}_{\hatcal{A}_r} ; \hat{X}_{\text{b}}, \hat{\bm{X}}_{\hatcal{B}_r}) &= H(\hat{\bm{X}}_{\hatcal{A}_r}) - H(\hat{\bm{X}}_{\hatcal{A}_{\hat{r}}} | \hat{X}_{\text{b}}, \hat{\bm{X}}_{\hatcal{B}_{\hat{r}}}) \\
    &= H(\hat{\bm{X}}_{\hatcal{A}_r}) - H(\bm{X}_{\mathcal{A}_{\hat{r}}}|\bm{X}_{\mathcal{B}_{\hat{r}}}) \label{eq:change-var-3}\\
    &= H(\bm{X}_{\mathcal{A}_r}) - H(\bm{X}_{\mathcal{A}_{\hat{r}}}|\bm{X}_{\mathcal{B}_{\hat{r}}})\label{eq:change-var-4}\\
    &= I(\bm{X}_{\mathcal{A}_r} ; \bm{X}_{\mathcal{B}_r}) < \eta,
  \end{align}
\end{subequations}
where \eqref{eq:change-var-3} follows from \eqref{eq:change-var-2}, and \eqref{eq:change-var-4} follows from $\bm{X}_{\mathcal{S}} \stackrel{\text{d}}{=} \hat{\bm{X}}_{\mathcal{S}}$.
    So, the index code is $(\mathcal{S}, (p_{\hat{X}_s}:s \in \mathcal{S}),\epsilon, \eta ,\hat{n})$-feasible.

      \section{Proof of Theorem~\ref{theorem:network-to-index} -- Part 2 (the backward direction)}\label{proof:2-2}
    We will now prove Proof of Part 2 in Theorem~\ref{theorem:network-to-index}, that is, when $\mathbb{I}$ is 
$(\mathcal{S}, (p_{\hat{X}_s}:s \in \mathcal{S}),\epsilon, \eta ,\hat{n})$-feasible. %$\Rightarrow$  $\mathbb{N}$ is $(\mathcal{S}, (p_{X_s}:s \in \mathcal{S}),\epsilon, \eta ,n)$-feasible}

   Recall that $\hat{\bm{X}}_{\mathcal{S}' \cup \mathcal{E}'}$ are independent and uniformly distributed. %\stackrel{\text{d}}{=} \bm{X}'_{\mathcal{S}'} \stackrel{\text{d}}{=} (\bm{X}_{\mathcal{S}},\bm{Z}_{\mathcal{V}})$, and $\hat{X}_e$, for each $e \in \mathcal{E}'$, to be uniformly distributed over $[2^{\lfloor c'_e n \rfloor}]$. We have assumed here that $X_s$, for each $s \in \mathcal{S}$, is uniformly distributed over $[M_s]$.
  %  We will show that if $\mathbb{I}$ is $(\mathcal{S}, (p_{\hat{X}_s}:s \in \mathcal{S}),\epsilon, \eta ,\hat{n})$-feasible, then $\mathbb{N}'$ is $(\mathcal{S}, (p_{X'_s}:s \in \mathcal{S}),\epsilon, \eta ,n)$-feasible.  The latter implies that $\mathbb{N}$ is $(\mathcal{S}, (p_{X_s}:s \in \mathcal{S}),\epsilon, \eta ,n)$-feasible.
%\noindent \textit{Decoding criteria:}   
   We will again use the network-code construction proposed by Effros et al.~\cite{effrosrouayheblangberg15}.

   \subsection{Code construction}
   
   We first show some preliminary results required for decodability.
   Define the following:
   \begin{definition}
     Consider $\mathbb{I}$. For any realisation $\hat{\bm{x}}_{\mathcal{S}'}$, let $\mathcal{Y}_{\hat{\bm{x}}_{\mathcal{S}'}}$ denote the set of realisations $\hat{\bm{x}}_{\mathcal{E}'}$ such that if the message tuple $(\hat{\bm{x}}_{\mathcal{S}'},\hat{\bm{x}}_{\mathcal{E}'})$ for any $\hat{\bm{x}}_{\mathcal{E}'} \in \mathcal{Y}_{\hat{\bm{x}}_{\mathcal{S}'}}$, then  all receivers can decode their required messages correctly.
   \end{definition}
   
   This means for any $(\hat{\bm{x}}_{\mathcal{S}'},\hat{\bm{x}}_{\mathcal{E}'})$ such that $\hat{\bm{x}}_{\mathcal{E}'} \in \mathcal{Y}_{\hat{\bm{x}}_{\mathcal{S}'}}$, we have
    \begin{subequations}
      \begin{align}
        \hat{\mathsf{d}}_{\hat{t}_i}(\hat{x}_{\text{b}},\hat{\bm{x}}_{\hatcal{{H}}_{\hat{t}_i}})
        &= \hat{\mathsf{d}}_{\hat{t}_i}(\hat{x}_{\text{b}},\hat{\bm{x}}_{\inset{i} \cup  O'^{-1}(i)}) \label{eq:generate-dest}\\
        &= \hat{\bm{x}}_{\hatcal{{W}}_{\hat{t}_i}} = \hat{\bm{x}}_{\{s \in [S]: i \in \mathcal{D}'(s)\}},
      \end{align}
    \end{subequations}
    for receiver~$\hat{t}_i$, for each $i \in \mathcal{U}'$,
    \begin{subequations}
      \begin{align}
        \hat{\mathsf{d}}_{\hat{t}_e}(\hat{x}_{\text{b}},\hat{\bm{x}}_{\hatcal{{H}}_{\hat{t}_e}})
        &= \hat{\mathsf{d}}_{\hat{t}_e}(\hat{x}_{\text{b}},\hat{\bm{x}}_{\inset{\tail{e}} \cup O'^{-1}(\tail{e})})\label{eq:generate-edge-2}\\
        &= \hat{\bm{x}}_{\hatcal{{W}}_{\hat{t}_e}}= \hat{x}_{e} \in [ 2^{
          \lfloor c'_e n \rfloor}]. \label{eq:generate-edge}
      \end{align}
    \end{subequations}
    and receiver~$\hat{t}_e$, for each $e \in \mathcal{E}'$,
    
    In the secure index-coding instance $\mathbb{I}$, messages $\hat{\bm{X}}_{\mathcal{E}'}$ are independent of messages $\hat{\bm{X}}_{\mathcal{S}'}$, and the broadcast message $\hat{X}_{\text{b}}$ is a function of these messages $\hat{\mathsf{e}}(\hat{\bm{X}}_{\hatcal{S}})$, which is computed by the sender.

    We would like to use the decoding functions \eqref{eq:generate-dest} and \eqref{eq:generate-edge-2} for the network-coding equivalence $\mathbb{N}'$. But, in $\mathbb{N}'$, there is no centralised node to calculate $\hat{X}_{\text{b}}$. To deal with this problem, it is proposed~\cite{effrosrouayheblangberg15} that the value of $\hat{x}_{\text{b}}$ in these functions be fixed to some constant $\sigma \in [2^{\hat{n}}]$. In other words, in contrast to $\mathbb{I}$ where $\hat{X}_{\text{b}}$ varies with $\hat{\bm{X}}_{\mathcal{S}'}$, we fix this value for $\mathbb{N}'$. Then, we set the local encoding function of each edge $e \in \mathcal{E}'$ to be
    \ifx\doublecolumn\undefined
    % ==== put single-column equations here
    \begin{equation}
        \mathsf{e}_e(\bm{x}'_{\inset{\tail{e}}}, \bm{x}'_{O'^{-1}(\tail{e})}) = \hat{\mathsf{d}}_{\hat{t}_e}(\sigma,\bm{x}'_{\inset{\tail{e}}} \bm{x}'_{ O'^{-1}(\tail{e})}) \in [ 2^{
          \lfloor c'_e n \rfloor}], \label{eq:ic2nc1}
      \end{equation}
\else
% ==== put double-column equations here
\begin{align}
        \mathsf{e}_e(\bm{x}'_{\inset{\tail{e}}}, \bm{x}'_{O'^{-1}(\tail{e})}) &= \hat{\mathsf{d}}_{\hat{t}_e}(\sigma,\bm{x}'_{\inset{\tail{e}}} \bm{x}'_{ O'^{-1}(\tail{e})})\nonumber \\ &\in [ 2^{
          \lfloor c'_e n \rfloor}], \label{eq:ic2nc1}
      \end{align}
\fi
      and the decoding function of each destination node $i \in \mathcal{U}'$ to be
      \begin{equation}
        \mathsf{d}_i(\bm{x}'_{\inset{i}}, \bm{x}'_{O'^{-1}(i)}) = \hat{\mathsf{d}}_{\hat{t}_i}(\sigma,\bm{x}'_{\inset{i}}, \bm{x}'_{ O'^{-1}(i)}). \label{eq:ic2nc2}
      \end{equation}
 
      The idea is that for each edge $e \in \mathcal{E}'$ in $\mathbb{N}'$, its tail node $\tail{e}$ can generate the correct outgoing edge messages $x'_e$ from the incoming messages $\bm{x}'_{\inset{\tail{e}}}$, the source messages $\bm{x}'_{O'^{-1}(\tail{e})}$ originating from the node,  and the chosen $\sigma$ via \eqref{eq:generate-edge-2} (or equivalently, \eqref{eq:ic2nc1}), and consequently, all destination nodes can recover their required messages via \eqref{eq:generate-dest} (or equivalently, \eqref{eq:ic2nc2}). For a chosen $\sigma$, define a function $\phi_{\sigma}(\cdot)$ to be the collection of global edge encoding functions \eqref{eq:ic2nc1}, that is, $\phi_{\sigma}(\bm{x}'_{\mathcal{S}'})  \equalbydef (\mathsf{g}_e: e \in \mathcal{E}') = \bm{x}'_{\mathcal{E}'}$.

      The challenge here is to select a suitable $\sigma$ for $\mathbb{N}'$. A suitable $\sigma$ exists to guarantee decodability~\cite{effrosrouayheblangberg15}. In this paper, we need to further show that a suitable $\sigma$ exists to guarantee both decodability and security.

%      We start with some useful properties:

 %        Now, for $\mathbb{N}'$, we select and fix an appropriate $\sigma \in [2^{\hat{n}}]$ (its existence is proven in Proposition~\ref{prop:sigma}), 
%      We can then use the same argument as that in Effros et al.\ to show that $P_\text{e} \leq \epsilon$ for both $\mathbb{N}'$ and $\mathbb{N}$.
    % However, given $\hat{X}_{\text{b}}$, the messages $\hat{\bm{X}}_{\mathcal{E}'}$ and $\hat{\bm{X}}_{\mathcal{S}'}$ are dependent.

      \subsection{Some decodability properties}
      
    We start with the following proposition:
    \begin{proposition} \label{prop:only one-xe}
      For any choice of $\sigma \in [2^{\hat{n}}]$ and any realisation of $\hat{\bm{x}}_{\mathcal{S}'}$, there is at most one $\hat{\bm{x}}_{\mathcal{E}'} \in \mathcal{Y}_{\hat{\bm{x}}_{\mathcal{S}'}}$ for which $\hat{\mathsf{e}}(\hat{\bm{x}}_{\hatcal{S}}) = \hat{\mathsf{e}}(\hat{\bm{x}}_{\mathcal{S}'},\hat{\bm{x}}_{\mathcal{E}'}) = \sigma$.
    \end{proposition}
    
    Effros et al.~\cite[Claim~1]{effrosrouayheblangberg15} have proven this for a slightly different network-to-index instance mapping, where there is an additional receiver for the index-coding equivalence called $\hat{t}_\text{all}$ that has $\hat{\bm{X}}_{\mathcal{S}'}$ and wants $\hat{\bm{X}}_{\mathcal{E}'}$. Their proof relies mainly on the existence of the additional receiver. We will prove Proposition~\ref{prop:only one-xe} without this additional receiver.
    \begin{IEEEproof}[Proof of Proposition~\ref{prop:only one-xe}]
      Pick any realisation $\hat{\bm{x}}_{\mathcal{S}'}$ of $\hat{\bm{X}}_{\mathcal{S}'}$. Suppose to the contradiction that there exists two distinct realisations $\hat{\bm{x}}_{\mathcal{E}'}$ and $\hat{\bm{x}}_{\mathcal{E}'}'$ of $\hat{\bm{X}}_{\mathcal{E}'}$ such that \textit{(i)} $\hat{\mathsf{e}}(\hat{\bm{x}}_{\mathcal{S}'},\hat{\bm{x}}_{\mathcal{E}'}) = \hat{\mathsf{e}}(\hat{\bm{x}}_{\mathcal{S}'},\hat{\bm{x}}_{\mathcal{E}'}) = \sigma$, and \textit{(ii)} $\hat{\bm{x}}_{\mathcal{E}'}, \hat{\bm{x}}_{\mathcal{E}'}' \in \mathcal{Y}_{\hat{\bm{x}}_{\mathcal{S}'}}$. Now, as $\mathbb{I}$ is constructed from an acyclic network-coding instance $\mathbb{N}'$, it follows that given a deterministic index code $(\hat{\mathsf{e}},\hat{\mathsf{D}})$ (where $\hat{\mathsf{D}} =\{\hat{\mathsf{d}}_t: t \in \hatcal{{T}} \})$), the messages $\hat{\bm{x}}_{\mathcal{S}'}$, and the broadcast message $\sigma$, we can completely determine the messages  $\hat{\bm{x}}_{\mathcal{E}'}$. To see this, start from a vertex $i$ with in-degree zero in $\mathbb{N}'$, all receivers $\hat{t}_e$ in $\mathbb{I}$ where $e \in \outset{i}$ must decode $\hat{x}_e$ solely from $\hat{\bm{x}}_{O'^{-1}(i)}$ and $\sigma$. By starting from all vertices with zero in-degree (also known as root or source vertices) and traversing the edges $e \in \mathcal{E}'$ in the graph of $\mathbb{N}'$, we can identify all corresponding receiver $\hat{t}_e$ in $\mathbb{I}$, who must decode $\hat{x}_e$ solely from the broadcast message $\sigma$, part of $\{\hat{x}_i: i \in \mathcal{S}'\}$, and part of $\{\hat{x}_e: e \in \mathcal{E}'\}$ that we have obtained from previous steps. Now, since the messages $\hat{\bm{x}}_{\mathcal{E}'}$ is a deterministic function of $(\hat{\mathsf{D}}, \hat{\bm{x}}_{\mathcal{S}'},\sigma)$, some receiver $\hat{t}_e$, $e \in \mathcal{E}'$ must decode its required message $\hat{x}_e$ wrongly in either one of the two realisations of $\hat{\bm{X}}_{\mathcal{E}'}$, namely, $\hat{\bm{x}}_{\mathcal{E}'}$ and $\hat{\bm{x}}_{\mathcal{E}'}'$. This contradictions the definition of $\mathcal{Y}_{\hat{\bm{x}}_{\mathcal{S}'}}$.
    \end{IEEEproof}
    
    Next, we state a proposition due to Effros et al.
    \begin{proposition} \label{prop:sigma}
    	({\cite[Claim 2]{effrosrouayheblangberg15}}) If each $\hat{X}_i, i \in \mathcal{S}'$, is uniformly distributed, then there exists a $\sigma \in [2^{\hat{n}}]$ such that at least $(1-\epsilon)$ of the source realisations $\hat{\bm{x}}_{\mathcal{S}'}$ of $\hat{\bm{X}}_{\mathcal{S}'}$ satisfy $\hat{\mathsf{e}}(\hat{\bm{x}}_{\mathcal{S}'},\hat{\bm{x}}_{\mathcal{E}'})=\sigma$ for some $\hat{\bm{x}}_{\mathcal{E}'} \in \mathcal{Y}_{\hat{\bm{x}}_{\mathcal{S}'}}$.
      \end{proposition}

      Note that due our use of a slightly different mapping \eqref{eq:normalisation-nc}, the above lemma does not require the assumption that all $\{2^{c_e n}: e \in \mathcal{E}'\}$ being integers, an assumption that resulted in some slight mismatch in rates~\cite[p.~2484]{effrosrouayheblangberg15}. % (in addition, $2^{\hat{c}_{\text{b}}}$ also needs to be an integer in Effros et al.) %However, the proof requires that each individual messages of $\hat{\bm{X}}_{\hatcal{S}}$ to be uniformly distributed. This ensures that there are sufficiently many source realisations such that $\hat{\mathsf{e}}(\hat{\bm{x}}_{\mathcal{S}'},\hat{\bm{x}}_{\mathcal{E}'})=\sigma$. 

  \subsection{Some security properties}
      
%\noindent \textit{Security criteria:}   

  Since $\mathbb{I}$ is $(\mathcal{S}, (p_{\hat{X}_s}:s \in \mathcal{S}),\epsilon, \eta ,\hat{n})$-feasible, we have
% \begin{subequations}
%   \begin{align}
%     I(\bm{X}_{\mathcal{A}_r}; \bm{X}_{\mathcal{B}_r}) &= I(\bm{X}'_{\mathcal{A}_r}; \bm{X}'_{\mathcal{B}_r})\\
%     &= I(\bm{X}'_{\mathcal{A}_r}; \{\mathsf{e}_e(\bm{X}'_{\inset{\tail{e}}}, \bm{X}'_{O'^{-1}(\tail{e})}): e \in \mathcal{B}_r\})
%   \end{align}
% \end{subequations}
\begin{subequations}
  \begin{align}
    \eta &\geq I(\hat{\bm{X}}_{\hatcal{A}_r} ; \hat{X}_{\text{b}}, \hat{\bm{X}}_{\hatcal{B}_r}) \\
    %      &= H(\hat{X}_{\text{b}}, \hat{\bm{X}}_{\hatcal{B}_r}) - H(\hat{X}_{\text{b}}, \hat{\bm{X}}_{\hatcal{B}_r}|\hat{\bm{X}}_{\hatcal{A}_r})\\
    % &= H(\hat{\bm{X}}_{\hatcal{B}_r}) - H(\hat{\bm{X}}_{\hatcal{B}_r} |\hat{X}_{\text{b}} ) - H(\hat{\bm{X}}_{\hatcal{B}_r}|\hat{\bm{X}}_{\hatcal{A}_r} ) + H (\hat{\bm{X}}_{\hatcal{B}_r} |\hat{X}_{\text{b}},\hat{\bm{X}}_{\hatcal{A}_r} )\\
    % &= H(\hat{\bm{X}}_{\hatcal{B}_r}) - H(\hat{\bm{X}}_{\hatcal{B}_r} | \hat{\bm{X}}_{\hatcal{A}_r},\hat{X}_{\text{b}}) \\
    % &= H(\hat{\bm{X}}_{\hatcal{B}_r}) - \sum_{\sigma' \in [2^{\hat{n}}]} p_{\hat{X}_\text{b}}(\sigma') H(\hat{\bm{X}}_{\hatcal{B}_r} | \hat{\bm{X}}_{\hatcal{A}_r}, \hat{X}_{\text{b}} = \sigma' ) \label{eq:ave-bounded-eta-half}\\
%         &= H(\hat{\bm{X}}_{\hatcal{A}_r}) - H(\hat{\bm{X}}_{\hatcal{A}_r} | \hat{X}_{\text{b}}, \hat{\bm{X}}_{\hatcal{B}_r}) \\
 %    &= H(\hat{\bm{X}}_{\hatcal{A}_r}) - \sum_{\sigma' \in [2^{\hat{n}}]} p_{\hat{X}_\text{b}}(\sigma') H(\hat{\bm{X}}_{\hatcal{A}_r} | \hat{\bm{X}}_{\hatcal{B}_r}, \hat{X}_{\text{b}} = \sigma' ) \\
% %         &\geq H(\hat{\bm{X}}_{\hatcal{A}_r}) - H(\hat{\bm{X}}_{\hatcal{A}_r} | \hat{X}_{\text{b}})\\
    &\geq 
    I(\hat{\bm{X}}_{\hatcal{A}_r} ;  \hat{\bm{X}}_{\hatcal{B}_r} | \hat{X}_{\text{b}} ) \label{eq:nc-ave-security}\\
    % I(\hat{\bm{X}}_{\hatcal{A}_r} ;  \hat{\bm{X}}_{\hatcal{B}_r}) &\leq \eta\\
    & =\sum_{\sigma \in [2^{\hat{n}}]} p_{\hat{X}_\text{b}}(\sigma) I(\hat{\bm{X}}_{\hatcal{A}_r} ;  \hat{\bm{X}}_{\hatcal{B}_r} | \hat{X}_{\text{b}} = \sigma ). \label{eq:ave-bounded-eta}
  \end{align}
\end{subequations}

It turns out that there may not exist a suitable $\sigma$ that gives both matching error and leakage criteria in $\mathbb{N}$' in general. Nonetheless, we are able to prove its existence when $\epsilon=0$ (i.e., perfect decodability). Otherwise when $\epsilon > 0$, we prove a weaker form of equivalence where the error probability and the leakage of $\mathbb{N}'$ do not exactly match those for $\mathbb{I}$.

\subsection{Proof of Part 2a in Theorem~\ref{theorem:network-to-index}: $\epsilon=0$}

\subsubsection{Decoding criteria}
Under this condition, for any message realisation $(\hat{\bm{x}}_{\mathcal{S}'},\hat{\bm{x}}_{\mathcal{E}'})$, all receivers in $\mathbb{I}$ can decode their required messages correctly. From the definition of $\mathcal{Y}_{\hat{\bm{x}}_{\mathcal{S}'}}$, we have that $\mathcal{Y}_{\hat{\bm{x}}_{\mathcal{S}'}} = \prod_{e=\mathcal{E}'}[2^{\lfloor c_e' n \rfloor}] = \mathcal{X}_{\mathcal{E}'}$ is the set of all realisations of $\hat{\bm{x}}_{\mathcal{E}'}$, for any $\hat{\bm{x}}_{\mathcal{S}'}$. Due to the normalisation of the edge capacities, $|\mathcal{Y}_{\hat{\bm{x}}_{\mathcal{S}'}}| = |\prod_{e=\mathcal{E}'}[2^{\lfloor c'_e n \rfloor}]| = 2^{\hat{n}}$. From Proposition~\ref{prop:only one-xe}, we know that if we pick any $\hat{\bm{x}}_{\mathcal{S}'}$, there is a bijective map between $\hat{\bm{x}}_{\mathcal{E}'}$ and $\hat{x}_\text{b}$.

Now, we select any $\sigma \in [2^{\hat{n}}]$. For every $\hat{\bm{x}}_{\mathcal{S}'}$, we can find exactly one $\hat{\bm{x}}_{\mathcal{E}'} \in \mathcal{Y}_{\hat{\bm{x}}_{\mathcal{S}'}} = \mathcal{X}_{\mathcal{S}'}$ for which $\hat{\mathsf{e}}(\hat{\bm{x}}_{\mathcal{S}'},\hat{\bm{x}}_{\mathcal{E}'}) = \sigma$.
So, for $\mathbb{N}'$, by selecting any $\sigma \in [2^{\hat{n}}]$ in \eqref{eq:ic2nc1} and \eqref{eq:ic2nc2}, decoding in $\mathbb{N}'$ will succeed, giving $P_\text{e} = 0$.

\subsubsection{Security criteria}

% We select $\sigma'$ in \eqref{eq:ic2nc1} and \eqref{eq:ic2nc2}  for $\mathbb{N}'$. This choice of $\sigma'$ satisfies the requirement in Proposition~\ref{prop:sigma}, meaning that for all source realisations (note that $\epsilon=0$), we know that there exists some (in fact, a unique one) $\hat{\bm{x}}_{\mathcal{E}'} \in \mathcal{Y}_{\hat{\bm{x}}_{\mathcal{S}'}}$ for which $\hat{\mathsf{e}}(\hat{\bm{x}}_{\mathcal{S}'},\hat{\bm{x}}_{\mathcal{E}'})=\sigma$. So, the decoding criterion for $\mathbb{N}'$ is preserved. In fact, we can choose $\sigma$ to be any value in $[2^{\hat{n}}]$ to satisfy the decoding criteria if $\epsilon = 0$.

Since $(\hat{\bm{X}}_{\mathcal{S}'},\hat{\bm{X}}_{\mathcal{E}'})$ are independent  and uniformly distributed, $p_{\hat{\bm{X}}_{\mathcal{S}'},\hat{\bm{X}}_{\mathcal{E}'}}(\hat{\bm{x}}_{\mathcal{S}'},\hat{\bm{x}}_{\mathcal{E}'}) = 1 /(|\mathcal{X}_{\mathcal{S}'}||\mathcal{X}_{\mathcal{E}'}|)$ and $p_{\hat{\bm{X}}_{\mathcal{S}'}}(\hat{\bm{x}}_{\mathcal{S}'}) = 1 /|\mathcal{X}_{\mathcal{S}'}|$, for all $\hat{\bm{x}}_{\mathcal{S}'}$ and $\hat{\bm{x}}_{\mathcal{E}'}$. As there is a bijective map between $\hat{\bm{x}}_{\mathcal{E}'}$ and $\hat{x}_\text{b}$ given any $\hat{\bm{x}}_{\mathcal{S}'}$, we have the following for every $\hat{\bm{x}}_{\mathcal{S}'}$ and $\hat{x}_\text{b}$:
\begin{subequations}
  \begin{align} p_{\hat{X}_\text{b}|\hat{\bm{X}}_{\mathcal{S}'},}(\hat{x}_\text{b}|\hat{\bm{x}}_{\mathcal{S}'}) &= p_{\hat{\bm{X}}_{\mathcal{E}'}|\hat{\bm{X}}_{\mathcal{S}'},}(\hat{\bm{x}}_{\mathcal{E}'}|\hat{\bm{x}}_{\mathcal{S}'}) \label{eq:bijective-map}\\
    &= \frac{p_{\hat{\bm{X}}_{\mathcal{S}'},\hat{\bm{X}}_{\mathcal{E}'}}(\hat{\bm{x}}_{\mathcal{S}'},\hat{\bm{x}}_{\mathcal{E}'})}{p_{\hat{\bm{X}}_{\mathcal{S}'}}(\hat{\bm{x}}_{\mathcal{S}'})} = \frac{1}{|\mathcal{X}_{\mathcal{E}'}|},
  \end{align}
\end{subequations}
and
\begin{subequations}
  \begin{align}
    p_{\hat{X}_\text{b}}(\hat{x}_\text{b})
    &=\sum_{\hat{\bm{x}}_{\mathcal{S}'},\hat{\bm{x}}_{\mathcal{E}'}} p_{\hat{\bm{X}}_{\mathcal{S}'},\hat{\bm{X}}_{\mathcal{E}'}}(\hat{\bm{x}}_{\mathcal{S}'},\hat{\bm{x}}_{\mathcal{E}'})\bm{1}(\hat{\mathsf{e}}(\hat{\bm{x}}_{\mathcal{S}'},\hat{\bm{x}}_{\mathcal{E}'}) = \sigma)\\
    &= \frac{1}{|\mathcal{X}_{\mathcal{S}'}||\mathcal{X}_{\mathcal{E}'}|} \sum_{\hat{\bm{x}}_{\mathcal{S}'}} 1  = \frac{1}{|\mathcal{X}_{\mathcal{E}'}|}, \label{eq:one-s}
  \end{align}
\end{subequations}
where $\bm{1}(E)$ is the indicator function, which returns 1 is $E$ is true, and 0 otherwise, and \eqref{eq:one-s} is obtained as there is exactly one $\hat{\bm{x}}_{\mathcal{E}'}$ for which $\hat{\mathsf{e}}(\hat{\bm{x}}_{\mathcal{S}'},\hat{\bm{x}}_{\mathcal{E}'}) = \sigma$.

So,
\begin{subequations}
  \begin{align}
    p_{\hat{\bm{X}}_{\mathcal{S}'} | \hat{X}_{\text{b}}}(\hat{\bm{x}}_{\mathcal{S}'}|\hat{x}_{\text{b}}) &= \frac{p_{\hat{\bm{X}}_{\mathcal{S}'}}(\hat{\bm{x}}_{\mathcal{S}'}) p_{\hat{X}_\text{b}|\hat{\bm{X}}_{\mathcal{S}'}}(\hat{x}_\text{b}|\hat{\bm{x}}_{\mathcal{S}'})}{  p_{\hat{X}_\text{b}}(\hat{x}_\text{b})}\\
    &= \frac{1}{|\mathcal{X}_{\mathcal{S}'}|}, \label{eq:s-given-b-uniform}
  \end{align}
\end{subequations}
for all $\hat{\bm{x}}_{\mathcal{S}'}$ and $\hat{x}_\text{b}$.

For $\mathbb{N}'$, by choosing any $\sigma$ in \eqref{eq:ic2nc1} and \eqref{eq:ic2nc2}), we  have  $\bm{X}_{\mathcal{E}'}' = \phi_{\sigma}(\bm{X}_{\mathcal{S}'}')$, and thus giving
\begin{equation}
  p_{\bm{X}_{\mathcal{E}'}'|\bm{X}_{\mathcal{S}'}'}(\bm{b}|\bm{a}) = p_{\hat{\bm{X}}_{\mathcal{E}'}|\hat{\bm{X}}_{\mathcal{S}'}\hat{X}_{\text{b}}}(\bm{b}|\bm{a},\sigma), \label{eq:zero-error-equality}
\end{equation}
and thus
\begin{subequations}
  \begin{align}
    p_{\bm{X}_{\mathcal{E}'}',\bm{X}_{\mathcal{S}'}'}(\bm{b},\bm{a}) &= p_{\bm{X}_{\mathcal{E}'}'|\bm{X}_{\mathcal{S}'}'}(\bm{b}|\bm{a}) p_{\bm{X}_{\mathcal{S}'}'}(\bm{a})\\
&= p_{\bm{X}_{\mathcal{E}'}'|\bm{X}_{\mathcal{S}'}'}(\bm{b}|\bm{a}) \frac{1}{|\mathcal{X}_{\mathcal{S}'}|}\\
                                                                     &= p_{\hat{\bm{X}}_{\mathcal{E}'}|\hat{\bm{X}}_{\mathcal{S}'}\hat{X}_{\text{b}}}(\bm{b}|\bm{a},\sigma) p_{\hat{\bm{X}}_{\mathcal{S}'} | \hat{X}_{\text{b}}}(\bm{a}|\sigma) \label{eq:uniform-a-given-sigma}\\
    &= p_{\hat{\bm{X}}_{\mathcal{E}'},\hat{\bm{X}}_{\mathcal{S}'}|\hat{X}_{\text{b}}}(\bm{b},\bm{a}|\sigma), \label{eq:equal-distribution}
  \end{align}
\end{subequations}
where \eqref{eq:uniform-a-given-sigma} follows from \eqref{eq:s-given-b-uniform}.

From \eqref{eq:ave-bounded-eta}, since the weighted average of $I(\hat{\bm{X}}_{\hatcal{A}_r} ;  \hat{\bm{X}}_{\hatcal{B}_r} | \hat{X}_{\text{b}} = \sigma )$ over all $\sigma \in [2^{\hat{n}}]$ is not greater than $\eta$, there exists one $\sigma' \in [2^{\hat{n}}]$ for which $I(\hat{\bm{X}}_{\hatcal{A}_r} ;  \hat{\bm{X}}_{\hatcal{B}_r} | \hat{X}_{\text{b}} = \sigma' ) \leq \eta$.
This means, using this chosen $\sigma'$ for $\mathbb{N}'$ and $\mathbb{N}$,
\begin{equation}
  I(\bm{X}_{\mathcal{A}_r}; \bm{X}_{\mathcal{B}_r}) = I(\bm{X}_{\mathcal{A}'_r}'; \bm{X}_{\mathcal{B}'_r}')  \stackrel{(a)}{=} I(\hat{\bm{X}}_{\hatcal{A}_r} ;  \hat{\bm{X}}_{\hatcal{B}_r} | \hat{X}_{\text{b}} = \sigma') \leq  \eta, \label{eq:zero-decoding-error-case}
\end{equation}
where $(a)$ follows from \eqref{eq:equal-distribution}.

So, if $\epsilon = 0$, then that $\mathbb{I}$ is $(\mathcal{S}, (p_{\hat{X}_s}:s \in \mathcal{S}),0, \eta ,\hat{n})$-feasible  implies that  $\mathbb{N}$ is $(\mathcal{S}, (p_{X_s}:s \in \mathcal{S}),0, \eta ,n)$-feasible.

\subsection{Proof of Part 2b in Theorem~\ref{theorem:network-to-index}: $\epsilon > 0$}

\subsection*{Issues:}

Unfortunately, the results for the perfect-decoding case does not extend straightforwardly to the case of imperfect decoding due to the following reasons:
\begin{enumerate}
\item When $\epsilon=0$, choosing any $\sigma$ for $\mathbb{N}'$ guarantees perfect decoding for $\mathbb{N}'$, and we only need to select a good $\sigma$ to guarantee the security criterion. However, when $\epsilon > 0$, we need to choose a good $\sigma$ that simultaneously guarantees the decodability and the security criteria.
\item When $\epsilon > 0$, the random variables in the two instances $\mathbb{I}$ and $\mathbb{N}'$ do not necessarily have the same distribution as in \eqref{eq:equal-distribution}. This is because if some message realisation results in decoding error, then $|\mathcal{Y}_{\hat{\bm{x}}_{\mathcal{S}'}}| < 2^{\hat{n}}$ for some $\hat{\bm{x}}_{\mathcal{S}'}$. This means for this $\hat{\bm{x}}_{\mathcal{S}'}$, there could be multiple distinct $\hat{\bm{x}}^{(1)}_{\mathcal{E}'}, \hat{\bm{x}}^{(2)}_{\mathcal{E}'}  \in \mathcal{X}_{\mathcal{S}'}$, $\hat{\bm{x}}^{(1)}_{\mathcal{E}'} \neq \hat{\bm{x}}^{(2)}_{\mathcal{E}'}$, for which $\hat{\mathsf{e}}(\hat{\bm{x}}_{\mathcal{S}'},\hat{\bm{x}}^{(1)}_{\mathcal{E}'}) = \hat{\mathsf{e}}(\hat{\bm{x}}_{\mathcal{S}'},\hat{\bm{x}}^{(2)}_{\mathcal{E}'})$. This leads to the following:
  \begin{enumerate}
      \item As there may not be a bijective map between $\hat{\bm{x}}_{\mathcal{E}'}$ and $\hat{x}_\text{b}$ for this $\hat{\bm{x}}_{\mathcal{S}'}$, \eqref{eq:bijective-map} may not be true.
  \item  \eqref{eq:one-s} may not hold.
  \item  \eqref{eq:zero-error-equality} may not hold.
  \end{enumerate}
  As a results, we cannot guarantee (\ref{eq:zero-decoding-error-case}.a).
\end{enumerate}

\subsection*{Our proposed solution:}

We will resolve the above issues through the following steps:
\begin{enumerate}[label=\textsf{S.\arabic*}]
\item\label{step-1} Relate security expressions for $\mathbb{N}'$ to that for $\mathbb{I}$.
\item\label{step-2} Express  security  in $\mathbb{I}$ in terms of expressions obtained in \ref{step-1} averaged over $\hat{X}_\text{b}$.
\item\label{step-3} Relate the decoding criterion in $\mathbb{N}'$ to that in $\mathbb{I}$.
\item\label{step-4} Express decodability in $\mathbb{I}$ as an average over $\hat{X}_\text{b}$ using \ref{step-3}.
\item\label{step-5} Combine the results from steps \ref{step-2} and \ref{step-4} to find a $\sigma = \hat{x}_\text{b}$ that is simultaneously good for security and decodability.
%  \item\label{step-6} Obtain bounds of decoding error and leakage for the chosen $\sigma$.
  \end{enumerate}

We now present the steps in detail:

\subsubsection{\textnormal{\ref{step-1}}: Relate security expressions for $\mathbb{N}'$ to that for $\mathbb{I}$}

Note that the edge messages $\bm{X}_{\mathcal{E}'}'$ in $\mathbb{N}'$ are generated by choosing a specific $\sigma$ for the network code \eqref{eq:ic2nc1}--\eqref{eq:ic2nc2}, which are the decoding function in $\mathbb{I}$. So, if decoding in $\mathbb{I}$ is correct and $\hat{x}_\text{b} = \hat{\mathsf{e}}(\hat{\bm{x}}_{\mathcal{S}'},\hat{\bm{x}}_{\mathcal{E}'}) = \sigma$, then choosing $\sigma$ for $\mathbb{N}'$, we have $\bm{x}_{\mathcal{E}'}' = \phi_\sigma(\bm{x}_{\mathcal{S}'}')$ for $(\bm{x}_{\mathcal{S}'}',\bm{x}_{\mathcal{E}'}')=(\hat{\bm{x}}_{\mathcal{S}'},\hat{\bm{x}}_{\mathcal{E}'})$.

For this reason, we define the following for $\mathbb{I}$:
\begin{equation}
  \hat{D} =
  \begin{cases}
    1, & \text{ if decoding of all receivers in $\mathbb{I}$ is correct},\\
    0, & \text{ otherwise.}
  \end{cases}
\end{equation}
This means
\ifx\doublecolumn\undefined
% ==== put single-column equations here
\begin{equation}
  p_{\bm{X}_{\mathcal{E}'}'|\bm{X}_{\mathcal{S}'}'}(\bm{b}|\bm{a}) = p_{\hat{\bm{X}}_{\mathcal{E}'}|\hat{\bm{X}}_{\mathcal{S}'},\hat{D},\hat{X}_{\text{b}}}(\bm{b}|\bm{a},1,\sigma), \text{ for all } \bm{a}  \text{ where } p_{\hat{\bm{X}}_{\mathcal{S}'},\hat{D},\hat{X}_{\text{b}}}(\bm{a},1,\sigma) > 0. \label{eq:combine-000}
\end{equation}
\else
% ==== put double-column equations here
\begin{multline}
   p_{\bm{X}_{\mathcal{E}'}'|\bm{X}_{\mathcal{S}'}'}(\bm{b}|\bm{a}) = p_{\hat{\bm{X}}_{\mathcal{E}'}|\hat{\bm{X}}_{\mathcal{S}'},\hat{D},\hat{X}_{\text{b}}}(\bm{b}|\bm{a},1,\sigma), \\
  \text{ for all } \bm{a}  \text{ where } p_{\hat{\bm{X}}_{\mathcal{S}'},\hat{D},\hat{X}_{\text{b}}}(\bm{a},1,\sigma) > 0, \label{eq:combine-000}
\end{multline}
\fi
which is similar to \eqref{eq:zero-error-equality} for the perfect-decoding case.

Now, in order to restrict the leakage $I(\bm{X}'_{\mathcal{A}'_r}; \bm{X}'_{\mathcal{B}'_r})$ in $\mathbb{N}'$, we will relate $I(\bm{X}'_{\mathcal{A}'_r}; \bm{X}'_{\mathcal{B}'_r})$ to $I(\hat{\bm{X}}_{\hatcal{A}_r} ;  \hat{\bm{X}}_{\hatcal{B}_r} | \hat{D}=1, \hat{X}_{\text{b}} = \sigma)$, similar to (\ref{eq:zero-decoding-error-case}.a) for the perfect-decoding case.

To this end, we define the following:
\begin{definition}
\begin{equation*}
    \mathcal{G}_{\sigma} \equalbydef \{  \hat{\bm{x}}_{\mathcal{S}'} \in \mathcal{X}_{\mathcal{S}'}: \hat{d} = 1 \text{ and } \hat{x}_\text{b} = \sigma \}.
  \end{equation*}
\end{definition}
$\mathcal{G}_{\sigma}$ is the set of all message realisations $\hat{\bm{x}}_{\mathcal{S}'}$ that result in both \textit{(i)} correct decoding in $\mathbb{I}$ (meaning that $\hat{d}=1$) and \textit{(ii)} the broadcast message $\hat{x}_\text{b} = \hat{\mathsf{e}}(\hat{\bm{x}}_{\mathcal{S}'},\hat{\bm{x}}_{\mathcal{E}'}) = \sigma$, for some $\hat{\bm{x}}_{\mathcal{E}'}$.
  Also define
  \begin{align}
    (1-\epsilon') &\equalbydef \frac{|\mathcal{G}_\sigma|}{ |\mathcal{X}_{\mathcal{S}'}|}\\
    \mathcal{G}^\text{c}_\sigma &\equalbydef \mathcal{X}_{\mathcal{S}'} \setminus \mathcal{G}_\sigma.
  \end{align}
We will bound $\epsilon'$ later.

Step~1 is complete with the following lemma:
  \begin{lemma} \label{lemma:mutual-difference}
    For any $\sigma \in [ 2^{\hat{n}}]$,
    \ifx\doublecolumn\undefined
    % ==== put single-column equations here
    \begin{align}
      I(\bm{X}'_{\mathcal{A}'_r}; \bm{X}'_{\mathcal{B}'_r}) &\leq I(\hat{\bm{X}}_{\hatcal{A}_r} ;  \hat{\bm{X}}_{\hatcal{B}_r} | \hat{D}=1, \hat{X}_{\text{b}} = \sigma) + \epsilon' \log |\mathcal{X}_{\mathcal{S}'}| - \log (1 - \epsilon') + \frac{\epsilon'}{1-\epsilon'} \left(  \log e + \hat{n}  \right)).
    \end{align}
\else
% ==== put double-column equations here
\begin{align}
      I(\bm{X}'_{\mathcal{A}'_r}; \bm{X}'_{\mathcal{B}'_r}) &\leq I(\hat{\bm{X}}_{\hatcal{A}_r} ;  \hat{\bm{X}}_{\hatcal{B}_r} | \hat{D}=1, \hat{X}_{\text{b}} = \sigma)\nonumber \\ & \quad + \epsilon' \log |\mathcal{X}_{\mathcal{S}'}| - \log (1 - \epsilon') + \frac{\epsilon'}{1-\epsilon'} \left(  \log e + \hat{n}  \right)).
    \end{align}
\fi
\end{lemma}

\begin{IEEEproof}
  The proof of Lemma~\ref{lemma:mutual-difference} can be found in Appendix~\ref{appendix:proof-lemma}
  \end{IEEEproof}

\subsubsection{\textnormal{\ref{step-2}}: Express security in $\mathbb{I}$ in terms of expressions obtained in \textnormal{\ref{step-1}} averaged over $\hat{X}_\text{b}$}

  In this step, we will relate $I(\hat{\bm{X}}_{\hatcal{A}_r} ;  \hat{\bm{X}}_{\hatcal{B}_r} | \hat{D}=1, \hat{X}_{\text{b}} = \sigma)$ (in Lemma~\ref{lemma:mutual-difference}) to $I(\hat{\bm{X}}_{\hatcal{A}_r} ;  \hat{\bm{X}}_{\hatcal{B}_r}| \hat{X}_{\text{b}})$ (which is the leakage in $\mathbb{I}$) and then to the security criteria $\eta$ in $\mathbb{I}$.

% For $\mathbb{N}'$, we will choose a $\sigma$ for the local encoding function \eqref{eq:ic2nc1} such that the probability of decoding error is upper bounded by some $\epsilon'$ to be determined later.

From the identity
\ifx\doublecolumn\undefined
% ==== put single-column equations here
\begin{equation}
  I(\hat{\bm{X}}_{\hatcal{A}_r} ;  \hat{\bm{X}}_{\hatcal{B}_r}| \hat{X}_{\text{b}}) + I(\hat{\bm{X}}_{\hatcal{A}_r} ; \hat{D} | \hat{\bm{X}}_{\hatcal{B}_r}, \hat{X}_{\text{b}}) = I(\hat{\bm{X}}_{\hatcal{A}_r} ;  \hat{D}| \hat{X}_{\text{b}}) + I(\hat{\bm{X}}_{\hatcal{A}_r} ;  \hat{\bm{X}}_{\hatcal{B}_r} | \hat{D}, \hat{X}_{\text{b}}),
\end{equation}
\else
% ==== put double-column equations here
\begin{multline}
  I(\hat{\bm{X}}_{\hatcal{A}_r} ;  \hat{\bm{X}}_{\hatcal{B}_r}| \hat{X}_{\text{b}}) + I(\hat{\bm{X}}_{\hatcal{A}_r} ; \hat{D} | \hat{\bm{X}}_{\hatcal{B}_r}, \hat{X}_{\text{b}})\\ = I(\hat{\bm{X}}_{\hatcal{A}_r} ;  \hat{D}| \hat{X}_{\text{b}}) + I(\hat{\bm{X}}_{\hatcal{A}_r} ;  \hat{\bm{X}}_{\hatcal{B}_r} | \hat{D}, \hat{X}_{\text{b}}),
\end{multline}
\fi
we get
\ifx\doublecolumn\undefined
% ==== put single-column equations here
\begin{subequations}
  \begin{align}
    I(\hat{\bm{X}}_{\hatcal{A}_r} ;  \hat{\bm{X}}_{\hatcal{B}_r}| \hat{X}_{\text{b}}) &= I(\hat{\bm{X}}_{\hatcal{A}_r} ;  \hat{D}| \hat{X}_{\text{b}}) - I(\hat{\bm{X}}_{\hatcal{A}_r} ; \hat{D} | \hat{\bm{X}}_{\hatcal{B}_r}, \hat{X}_{\text{b}}) + I(\hat{\bm{X}}_{\hatcal{A}_r} ;  \hat{\bm{X}}_{\hatcal{B}_r} | \hat{D}, \hat{X}_{\text{b}})\\
    & = I(\hat{\bm{X}}_{\hatcal{A}_r} ;  \hat{D}| \hat{X}_{\text{b}}) - I(\hat{\bm{X}}_{\hatcal{A}_r} ; \hat{D} | \hat{\bm{X}}_{\hatcal{B}_r}, \hat{X}_{\text{b}})\nonumber \\ &\quad + p_{\hat{D}}(1)I(\hat{\bm{X}}_{\hatcal{A}_r} ;  \hat{\bm{X}}_{\hatcal{B}_r} | \hat{D} = 1, \hat{X}_{\text{b}}) + p_{\hat{D}}(0) I(\hat{\bm{X}}_{\hatcal{A}_r} ;  \hat{\bm{X}}_{\hatcal{B}_r} | \hat{D} = 0, \hat{X}_{\text{b}}). \label{eq:expanded}
  \end{align}
\end{subequations}
\else
% ==== put double-column equations here
\begin{subequations}
  \begin{align}
    &I(\hat{\bm{X}}_{\hatcal{A}_r} ;  \hat{\bm{X}}_{\hatcal{B}_r}| \hat{X}_{\text{b}})\nonumber \\ &= I(\hat{\bm{X}}_{\hatcal{A}_r} ;  \hat{D}| \hat{X}_{\text{b}}) - I(\hat{\bm{X}}_{\hatcal{A}_r} ; \hat{D} | \hat{\bm{X}}_{\hatcal{B}_r}, \hat{X}_{\text{b}}) + I(\hat{\bm{X}}_{\hatcal{A}_r} ;  \hat{\bm{X}}_{\hatcal{B}_r} | \hat{D}, \hat{X}_{\text{b}})\\
    & = I(\hat{\bm{X}}_{\hatcal{A}_r} ;  \hat{D}| \hat{X}_{\text{b}}) - I(\hat{\bm{X}}_{\hatcal{A}_r} ; \hat{D} | \hat{\bm{X}}_{\hatcal{B}_r}, \hat{X}_{\text{b}})\nonumber \\ &\quad + p_{\hat{D}}(1)I(\hat{\bm{X}}_{\hatcal{A}_r} ;  \hat{\bm{X}}_{\hatcal{B}_r} | \hat{D} = 1, \hat{X}_{\text{b}})\nonumber \\ &\quad + p_{\hat{D}}(0) I(\hat{\bm{X}}_{\hatcal{A}_r} ;  \hat{\bm{X}}_{\hatcal{B}_r} | \hat{D} = 0, \hat{X}_{\text{b}}). \label{eq:expanded}
  \end{align}
\end{subequations}
\fi

Substituting \eqref{eq:expanded} into \eqref{eq:ave-bounded-eta}, we have
\ifx\doublecolumn\undefined
% ==== put single-column equations here
\begin{multline}
    \sum_{\sigma \in [2^{\hat{n}}]} p_{\hat{X}_\text{b}}(\sigma) \Big[ I(\hat{\bm{X}}_{\hatcal{A}_r} ;  \hat{D}| \hat{X}_{\text{b}}=\sigma) - I(\hat{\bm{X}}_{\hatcal{A}_r} ; \hat{D} | \hat{\bm{X}}_{\hatcal{B}_r}, \hat{X}_{\text{b}}=\sigma) \\ + p_{\hat{D}}(1)I(\hat{\bm{X}}_{\hatcal{A}_r} ;  \hat{\bm{X}}_{\hatcal{B}_r} | \hat{D} = 1, \hat{X}_{\text{b}}=\sigma) + p_{\hat{D}}(0) I(\hat{\bm{X}}_{\hatcal{A}_r} ;  \hat{\bm{X}}_{\hatcal{B}_r} | \hat{D} = 0, \hat{X}_{\text{b}}=\sigma)  \Big] \leq \eta,
  \end{multline}
\else
% ==== put double-column equations here
\begin{align}
    \sum_{\sigma \in [2^{\hat{n}}]} p_{\hat{X}_\text{b}}(\sigma) \Big[ & I(\hat{\bm{X}}_{\hatcal{A}_r} ;  \hat{D}| \hat{X}_{\text{b}}=\sigma) - I(\hat{\bm{X}}_{\hatcal{A}_r} ; \hat{D} | \hat{\bm{X}}_{\hatcal{B}_r}, \hat{X}_{\text{b}}=\sigma)\nonumber \\ & + p_{\hat{D}}(1)I(\hat{\bm{X}}_{\hatcal{A}_r} ;  \hat{\bm{X}}_{\hatcal{B}_r} | \hat{D} = 1, \hat{X}_{\text{b}}=\sigma)\nonumber \\ &  + p_{\hat{D}}(0) I(\hat{\bm{X}}_{\hatcal{A}_r} ;  \hat{\bm{X}}_{\hatcal{B}_r} | \hat{D} = 0, \hat{X}_{\text{b}}=\sigma)  \Big] \leq \eta,
  \end{align}
\fi
for each eavesdropper $r \in \mathcal{R}$. Summing it for all eavesdroppers and swapping the summation order, we get
\ifx\doublecolumn\undefined
% ==== put single-column equations here
\begin{subequations}
    \begin{align}
      |\mathcal{R}| \eta &\geq  \sum_{r \in \mathcal{R}} \sum_{\sigma \in [2^{\hat{n}}]} p_{\hat{X}_\text{b}}(\sigma) \Big[ I(\hat{\bm{X}}_{\hatcal{A}_r} ;  \hat{D}| \hat{X}_{\text{b}}=\sigma) - I(\hat{\bm{X}}_{\hatcal{A}_r} ; \hat{D} | \hat{\bm{X}}_{\hatcal{B}_r}, \hat{X}_{\text{b}}=\sigma) \\ & \quad  + p_{\hat{D}}(1)I(\hat{\bm{X}}_{\hatcal{A}_r} ;  \hat{\bm{X}}_{\hatcal{B}_r} | \hat{D} = 1, \hat{X}_{\text{b}}=\sigma) + p_{\hat{D}}(0) I(\hat{\bm{X}}_{\hatcal{A}_r} ;  \hat{\bm{X}}_{\hatcal{B}_r} | \hat{D} = 0, \hat{X}_{\text{b}}=\sigma)  \Big]\\
    & =   \sum_{\sigma \in [2^{\hat{n}}]} p_{\hat{X}_\text{b}}(\sigma) \sum_{r \in \mathcal{R}} \Big[ I(\hat{\bm{X}}_{\hatcal{A}_r} ;  \hat{D}| \hat{X}_{\text{b}}=\sigma) - I(\hat{\bm{X}}_{\hatcal{A}_r} ; \hat{D} | \hat{\bm{X}}_{\hatcal{B}_r}, \hat{X}_{\text{b}}=\sigma) \\ & \quad  + p_{\hat{D}}(1)I(\hat{\bm{X}}_{\hatcal{A}_r} ;  \hat{\bm{X}}_{\hatcal{B}_r} | \hat{D} = 1, \hat{X}_{\text{b}}=\sigma) + p_{\hat{D}}(0) I(\hat{\bm{X}}_{\hatcal{A}_r} ;  \hat{\bm{X}}_{\hatcal{B}_r} | \hat{D} = 0, \hat{X}_{\text{b}}=\sigma)  \Big]. \label{eq:sigma-sec}
    \end{align}
  \end{subequations}
\else
% ==== put double-column equations here
\begin{subequations}
    \begin{align}
      |\mathcal{R}| \eta &\geq  \sum_{r \in \mathcal{R}} \sum_{\sigma \in [2^{\hat{n}}]} p_{\hat{X}_\text{b}}(\sigma) \Big[ I(\hat{\bm{X}}_{\hatcal{A}_r} ;  \hat{D}| \hat{X}_{\text{b}}=\sigma) \nonumber \\ & \quad - I(\hat{\bm{X}}_{\hatcal{A}_r} ; \hat{D} | \hat{\bm{X}}_{\hatcal{B}_r}, \hat{X}_{\text{b}}=\sigma) \nonumber \\ & \quad  + p_{\hat{D}}(1)I(\hat{\bm{X}}_{\hatcal{A}_r} ;  \hat{\bm{X}}_{\hatcal{B}_r} | \hat{D} = 1, \hat{X}_{\text{b}}=\sigma) \nonumber \\ & \quad + p_{\hat{D}}(0) I(\hat{\bm{X}}_{\hatcal{A}_r} ;  \hat{\bm{X}}_{\hatcal{B}_r} | \hat{D} = 0, \hat{X}_{\text{b}}=\sigma)  \Big]\\
    & =   \sum_{\sigma \in [2^{\hat{n}}]} p_{\hat{X}_\text{b}}(\sigma) \sum_{r \in \mathcal{R}} \Big[ I(\hat{\bm{X}}_{\hatcal{A}_r} ;  \hat{D}| \hat{X}_{\text{b}}=\sigma)\nonumber \\ & \quad - I(\hat{\bm{X}}_{\hatcal{A}_r} ; \hat{D} | \hat{\bm{X}}_{\hatcal{B}_r}, \hat{X}_{\text{b}}=\sigma) \nonumber \\ & \quad  + p_{\hat{D}}(1)I(\hat{\bm{X}}_{\hatcal{A}_r} ;  \hat{\bm{X}}_{\hatcal{B}_r} | \hat{D} = 1, \hat{X}_{\text{b}}=\sigma)\nonumber \\ & \quad + p_{\hat{D}}(0) I(\hat{\bm{X}}_{\hatcal{A}_r} ;  \hat{\bm{X}}_{\hatcal{B}_r} | \hat{D} = 0, \hat{X}_{\text{b}}=\sigma)  \Big]. \label{eq:sigma-sec}
    \end{align}
  \end{subequations}
\fi

We will use this and Lemma~\ref{lemma:mutual-difference} to bound the leakage in $\mathbb{N}'$ in step~\ref{step-5}. As we need to consider both decoding and security simultaneously, we will now consider the probability of decoding error.

\subsubsection{\textnormal{\ref{step-3}}: Relate the decoding criterion in $\mathbb{N}'$ to that in $\mathbb{I}$}

We first define some terminology:

In $\mathbb{I}$, we say that a realisation $(\hat{\bm{X}}_{\mathcal{S}'}, \hat{\bm{X}}_{\mathcal{E}'})$ is $\mathbb{I}$-good if and only if each receiver in $\mathbb{I}$ can decode its required messages correctly. By definition, there are at least $(1 - \epsilon)|\mathcal{X}_{\mathcal{S}'}||\mathcal{X}_{\mathcal{E}'}| = (1 - \epsilon)|\mathcal{X}_{\mathcal{S}'}| 2^{\hat{n}}$ good realisations. We say that a message realisation is $\mathbb{I}$-bad if and only if it is not $\mathbb{I}$-good.

Now, consider $\mathbb{N}'$ using the network code defined in \eqref{eq:ic2nc1} and \eqref{eq:ic2nc2}. We say that a realisation of messages $\bm{X}'_{\mathcal{S}'}$ is $\mathbb{N}'$-good if and only if every receiver in $\mathbb{N}'$ can decode its required messages correctly. By code construction, if $(\hat{\bm{x}}_{\mathcal{S}'}, \hat{\bm{x}}_{\mathcal{E}'})$ is $\mathbb{I}$-good for $\mathbb{I}$, then $\bm{x}'_{\mathcal{S}'} = \hat{\bm{x}}_{\mathcal{S}'}$ is $\mathbb{N}'$-good for $\mathbb{N}'$ using $\sigma = \hat{\mathsf{e}}(\hat{\bm{x}}_{\mathcal{S}'}, \hat{\bm{x}}_{\mathcal{E}'})$ for the network code.

For a specific $\sigma \in [2^{\hat{n}}]$, the set of $\mathbb{I}$-good realisations are defined as follows:
\begin{definition}
  \begin{align}
    \mathcal{Z}_{\sigma} \equalbydef \{  (\hat{\bm{x}}_{\mathcal{S}'}, \hat{\bm{x}}_{\mathcal{E}'}) \in \mathcal{X}_{\mathcal{S}'} \times \mathcal{X}_{\mathcal{E}'}: & \hat{\mathsf{e}}(\hat{\bm{x}}_{\mathcal{S}'}, \hat{\bm{x}}_{\mathcal{E}'}) = \sigma \text{ and }\\ & (\hat{\bm{x}}_{\mathcal{S}'}, \hat{\bm{x}}_{\mathcal{E}'}) \text{ is $\mathbb{I}$-good} \} .
  \end{align}
\end{definition}

Summing over all $\sigma \in [2^{\hat{n}}]$, the total number of $\mathbb{I}$-good realisations in $\mathbb{I}$ is %$\sum_{\sigma \in [2^{\hat{n}}]} |\mathcal{Z}_{\sigma}| \geq (1 - \epsilon)|\mathcal{X}_{\mathcal{S}'}| 2^{\hat{n}}$
  % \begin{equation}
  %   \label{eq:z-average}
    $\sum_{\sigma \in [2^{\hat{n}}]} |\mathcal{Z}_{\sigma}| \geq  (1- \epsilon) |\mathcal{X}_{\mathcal{S}'}| 2^{\hat{n}},$ and the total number of $\mathbb{I}$-bad realisations must be $2^{\hat{n}}|\mathcal{X}_{\mathcal{S}'}| - \sum_{\sigma \in [2^{\hat{n}}]} |\mathcal{Z}_{\sigma}| \leq \epsilon|\mathcal{X}_{\mathcal{S}'}| 2^{\hat{n}}$.
  % \end{equation}
  % meaning that $\frac{1}{2^{\hat{n}}}\sum_{\sigma \in [2^{\hat{n}}]} \frac{|\mathcal{Z}_{\sigma}|}{|\mathcal{X}_{\mathcal{S}'}|} \geq 1 - \epsilon$,
  % or equivalently,

  Next, note that for any chosen $\sigma \in [2^{\hat{n}}]$, invoking Proposition~\ref{prop:only one-xe}, we have $|\mathcal{G}_\sigma| = |\mathcal{Z}_\sigma|$.
So,
  \begin{subequations}
    \begin{align}
      \epsilon &\geq 1- \frac{1}{2^{\hat{n}}}\sum_{\sigma \in [2^{\hat{n}}]} \frac{|\mathcal{G}_{\sigma}|}{|\mathcal{X}_{\mathcal{S}'}|} \label{eq:z-average}\\
               &= \frac{1}{2^{\hat{n}}} \sum_{\sigma \in [2^{\hat{n}}]} \left( \frac{|\mathcal{X}_{\mathcal{S}'}| - |\mathcal{G}_{\sigma}|}{|\mathcal{X}_{\mathcal{S}'}|}   \right)\\
               &= \frac{1}{2^{\hat{n}}} \sum_{\sigma \in [2^{\hat{n}}]} \frac{|\mathcal{G}^\text{c}_\sigma|}{|\mathcal{X}_{\mathcal{S}'}|}\label{eq:sigma-dec}\\
      &= \sum_{\sigma \in [2^{\hat{n}}]} \mathtt{unif}([2^{\hat{n}}]) \frac{|\mathcal{G}^\text{c}_\sigma|}{|\mathcal{X}_{\mathcal{S}'}|}.\label{eq:sigma-dec-3}
      %        &= \frac{1}{2^{\hat{n}}}\sum_{\sigma \in [2^{\hat{n}}]} P_{\text{e}, \sigma} \label{eq:sigma-dec}\\
      % &\geq \sum_{\sigma \in [2^{\hat{n}}]} \mathtt{unif}([2^{n}]) P_{\text{e}, \sigma}. \label{eq:sigma-dec-3}
    \end{align}
  \end{subequations}

  Also, note that by choosing $\sigma$ for the network code, at least $|\mathcal{G}_{\sigma}|$ realisations of $\bm{X}'_{\mathcal{S}'}$ in $\mathbb{N}'$ that are $\mathbb{N}'$-good.
  Since the messages are $\bm{X}'_{\mathcal{S}'}$ uniformly generated, the probability of decoding error in $\mathbb{N}'$ when $\sigma$ is chosen is
  \begin{equation}
    P_{\text{e}, \sigma} \leq  \frac{|\mathcal{G}^\text{c}_\sigma|}{|\mathcal{X}_{\mathcal{S}'}|}.
  \end{equation}

\subsubsection{\textnormal{\ref{step-4}}: Express decodability in $\mathbb{I}$ as an average over $\hat{X}_\text{b}$ using \ref{step-3}.}
  
  For decodability, we would choose a $\sigma$ that is has a low $\frac{|\mathcal{G}^\text{c}_\sigma|}{|\mathcal{X}_{\mathcal{S}'}|}$, which we can then use to upper bound $P_{\text{e},\sigma}$.
%   \begin{subequations}
%   \begin{align}
%     \epsilon &\geq 1- \frac{1}{2^{\hat{n}}}\sum_{\sigma \in [2^{\hat{n}}]} \frac{|\mathcal{Z}_{\sigma'}|}{|\mathcal{X}_{\mathcal{S}'}|} \label{eq:z-average}\\
%              & = \frac{1}{2^{\hat{n}}}\sum_{\sigma \in [2^{\hat{n}}]} 1 -\frac{1}{2^{\hat{n}}}\sum_{\sigma \in [2^{\hat{n}}]} \frac{|\mathcal{Z}_{\sigma'}|}{|\mathcal{X}_{\mathcal{S}'}|}\\
%              &\geq \frac{1}{2^{\hat{n}}}\sum_{\sigma \in [2^{\hat{n}}]}  1 - P_{\text{success}, \sigma}\\
%              &= \frac{1}{2^{\hat{n}}}\sum_{\sigma \in [2^{\hat{n}}]} P_{\text{e}, \sigma} \label{eq:sigma-dec}\\
%                &= \sum_{\sigma \in [2^{\hat{n}}]} \mathtt{unif}([2^{n}]) P_{\text{e}, \sigma},\label{eq:sigma-dec-3}
%   \end{align}
% \end{subequations}
% where $P_{\text{e}, \sigma}$ is the probability of decoding error in $\mathbb{N}'$ when $\sigma$ is chosen.
The difficulty in choosing a suitable $\sigma$ is caused by the different ways in which the leakage and the error probability in $\mathbb{I}$ are related to $\sigma$. See \eqref{eq:sigma-sec} where $\eta$ is related to $p_{\hat{X}_\text{b}}(\sigma)$,  and \eqref{eq:sigma-dec-3} where $\epsilon$ is related to the uniform distribution.

To circumvent this, we will now consider three ways of relating $p_{\hat{X}_\text{b}}$ to $\epsilon$:
%To circumvent this, we will now consider three ways of bounding $\sum_{\sigma \in [2^{\hat{n}}]} p_{\hat{X}_\text{b}}(\sigma) P_{\text{e}, \sigma}$ in terms of $\epsilon$:
\begin{enumerate}[label=(\roman*)]
\item Recall that the total variation distance between $p_{\hat{X}_\text{b}}$ and $\mathtt{unif}([2^{\hat{n}}])$ is defined as $\delta(p_{\hat{X}_\text{b}},\mathtt{unif}([2^{\hat{n}}])) = \frac{1}{2}\sum_{\sigma \in [2^{\hat{n}}]} | p_{\hat{X}_\text{b}}(\sigma) - 2^{-\hat{n}}|$. Also, note that $0 \leq \frac{|\mathcal{G}^\text{c}_{\sigma'}|}{|\mathcal{X}_{\mathcal{S}'}|} \leq 1$ for all $\sigma' \in [2^{\hat{n}}]$ by definition. This means
  \begin{equation}
    \frac{1}{2}\sum_{\sigma \in [2^{\hat{n}}]} \left[ | p_{\hat{X}_\text{b}}(\sigma) - 2^{-\hat{n}}|\frac{|\mathcal{G}^\text{c}_\sigma|}{|\mathcal{X}_{\mathcal{S}'}|} \right] \leq \delta(p_{\hat{X}_\text{b}},\mathtt{unif}([2^{\hat{n}}])),
  \end{equation}
  which implies 
  \begin{equation}
      \sum_{\sigma \in [2^{\hat{n}}]} \left[( p_{\hat{X}_\text{b}}(\sigma) - 2^{-\hat{n}})\frac{|\mathcal{G}^\text{c}_\sigma|}{|\mathcal{X}_{\mathcal{S}'}|} \right]  \leq 2 \delta(p_{\hat{X}_\text{b}},\mathtt{unif}([2^{\hat{n}}])),
    \end{equation}
    and this gives
    \begin{subequations}
    \begin{align}
    \sum_{\sigma \in [2^{\hat{n}}]} p_{\hat{X}_\text{b}}(\sigma)        \frac{|\mathcal{G}^\text{c}_\sigma|}{|\mathcal{X}_{\mathcal{S}'}|} &\leq   \sum_{\sigma \in [2^{\hat{n}}]} 2^{-\hat{n}} \frac{|\mathcal{G}^\text{c}_\sigma|}{|\mathcal{X}_{\mathcal{S}'}|}\nonumber\\ &\quad + 2 \delta(p_{\hat{X}_\text{b}},\mathtt{unif}([2^{\hat{n}}]))\\
      &\leq \epsilon + 2 \delta(p_{\hat{X}_\text{b}},\mathtt{unif}([2^{\hat{n}}])), \label{eq:sigma-dec-4}
    \end{align}
  \end{subequations}
where \eqref{eq:sigma-dec-4}  follows from \eqref{eq:sigma-dec}.

\item Recall again that in $\mathbb{I}$, for a specific $\sigma \in [2^{\hat{n}}]$, there are $|\mathcal{Z}_{\sigma}|$ $\mathbb{I}$-good source realisations that gives $\hat{x}_\text{b}=\sigma$. And, there at most $\epsilon |\mathcal{X}_{\mathcal{S}'}| 2^{\hat{n}}$ $\mathbb{I}$-bad realisations, there are at most $|\mathcal{Z}_{\sigma}| + \epsilon |\mathcal{X}_{\mathcal{S}'}| 2^{\hat{n}}$ realisations that lead to $\hat{x}_\text{b}=\sigma$. By definition,
\begin{subequations}
  \begin{align}
    p_{\hat{X}_\text{b}}(\sigma) &= \sum_{\hat{\bm{x}}_{\mathcal{S}'},\hat{\bm{x}}_{\mathcal{E}'}} p_{\hat{\bm{X}}_{\mathcal{S}'},\hat{\bm{X}}_{\mathcal{E}'},\hat{X}_\text{b}}(\hat{\bm{x}}_{\mathcal{S}'},\hat{\bm{x}}_{\mathcal{E}'},\sigma)\\
                                 &= \sum_{\hat{\bm{x}}_{\mathcal{S}'},\hat{\bm{x}}_{\mathcal{E}'}} p_{\hat{\bm{X}}_{\mathcal{S}'},\hat{\bm{X}}_{\mathcal{E}'}}(\hat{\bm{x}}_{\mathcal{S}'},\hat{\bm{x}}_{\mathcal{E}'})\bm{1}(\hat{\mathsf{e}}(\hat{\bm{x}}_{\mathcal{S}'},\hat{\bm{x}}_{\mathcal{E}'}) = \sigma)\\
    &= \frac{1}{|\mathcal{X}_{\mathcal{S}'}| 2^{\hat{n}}}\sum_{\hat{\bm{x}}_{\mathcal{S}'},\hat{\bm{x}}_{\mathcal{E}'}} \bm{1}(\hat{\mathsf{e}}(\hat{\bm{x}}_{\mathcal{S}'},\hat{\bm{x}}_{\mathcal{E}'}) = \sigma)\\
    &= \frac{\text{the number of realisations that give $\hat{x}_\text{b}=\sigma$}}{|\mathcal{X}_{\mathcal{S}'}| 2^{\hat{n}}}\\
                                 &\leq \frac{|\mathcal{Z}_{\sigma}| + \epsilon |\mathcal{X}_{\mathcal{S}'}| 2^{\hat{n}}}{|\mathcal{X}_{\mathcal{S}'}| 2^{\hat{n}}}\\
                                 &\leq \frac{|\mathcal{X}_{\mathcal{S}'}| + \epsilon |\mathcal{X}_{\mathcal{S}'}| 2^{\hat{n}}}{|\mathcal{X}_{\mathcal{S}'}| 2^{\hat{n}}} \label{eq:fewer-sigma-than-s'}\\
    &= \frac{1}{2^{\hat{n}}} + \epsilon,\label{eq:fewer-sigma-than-s'-2}
  \end{align}
\end{subequations}

Now,
\begin{subequations}
  \begin{align}
    \sum_{\sigma \in [2^{\hat{n}}]} p_{\hat{X}_\text{b}}(\sigma) \frac{|\mathcal{G}^\text{c}_\sigma|}{|\mathcal{X}_{\mathcal{S}'}|} &\leq  \frac{1}{2^{\hat{n}}}\sum_{\sigma \in [2^{\hat{n}}]} \frac{|\mathcal{G}^\text{c}_\sigma|}{|\mathcal{X}_{\mathcal{S}'}|} + \epsilon \sum_{\sigma \in [2^{\hat{n}}]}  \frac{|\mathcal{G}^\text{c}_\sigma|}{|\mathcal{X}_{\mathcal{S}'}|} \\
                     &\leq \epsilon + \epsilon (\epsilon 2^{\hat{n}}) \label{eq:sub-epsilon} \\
    &= \epsilon ( 1 + \epsilon 2^{\hat{n}}), \label{eq:sigma_dec-5}
  \end{align}
\end{subequations}
where \eqref{eq:sub-epsilon}  follows from \eqref{eq:sigma-dec}.
\item Also, since $\frac{|\mathcal{G}^\text{c}_\sigma|}{|\mathcal{X}_{\mathcal{S}'}|} \leq 1$ for all $\sigma$, we have
  \begin{equation}
    \sum_{\sigma \in [2^{\hat{n}}]} p_{\hat{X}_\text{b}}(\sigma) \frac{|\mathcal{G}^\text{c}_\sigma|}{|\mathcal{X}_{\mathcal{S}'}|} \leq 1. \label{eq:sigma-dec-2}
  \end{equation}
\end{enumerate}

% Define $\delta_\sigma \equalbydef p_{\hat{X}_\text{b}}(\sigma) - \frac{1}{2^{\hat{n}}}$. Since $p_{\hat{X}_\text{b}}(\sigma) \leq 1$, we have  So,
% \begin{align}
%   \epsilon &\geq \sum_{\sigma \in [2^{\hat{n}}]} \frac{1}{2^{\hat{n}}} P_{\text{e}, \sigma}\\
%            &= \sum_{\sigma \in [2^{\hat{n}}]} (p_{\hat{X}_\text{b}}(\sigma) - \delta_\sigma)  P_{\text{e}, \sigma}\\
%   \sum_{\sigma \in [2^{\hat{n}}]} p_{\hat{X}_\text{b}}(\sigma)  P_{\text{e}, \sigma} & \leq \epsilon + \sum_{\sigma \in [2^{\hat{n}}]} \delta_\sigma  P_{\text{e}, \sigma}\\
%            &\leq \epsilon + \sum_{\sigma \in [2^{\hat{n}}]} \delta_\sigma \color{red} \text{ this is not correct}\\
%            &= \epsilon + \sum_{\sigma \in [2^{\hat{n}}]} p_{\hat{X}_\text{b}}(\sigma) - \sum_{\sigma \in [2^{\hat{n}}]} \frac{1}{2^{\hat{n}}}\\
%   &= \epsilon.
% \end{align}

From \eqref{eq:sigma-dec-4}, \eqref{eq:sigma_dec-5}, and \eqref{eq:sigma-dec-2}, we have
\ifx\doublecolumn\undefined
% ==== put single-column equations here
\begin{equation}
    \sum_{\sigma \in [2^{\hat{n}}]} p_{\hat{X}_\text{b}}(\sigma) \frac{|\mathcal{G}^\text{c}_\sigma|}{|\mathcal{X}_{\mathcal{S}'}|} \leq  \min \Big\{ \epsilon [1 + 2^{\hat{n}+1} \delta(p_{\hat{X}_\text{b}},\mathtt{unif}([2^{n}])) ] , \,\, \epsilon [ 1 + \epsilon 2^{\hat{n}}],\,\, 1 \Big\} \equalbydef \zeta. \label{eq:zeta}
\end{equation}
\else
% ==== put double-column equations here
\begin{subequations}
  \begin{align}
    \sum_{\sigma \in [2^{\hat{n}}]} p_{\hat{X}_\text{b}}(\sigma) \frac{|\mathcal{G}^\text{c}_\sigma|}{|\mathcal{X}_{\mathcal{S}'}|} \leq  \min \Big\{ &\epsilon [1 + 2 \delta(p_{\hat{X}_\text{b}},\mathtt{unif}([2^{n}])) ] , \,\, \nonumber\\ &\epsilon [ 1 + \epsilon 2^{\hat{n}}],\,\, 1 \Big\} \equalbydef \zeta. \label{eq:zeta}
  \end{align}
\end{subequations}
\fi
  
% \begin{cases}
%     (1 + 2\alpha) \epsilon, & \text{ for sub-case~2a-i}\\
%     \min \{\epsilon ( 1 + \epsilon 2^{\hat{n}}), 1 \}, & \text{ for sub-case~2a-ii},
%   \end{cases}
% \end{equation}

\subsubsection{\textnormal{\ref{step-5}}: Combing the results from steps \ref{step-2} and \ref{step-4} to find a $\sigma = \hat{x}_\text{b}$ that is simultaneously good for security and decodability in $\mathbb{I}$}

Combining \eqref{eq:zeta} and \eqref{eq:sigma-sec}, we get
\ifx\doublecolumn\undefined
% ==== put single-column equations here
\begin{align}
  |\mathcal{R}| \eta + \zeta &\geq  \sum_{\sigma \in [2^{\hat{n}}]} p_{\hat{X}_\text{b}}(\sigma) \Big\{  \frac{|\mathcal{G}^\text{c}_\sigma|}{|\mathcal{X}_{\mathcal{S}'}|} + \sum_{r \in \mathcal{R}} \Big[ I(\hat{\bm{X}}_{\hatcal{A}_r} ;  \hat{D}| \hat{X}_{\text{b}}=\sigma) - I(\hat{\bm{X}}_{\hatcal{A}_r} ; \hat{D} | \hat{\bm{X}}_{\hatcal{B}_r}, \hat{X}_{\text{b}}=\sigma) \\ & \quad  + p_{\hat{D}}(1)I(\hat{\bm{X}}_{\hatcal{A}_r} ;  \hat{\bm{X}}_{\hatcal{B}_r} | \hat{D} = 1, \hat{X}_{\text{b}}=\sigma) + p_{\hat{D}}(0) I(\hat{\bm{X}}_{\hatcal{A}_r} ;  \hat{\bm{X}}_{\hatcal{B}_r} | \hat{D} = 0, \hat{X}_{\text{b}}=\sigma)  \Big] \Big\}.
\end{align}
\else
% ==== put double-column equations here
\begin{align}
  |\mathcal{R}| \eta + \zeta &\geq  \sum_{\sigma \in [2^{\hat{n}}]}p_{\hat{X}_\text{b}}(\sigma) \Bigg( \frac{|\mathcal{G}^\text{c}_\sigma|}{|\mathcal{X}_{\mathcal{S}'}|} +  \sum_{r \in \mathcal{R}} \Big[ I(\hat{\bm{X}}_{\hatcal{A}_r} ;  \hat{D}| \hat{X}_{\text{b}}=\sigma)\nonumber \\ & \quad - I(\hat{\bm{X}}_{\hatcal{A}_r} ; \hat{D} | \hat{\bm{X}}_{\hatcal{B}_r}, \hat{X}_{\text{b}}=\sigma)\nonumber \\ & \quad  + p_{\hat{D}}(1)I(\hat{\bm{X}}_{\hatcal{A}_r} ;  \hat{\bm{X}}_{\hatcal{B}_r} | \hat{D} = 1, \hat{X}_{\text{b}}=\sigma)\nonumber \\ & \quad + p_{\hat{D}}(0) I(\hat{\bm{X}}_{\hatcal{A}_r} ;  \hat{\bm{X}}_{\hatcal{B}_r} | \hat{D} = 0, \hat{X}_{\text{b}}=\sigma)  \Big] \Bigg).
\end{align}
\fi

So, there exists at least one $\sigma \in [2^{\hat{n}}]$ such that
\ifx\doublecolumn\undefined
% ==== put single-column equations here
    \begin{align}
      |\mathcal{R}| \eta + \zeta &\geq \frac{|\mathcal{G}^\text{c}_\sigma|}{|\mathcal{X}_{\mathcal{S}'}|} + \sum_{r \in \mathcal{R}} \Big[ I(\hat{\bm{X}}_{\hatcal{A}_r} ;  \hat{D}| \hat{X}_{\text{b}}=\sigma) - I(\hat{\bm{X}}_{\hatcal{A}_r} ; \hat{D} | \hat{\bm{X}}_{\hatcal{B}_r}, \hat{X}_{\text{b}}=\sigma) \\ & \quad  + p_{\hat{D}}(1)I(\hat{\bm{X}}_{\hatcal{A}_r} ;  \hat{\bm{X}}_{\hatcal{B}_r} | \hat{D} = 1, \hat{X}_{\text{b}}=\sigma) + p_{\hat{D}}(0) I(\hat{\bm{X}}_{\hatcal{A}_r} ;  \hat{\bm{X}}_{\hatcal{B}_r} | \hat{D} = 0, \hat{X}_{\text{b}}=\sigma)  \Big],
    \end{align}
\else
% ==== put double-column equations here
    \begin{align}
      |\mathcal{R}| \eta + \zeta &\geq \frac{|\mathcal{G}^\text{c}_\sigma|}{|\mathcal{X}_{\mathcal{S}'}|} + \sum_{r \in \mathcal{R}} \Big[ I(\hat{\bm{X}}_{\hatcal{A}_r} ;  \hat{D}| \hat{X}_{\text{b}}=\sigma)\nonumber \\ & \quad - I(\hat{\bm{X}}_{\hatcal{A}_r} ; \hat{D} | \hat{\bm{X}}_{\hatcal{B}_r}, \hat{X}_{\text{b}}=\sigma)\nonumber \\ & \quad  + p_{\hat{D}}(1)I(\hat{\bm{X}}_{\hatcal{A}_r} ;  \hat{\bm{X}}_{\hatcal{B}_r} | \hat{D} = 1, \hat{X}_{\text{b}}=\sigma)\nonumber \\ & \quad + p_{\hat{D}}(0) I(\hat{\bm{X}}_{\hatcal{A}_r} ;  \hat{\bm{X}}_{\hatcal{B}_r} | \hat{D} = 0, \hat{X}_{\text{b}}=\sigma)  \Big],
    \end{align}
\fi
from which we have the probability of decoding error in $\mathbb{N}'$ being bounded from above as
\begin{equation}
 P_{\text{e}, \sigma} \leq \frac{|\mathcal{G}^\text{c}_\sigma|}{|\mathcal{X}_{\mathcal{S}'}|} \leq  |\mathcal{R}| \eta + \zeta,
\end{equation}
and the following security constraint for $\mathbb{I}$:
  \begin{align}
    &\sum_{r \in \mathcal{R}} p_{\hat{D}}(1)I(\hat{\bm{X}}_{\hatcal{A}_r} ;  \hat{\bm{X}}_{\hatcal{B}_r} | \hat{D} = 1, \hat{X}_{\text{b}}=\sigma)\nonumber\\ & \leq |\mathcal{R}| \eta + \zeta + \sum_{r \in \mathcal{R}} \Big[- I(\hat{\bm{X}}_{\hatcal{A}_r} ;  \hat{D}| \hat{X}_{\text{b}}=\sigma) \nonumber \\ &\quad + I(\hat{\bm{X}}_{\hatcal{A}_r} ; \hat{D} | \hat{\bm{X}}_{\hatcal{B}_r}, \hat{X}_{\text{b}}=\sigma) \nonumber \\ &\quad  -p_{\hat{D}}(0) I(\hat{\bm{X}}_{\hatcal{A}_r} ;  \hat{\bm{X}}_{\hatcal{B}_r} | \hat{D} = 0, \hat{X}_{\text{b}}=\sigma)  \Big],
  \end{align}
  which implies that for every eavesdropper $r' \in \mathcal{R}$,
  \ifx\doublecolumn\undefined
  % ==== put single-column equations here
  \begin{align}
    &I(\hat{\bm{X}}_{\hatcal{A}_{r'}} ;  \hat{\bm{X}}_{\hatcal{B}_{r'}} | \hat{D} = 1, \hat{X}_{\text{b}}=\sigma) & \leq \frac{1}{p_{\hat{D}}(1)} \left[ |\mathcal{R}| \eta + \zeta + \sum_{r \in \mathcal{R}} I(\hat{\bm{X}}_{\hatcal{A}_r} ; \hat{D} | \hat{\bm{X}}_{\hatcal{B}_r}, \hat{X}_{\text{b}}=\sigma) \right]\\
   &\leq \frac{1}{1 - \epsilon} [|\mathcal{R}| ( \eta + H(\hat{D})) + \zeta]. \label{eq:final}
  \end{align}
\else
% ==== put double-column equations here
\begin{align}
    &I(\hat{\bm{X}}_{\hatcal{A}_{r'}} ;  \hat{\bm{X}}_{\hatcal{B}_{r'}} | \hat{D} = 1, \hat{X}_{\text{b}}=\sigma)\nonumber\\ & \leq \frac{1}{p_{\hat{D}}(1)} \left[ |\mathcal{R}| \eta + \zeta + \sum_{r \in \mathcal{R}} I(\hat{\bm{X}}_{\hatcal{A}_r} ; \hat{D} | \hat{\bm{X}}_{\hatcal{B}_r}, \hat{X}_{\text{b}}=\sigma) \right]\\
   &\leq \frac{1}{1 - \epsilon} [|\mathcal{R}| ( \eta + H(\hat{D})) + \zeta]. \label{eq:final}
  \end{align}
\fi
  Note that $p_{\hat{D}}(1) \geq 1 - \epsilon$, and we have assumed that $0 < \epsilon \leq 0.5$.

Lastly, recall by definition that $\epsilon' \equalbydef \frac{|\mathcal{G}^\text{c}_\sigma|}{ |\mathcal{X}_{\mathcal{S}'}|} \leq  |\mathcal{R}| \eta + \zeta$ for the chosen $\sigma$.
By substituting Lemma~\ref{lemma:mutual-difference} into \eqref{eq:final}, we have the following security constraint for $\mathbb{N}'$:
\ifx\doublecolumn\undefined
% ==== put single-column equations here
\begin{subequations}
\begin{align}
  &I(\bm{X}'_{\mathcal{A}'_{r'}}; \bm{X}'_{\mathcal{B}'_{r'}})\nonumber \\
  &\leq  \frac{1}{1 - \epsilon} [|\mathcal{R}| ( \eta + H_\text{b}(\epsilon)) + \zeta]\nonumber \\
  &\quad + (|\mathcal{R}| \eta + \zeta) \log |\mathcal{X}_{\mathcal{S}'}| - \log (1 - (|\mathcal{R}| \eta + \zeta))\nonumber\\ &\quad + \frac{(|\mathcal{R}| \eta + \zeta)}{1-(|\mathcal{R}| \eta + \zeta)} \left(  \log e + \hat{n}  \right)) \\
  & = (|\mathcal{R}| \eta + \zeta) \left( \frac{1}{1 - \epsilon} + \frac{\log e  + \hat{n}}{1-(|\mathcal{R}| \eta + \zeta)}+ \log |\mathcal{X}_{\mathcal{S}'}|  \right) \nonumber \\
      &\quad + \frac{1}{1 - \epsilon} |\mathcal{R}| H_\text{b}(\epsilon) - \log \left(1 - (|\mathcal{R}| \eta + \zeta)\right),
\end{align}
\end{subequations}
\else
% ==== put double-column equations here
\begin{subequations}
\begin{align}
  &I(\bm{X}'_{\mathcal{A}'_{r'}}; \bm{X}'_{\mathcal{B}'_{r'}})\nonumber \\
  &\leq  \frac{1}{1 - \epsilon} [|\mathcal{R}| ( \eta + H_\text{b}(\epsilon)) + \zeta]\nonumber \\
  &\quad + (|\mathcal{R}| \eta + \zeta) \log |\mathcal{X}_{\mathcal{S}'}| - \log (1 - (|\mathcal{R}| \eta + \zeta))\nonumber\\ &\quad + \frac{(|\mathcal{R}| \eta + \zeta)}{1-(|\mathcal{R}| \eta + \zeta)} \left(  \log e + \hat{n}  \right)) \\
  & = (|\mathcal{R}| \eta + \zeta) \left( \frac{1}{1 - \epsilon} + \frac{\log e  + \hat{n}}{1-(|\mathcal{R}| \eta + \zeta)}+ \log |\mathcal{X}_{\mathcal{S}'}|  \right) \nonumber \\
      &\quad + \frac{1}{1 - \epsilon} |\mathcal{R}| H_\text{b}(\epsilon) - \log \left(1 - (|\mathcal{R}| \eta + \zeta)\right),
\end{align}
\end{subequations}
\fi
for each $r' \in \mathcal{R}$ in $\mathbb{N}'$.

Finally, note that $I(\bm{X}'_{\mathcal{A}'_{r'}}; \bm{X}'_{\mathcal{B}'_{r'}}) \leq (\bm{X}'_{\mathcal{B}'_{r'}}) \leq H(\bm{X}'_{\mathcal{E}'}) \leq \hat{n}$.
\hfill $\blacksquare$

  \bibliography{../bib}

  \appendices

  \section{Proof of Lemma~\ref{lemma:mutual-difference}} \label{appendix:proof-lemma}
  
  Lemma~\ref{lemma:mutual-difference} follows directly from the following two proposition:
\begin{proposition} \label{lemma:hb}
  For any $\sigma \in [ 2^{\hat{n}}]$,
  \ifx\doublecolumn\undefined
  % ==== put single-column equations here
  $$H(\bm{X}'_{\mathcal{B}'_r}) - H(\hat{\bm{X}}_{\hatcal{B}_r} | \hat{D}=1, \hat{X}_{\text{b}} = \sigma) \leq \epsilon' \log |\mathcal{X}_{\mathcal{S}'}| - \log (1 - \epsilon').$$
\else
% ==== put double-column equations here
\begin{align}
  & H(\bm{X}'_{\mathcal{B}'_r}) - H(\hat{\bm{X}}_{\hatcal{B}_r} | \hat{D}=1, \hat{X}_{\text{b}} = \sigma)\nonumber \\ &\quad  \leq \epsilon' \log |\mathcal{X}_{\mathcal{S}'}| - \log (1 - \epsilon').
\end{align}
\fi
\end{proposition}

\begin{proposition} \label{lemma:hb-2}
  For any $\sigma \in [ 2^{\hat{n}}]$,
  \ifx\doublecolumn\undefined
  % ==== put single-column equations here
  \begin{equation}
     H( \hat{\bm{X}}_{\hatcal{B}_r} | \hat{\bm{X}}_{\hatcal{A}_r},\hat{D}=1, \hat{X}_{\text{b}} = \sigma) - H(\bm{X}'_{\mathcal{B}'_r} | \bm{X}'_{\mathcal{A}'_r}) \leq \frac{\epsilon'}{1-\epsilon'} \left(  \log e + \hat{n}  \right).
  \end{equation}
\else
% ==== put double-column equations here
\begin{align}
     &H( \hat{\bm{X}}_{\hatcal{B}_r} | \hat{\bm{X}}_{\hatcal{A}_r},\hat{D}=1, \hat{X}_{\text{b}} = \sigma) - H(\bm{X}'_{\mathcal{B}'_r} | \bm{X}'_{\mathcal{A}'_r})\nonumber \\ &\quad \leq \frac{\epsilon'}{1-\epsilon'} \left(  \log e + \hat{n}  \right)).
  \end{align}
\fi
\end{proposition}

With the above proposition we have Lemma~\ref{lemma:mutual-difference}, as follows:
\begin{subequations}
  \begin{align*}
    &I(\bm{X}'_{\mathcal{A}'_r}; \bm{X}'_{\mathcal{B}'_r}) - I(\hat{\bm{X}}_{\hatcal{A}_r} ;  \hat{\bm{X}}_{\hatcal{B}_r} | \hat{D}=1, \hat{X}_{\text{b}} = \sigma) \nonumber\\
    & = H(\bm{X}'_{\mathcal{B}'_r}) - H(\bm{X}'_{\mathcal{B}'_r} | \bm{X}'_{\mathcal{A}'_r}) - [ H(\hat{\bm{X}}_{\hatcal{B}_r} | \hat{D}=1, \hat{X}_{\text{b}} = \sigma)\nonumber \\ &\quad -  H( \hat{\bm{X}}_{\hatcal{B}_r} | \hat{\bm{X}}_{\hatcal{A}_r},\hat{D}=1, \hat{X}_{\text{b}} = \sigma)]\\
    &\leq \epsilon' \log |\mathcal{X}_{\mathcal{S}'}| - \log (1 - \epsilon') + \frac{\epsilon'}{1-\epsilon'} \left(  \log e + \hat{n}  \right)). &\hfill \blacksquare
  \end{align*}
\end{subequations}

In the following, we omit the subscript of probability mass functions. The reader can easily infer the subscript from the argument. %We use $\hat{p}(\cdot)$ for $\mathbb{I}$ and $p'(\cdot)$ for $\mathbb{N}'$.

\subsection{Proof of Proposition~\ref{lemma:hb}}
  Recall that the messages $(\hat{\bm{X}}_{\mathcal{S}'},\hat{\bm{X}}_{\mathcal{E}'})$ for $\mathbb{I}$ and $\bm{X}'_{\mathcal{S}'}$ for $\mathbb{N}'$ are both uniformly distributed.
 % As the probability of decoding error in $\mathbb{I}$ is upper bounded by $\epsilon$, there are at most $\epsilon |\mathcal{X}_{\mathcal{S}'}| |\mathcal{X}_{\mathcal{E}'}|$ message realisations $(\hat{\bm{x}}_{\mathcal{S}'},\hat{\bm{x}}_{\mathcal{E}'})$ that result in error decoding.
 From Proposition~\ref{prop:only one-xe}, if decoding is successful, that is, $\hat{d}=1$, we know that for each $\hat{\bm{x}}_{\mathcal{S}'} \in \mathcal{G}_\sigma$, there is only one unique $\hat{\bm{x}}_{\mathcal{E}'}$ for which $\hat{\mathsf{e}}(\hat{\bm{x}}_{\mathcal{S}'},\hat{\bm{x}}_{\mathcal{E}'}) = \sigma$. %This means $|\mathcal{G}_\sigma| \geq |\mathcal{X}_{\mathcal{S}'}| - \epsilon |\mathcal{X}_{\mathcal{S}'}| |\mathcal{X}_{\mathcal{E}'}| = (1 - \epsilon 2^{\hat{n}}) |\mathcal{X}_{\mathcal{S}'}|$.
 % As $(\hat{\bm{X}}_{\mathcal{S}'},\hat{\bm{X}}_{\mathcal{E}'})$ and $\bm{X}'_{\mathcal{S}'}$ are both uniformly distributed, we have that
 This implies
\begin{align}
  p(\hat{\bm{x}}_{\mathcal{S}'}|1,\sigma) &= \begin{cases}
    \displaystyle \frac{1}{|\mathcal{G}_\sigma|}, & \text{ if } \hat{\bm{x}}_{\mathcal{S}'} \in \mathcal{G}_\sigma, \\
    0, & \text{ otherwise};
  \end{cases} \label{eq:uniform-001}\\
  \intertext{and }
  p(\bm{x}'_{\mathcal{S}'}) &= \frac{1}{|\mathcal{X}_{\mathcal{S}'}|}, \quad \text{ for all } \bm{x}'_{\mathcal{S}'} \in \mathcal{X}_{\mathcal{S}'}.
\end{align}

Note that when decoding is correct, $\bm{\hat{x}}_{\mathcal{E}'} = \phi_\sigma(\hat{\bm{x}}_{\mathcal{S}'})$ is a deterministic function of $\hat{\bm{x}}_{\mathcal{S}'}$ and $\sigma$. So, 
\begin{subequations}
  \begin{align}
    p (\hat{\bm{x}}_{\hatcal{B}_r}| 1, \sigma) &= \sum_{\hat{\bm{x}}_{\mathcal{S}'}} p(\hat{\bm{x}}_{\mathcal{S}'},\hat{\bm{x}}_{\hatcal{B}_r}| 1, \sigma)\\
                                               &= \sum_{\hat{\bm{x}}_{\mathcal{S}'}} p(\hat{\bm{x}}_{\hatcal{B}_r}|\hat{\bm{x}}_{\mathcal{S}'}, 1, \sigma) p(\hat{\bm{x}}_{\mathcal{S}'}| 1, \sigma)\\
    &= \sum_{\hat{\bm{x}}_{\mathcal{S}'}} \bm{1}([\phi_\sigma(\hat{\bm{x}}_{\mathcal{S}'})]_{\hatcal{B}_r} =\hat{\bm{x}}_{\hatcal{B}_r}) p(\hat{\bm{x}}_{\mathcal{S}'}| 1, \sigma)\\
    &= \sum_{\substack{\hat{\bm{x}}_{\mathcal{S}'} \in \mathcal{X}_{\mathcal{S}'}\\\text{s.t. } [\phi_\sigma(\hat{\bm{x}}_{\mathcal{S}'})]_{\hatcal{B}_r} =\hat{\bm{x}}_{\hatcal{B}_r}}} p(\hat{\bm{x}}_{\mathcal{S}'}| 1, \sigma), \label{eq:sub-001}
  \end{align}
\end{subequations}
where we have used the notation $[\bm{a}_\mathcal{A}]_{\mathcal{B}} \equalbydef \bm{a}_\mathcal{B}$ to denote a sub-vector, for some $\mathcal{B} \subseteq \mathcal{A}$. Similarly,
\begin{subequations}
  \begin{align}
    p(\bm{x}'_{\mathcal{B}'_r}) &= \sum_{\bm{x}'_{\mathcal{S}'}} p(\bm{x}'_{\mathcal{S}'},\bm{x}'_{\mathcal{B}'_r})\\
                                &= \sum_{\substack{\bm{x}'_{\mathcal{S}'} \in \mathcal{X}_{\mathcal{S}'}\\\text{s.t. } [\phi_\sigma(\bm{x}'_{\mathcal{S}'})]_{\mathcal{B}'_r} =\bm{x}'_{\mathcal{B}'_r}}} p(\bm{x}'_{\mathcal{S}'})\\
    &= \sum_{\substack{\bm{x}'_{\mathcal{S}'} \in \mathcal{X}_{\mathcal{S}'}\\\text{s.t. } [\phi_\sigma(\bm{x}'_{\mathcal{S}'})]_{\mathcal{B}'_r} =\bm{x}'_{\mathcal{B}'_r}}} \frac{1}{|\mathcal{X}'_{\mathcal{S}'}|}. \label{eq:sub-002}
  \end{align}
\end{subequations}

Now, using \eqref{eq:sub-001},
\ifx\doublecolumn\undefined
% ==== put single-column equations here
\begin{subequations}
  \begin{align}
    &H(\hat{\bm{X}}_{\hatcal{B}_r} | \hat{D}=1, \hat{X}_{\text{b}} = \sigma) \nonumber\\
    & = - \sum_{\hat{\bm{x}}_{\hatcal{B}_r}} p (\hat{\bm{x}}_{\hatcal{B}_r}| 1, \sigma) \log p (\hat{\bm{x}}_{\hatcal{B}_r}| 1, \sigma)\\
    &= - \sum_{\hat{\bm{x}}_{\hatcal{B}_r}} \left(\sum_{\substack{\hat{\bm{x}}_{\mathcal{S}'} \in \mathcal{X}_{\mathcal{S}'}\\\text{s.t. } [\phi_\sigma(\hat{\bm{x}}_{\mathcal{S}'})]_{\hatcal{B}_r} =\hat{\bm{x}}_{\hatcal{B}_r}}} p(\hat{\bm{x}}_{\mathcal{S}'}| 1, \sigma)  \right) \log \left(\sum_{\substack{\hat{\bm{x}}_{\mathcal{S}'} \in \mathcal{X}_{\mathcal{S}'}\\\text{s.t. } [\phi_\sigma(\hat{\bm{x}}_{\mathcal{S}'})]_{\hatcal{B}_r} =\hat{\bm{x}}_{\hatcal{B}_r}}} p(\hat{\bm{x}}_{\mathcal{S}'}| 1, \sigma)  \right)\\
    &= - \sum_{\hat{\bm{x}}_{\hatcal{B}_r}} \left(\sum_{\substack{\hat{\bm{x}}_{\mathcal{S}'} \in \mathcal{G}_\sigma \\ \text{s.t. } [\phi_\sigma(\hat{\bm{x}}_{\mathcal{S}'})]_{\hatcal{B}_r} =\hat{\bm{x}}_{\hatcal{B}_r}}} \frac{1}{|\mathcal{G}_\sigma|}  \right) \log \left(\sum_{\substack{\hat{\bm{x}}_{\mathcal{S}'} \in \mathcal{G}_\sigma \\ \text{s.t. } [\phi_\sigma(\hat{\bm{x}}_{\mathcal{S}'})]_{\hatcal{B}_r} =\hat{\bm{x}}_{\hatcal{B}_r}}} \frac{1}{|\mathcal{G}_\sigma|}  \right)\\
    &= - \sum_{\hat{\bm{x}}_{\hatcal{B}_r}} \left(\sum_{\substack{\hat{\bm{x}}_{\mathcal{S}'} \in \mathcal{G}_\sigma \\ \text{s.t. } [\phi_\sigma(\hat{\bm{x}}_{\mathcal{S}'})]_{\hatcal{B}_r} =\hat{\bm{x}}_{\hatcal{B}_r}}} \frac{1}{(1-\epsilon') |\mathcal{X}_{\mathcal{S}'}|}  \right) \log \left(\sum_{\substack{\hat{\bm{x}}_{\mathcal{S}'} \in \mathcal{G}_\sigma \\ \text{s.t. } [\phi_\sigma(\hat{\bm{x}}_{\mathcal{S}'})]_{\hatcal{B}_r} =\hat{\bm{x}}_{\hatcal{B}_r}}} \frac{1}{(1-\epsilon') |\mathcal{X}_{\mathcal{S}'}|}  \right)\\
    &= - \frac{1}{1 - \epsilon'} \sum_{\hat{\bm{x}}_{\hatcal{B}_r}} \left(\sum_{\substack{\hat{\bm{x}}_{\mathcal{S}'} \in \mathcal{G}_\sigma \\ \text{s.t. } [\phi_\sigma(\hat{\bm{x}}_{\mathcal{S}'})]_{\hatcal{B}_r} =\hat{\bm{x}}_{\hatcal{B}_r}}} \frac{1}{ |\mathcal{X}_{\mathcal{S}'}|}  \right) \log \left(\sum_{\substack{\hat{\bm{x}}_{\mathcal{S}'} \in \mathcal{G}_\sigma \\ \text{s.t. } [\phi_\sigma(\hat{\bm{x}}_{\mathcal{S}'})]_{\hatcal{B}_r} =\hat{\bm{x}}_{\hatcal{B}_r}}} \frac{1}{|\mathcal{X}_{\mathcal{S}'}|}  \right) + \log (1 - \epsilon') \label{eq:difference-1}
  \end{align}
\end{subequations}
\else
% ==== put double-column equations here
\begin{subequations}
  \begin{align}
    &H(\hat{\bm{X}}_{\hatcal{B}_r} | \hat{D}=1, \hat{X}_{\text{b}} = \sigma) \nonumber\\
    & = - \sum_{\hat{\bm{x}}_{\hatcal{B}_r}} p (\hat{\bm{x}}_{\hatcal{B}_r}| 1, \sigma) \log p (\hat{\bm{x}}_{\hatcal{B}_r}| 1, \sigma)\\
    &= - \sum_{\hat{\bm{x}}_{\hatcal{B}_r}} \left(\sum_{\substack{\hat{\bm{x}}_{\mathcal{S}'} \in \mathcal{X}_{\mathcal{S}'}\\\text{s.t. } [\phi_\sigma(\hat{\bm{x}}_{\mathcal{S}'})]_{\hatcal{B}_r} =\hat{\bm{x}}_{\hatcal{B}_r}}} p(\hat{\bm{x}}_{\mathcal{S}'}| 1, \sigma)  \right)\nonumber \\ &\quad\quad\quad  \log \left(\sum_{\substack{\hat{\bm{x}}_{\mathcal{S}'} \in \mathcal{X}_{\mathcal{S}'}\\\text{s.t. } [\phi_\sigma(\hat{\bm{x}}_{\mathcal{S}'})]_{\hatcal{B}_r} =\hat{\bm{x}}_{\hatcal{B}_r}}} p(\hat{\bm{x}}_{\mathcal{S}'}| 1, \sigma)  \right)\\
    &= - \sum_{\hat{\bm{x}}_{\hatcal{B}_r}} \left(\sum_{\substack{\hat{\bm{x}}_{\mathcal{S}'} \in \mathcal{G}_\sigma \\ \text{s.t. } [\phi_\sigma(\hat{\bm{x}}_{\mathcal{S}'})]_{\hatcal{B}_r} =\hat{\bm{x}}_{\hatcal{B}_r}}} \frac{1}{|\mathcal{G}_\sigma|}  \right) \nonumber \\ &\quad\quad\quad  \log \left(\sum_{\substack{\hat{\bm{x}}_{\mathcal{S}'} \in \mathcal{G}_\sigma \\ \text{s.t. } [\phi_\sigma(\hat{\bm{x}}_{\mathcal{S}'})]_{\hatcal{B}_r} =\hat{\bm{x}}_{\hatcal{B}_r}}} \frac{1}{|\mathcal{G}_\sigma|}  \right)\\
    &= - \sum_{\hat{\bm{x}}_{\hatcal{B}_r}} \left(\sum_{\substack{\hat{\bm{x}}_{\mathcal{S}'} \in \mathcal{G}_\sigma \\ \text{s.t. } [\phi_\sigma(\hat{\bm{x}}_{\mathcal{S}'})]_{\hatcal{B}_r} =\hat{\bm{x}}_{\hatcal{B}_r}}} \frac{1}{(1-\epsilon') |\mathcal{X}_{\mathcal{S}'}|}  \right)\nonumber \\ &\quad\quad\quad  \log \left(\sum_{\substack{\hat{\bm{x}}_{\mathcal{S}'} \in \mathcal{G}_\sigma \\ \text{s.t. } [\phi_\sigma(\hat{\bm{x}}_{\mathcal{S}'})]_{\hatcal{B}_r} =\hat{\bm{x}}_{\hatcal{B}_r}}} \frac{1}{(1-\epsilon') |\mathcal{X}_{\mathcal{S}'}|}  \right)\\
    &= - \frac{1}{1 - \epsilon'} \sum_{\hat{\bm{x}}_{\hatcal{B}_r}} \left(\sum_{\substack{\hat{\bm{x}}_{\mathcal{S}'} \in \mathcal{G}_\sigma \\ \text{s.t. } [\phi_\sigma(\hat{\bm{x}}_{\mathcal{S}'})]_{\hatcal{B}_r} =\hat{\bm{x}}_{\hatcal{B}_r}}} \frac{1}{ |\mathcal{X}_{\mathcal{S}'}|}  \right) \nonumber \\ &\quad\quad\quad  \log \left(\sum_{\substack{\hat{\bm{x}}_{\mathcal{S}'} \in \mathcal{G}_\sigma \\ \text{s.t. } [\phi_\sigma(\hat{\bm{x}}_{\mathcal{S}'})]_{\hatcal{B}_r} =\hat{\bm{x}}_{\hatcal{B}_r}}} \frac{1}{|\mathcal{X}_{\mathcal{S}'}|}  \right) + \log (1 - \epsilon') \label{eq:difference-1}
  \end{align}
\end{subequations}
\fi
where \eqref{eq:difference-1} follows from $\displaystyle \frac{1}{1 - \epsilon'} \sum_{\hat{\bm{x}}_{\hatcal{B}_r}} \sum_{\substack{\hat{\bm{x}}_{\mathcal{S}'} \in \mathcal{G}_\sigma \\ \text{s.t. } [\phi_\sigma(\hat{\bm{x}}_{\mathcal{S}'})]_{\hatcal{B}_r} =\hat{\bm{x}}_{\hatcal{B}_r}}} \frac{1}{ |\mathcal{X}_{\mathcal{S}'}|}   = \sum_{\hat{\bm{x}}_{\hatcal{B}_r}} p (\hat{\bm{x}}_{\hatcal{B}_r}| 1, \sigma) = 1$.

Next, using \eqref{eq:sub-002},
\ifx\doublecolumn\undefined
% ==== put single-column equations here
\begin{subequations}
  \begin{align}
    & H(\bm{X}'_{\mathcal{B}'_r}) \nonumber\\
    &= - \sum_{\bm{x}'_{\mathcal{B}'_r}} p(\bm{x}'_{\mathcal{B}'_r}) \log p(\bm{x}'_{\mathcal{B}'_r})\\
    &= - \sum_{\bm{x}'_{\mathcal{B}'_r}} \left(\sum_{\substack{\bm{x}'_{\mathcal{S}'} \in \mathcal{X}_{\mathcal{S}'}\\\text{s.t. } [\phi_\sigma(\bm{x}'_{\mathcal{S}'})]_{\mathcal{B}'_r} =\bm{x}'_{\mathcal{B}'_r}}} \frac{1}{|\mathcal{X}'_{\mathcal{S}'}|} \right) \log p(\bm{x}'_{\mathcal{B}'_r})\\
    &= - \sum_{\bm{x}'_{\mathcal{B}'_r}} \left(\sum_{\substack{\bm{x}'_{\mathcal{S}'} \in \mathcal{G}_\sigma\\\text{s.t. } [\phi_\sigma(\bm{x}'_{\mathcal{S}'})]_{\mathcal{B}'_r} =\bm{x}'_{\mathcal{B}'_r}}} \frac{1}{|\mathcal{X}'_{\mathcal{S}'}|} \right) \log \left(\sum_{\substack{\bm{x}'_{\mathcal{S}'} \in \mathcal{X}_{\mathcal{S}'}\\\text{s.t. } [\phi_\sigma(\bm{x}'_{\mathcal{S}'})]_{\mathcal{B}'_r} =\bm{x}'_{\mathcal{B}'_r}}} \frac{1}{|\mathcal{X}'_{\mathcal{S}'}|} \right) \nonumber \\ &\quad  - \sum_{\bm{x}'_{\mathcal{B}'_r}} \left(\sum_{\substack{\bm{x}'_{\mathcal{S}'} \in \mathcal{X}_{\mathcal{S}'} \setminus \mathcal{G}_\sigma\\\text{s.t. } [\phi_\sigma(\bm{x}'_{\mathcal{S}'})]_{\mathcal{B}'_r} =\bm{x}'_{\mathcal{B}'_r}}} \frac{1}{|\mathcal{X}'_{\mathcal{S}'}|} \right) \log \left(\sum_{\substack{\bm{x}'_{\mathcal{S}'} \in \mathcal{X}_{\mathcal{S}'}\\\text{s.t. } [\phi_\sigma(\bm{x}'_{\mathcal{S}'})]_{\mathcal{B}'_r} =\bm{x}'_{\mathcal{B}'_r}}} \frac{1}{|\mathcal{X}'_{\mathcal{S}'}|} \right)\\
    &\leq - \sum_{\bm{x}'_{\mathcal{B}'_r}} \left(\sum_{\substack{\bm{x}'_{\mathcal{S}'} \in \mathcal{G}_\sigma\\\text{s.t. } [\phi_\sigma(\bm{x}'_{\mathcal{S}'})]_{\mathcal{B}'_r} =\bm{x}'_{\mathcal{B}'_r}}} \frac{1}{|\mathcal{X}'_{\mathcal{S}'}|} \right)   \log \left(\sum_{\substack{\bm{x}'_{\mathcal{S}'} \in \mathcal{G}_\sigma \\\text{s.t. } [\phi_\sigma(\bm{x}'_{\mathcal{S}'})]_{\mathcal{B}'_r} =\bm{x}'_{\mathcal{B}'_r}}} \frac{1}{|\mathcal{X}'_{\mathcal{S}'}|} \right) - \frac{1}{|\mathcal{X}'_{\mathcal{S}'}|} \left( \sum_{\bm{x}'_{\mathcal{B}'_r}} \sum_{\substack{\bm{x}'_{\mathcal{S}'} \in \mathcal{X}_{\mathcal{S}'} \setminus \mathcal{G}_\sigma\\\text{s.t. } [\phi_\sigma(\bm{x}'_{\mathcal{S}'})]_{\mathcal{B}'_r} =\bm{x}'_{\mathcal{B}'_r}}}   \log \left( \frac{1}{ |\mathcal{X}_{\mathcal{S}'}|} \right)  \right) \label{eq:reduce-set}\\
    &\leq - \frac{1}{1 - \epsilon'}  \sum_{\bm{x}'_{\mathcal{B}'_r}} \left(\sum_{\substack{\bm{x}'_{\mathcal{S}'} \in \mathcal{G}_\sigma\\\text{s.t. } [\phi_\sigma(\bm{x}'_{\mathcal{S}'})]_{\mathcal{B}'_r} =\bm{x}'_{\mathcal{B}'_r}}} \frac{1}{|\mathcal{X}'_{\mathcal{S}'}|} \right)  \log \left(\sum_{\substack{\bm{x}'_{\mathcal{S}'} \in \mathcal{G}_\sigma \\\text{s.t. } [\phi_\sigma(\bm{x}'_{\mathcal{S}'})]_{\mathcal{B}'_r} =\bm{x}'_{\mathcal{B}'_r}}} \frac{1}{|\mathcal{X}'_{\mathcal{S}'}|} \right) + \epsilon' \log |\mathcal{X}_{\mathcal{S}'}|
    \label{eq:difference-2}
  \end{align}
\end{subequations}
\else
% ==== put double-column equations here
\begin{subequations}
  \begin{align}
    & H(\bm{X}'_{\mathcal{B}'_r}) \nonumber\\
    &= - \sum_{\bm{x}'_{\mathcal{B}'_r}} p(\bm{x}'_{\mathcal{B}'_r}) \log p(\bm{x}'_{\mathcal{B}'_r})\\
    &= - \sum_{\bm{x}'_{\mathcal{B}'_r}} \left(\sum_{\substack{\bm{x}'_{\mathcal{S}'} \in \mathcal{X}_{\mathcal{S}'}\\\text{s.t. } [\phi_\sigma(\bm{x}'_{\mathcal{S}'})]_{\mathcal{B}'_r} =\bm{x}'_{\mathcal{B}'_r}}} \frac{1}{|\mathcal{X}'_{\mathcal{S}'}|} \right) \log p(\bm{x}'_{\mathcal{B}'_r})\\
    &= - \sum_{\bm{x}'_{\mathcal{B}'_r}} \left(\sum_{\substack{\bm{x}'_{\mathcal{S}'} \in \mathcal{G}_\sigma\\\text{s.t. } [\phi_\sigma(\bm{x}'_{\mathcal{S}'})]_{\mathcal{B}'_r} =\bm{x}'_{\mathcal{B}'_r}}} \frac{1}{|\mathcal{X}'_{\mathcal{S}'}|} \right) \log \left(\sum_{\substack{\bm{x}'_{\mathcal{S}'} \in \mathcal{X}_{\mathcal{S}'}\\\text{s.t. } [\phi_\sigma(\bm{x}'_{\mathcal{S}'})]_{\mathcal{B}'_r} =\bm{x}'_{\mathcal{B}'_r}}} \frac{1}{|\mathcal{X}'_{\mathcal{S}'}|} \right) \nonumber \\ &\quad  - \sum_{\bm{x}'_{\mathcal{B}'_r}} \left(\sum_{\substack{\bm{x}'_{\mathcal{S}'} \in \mathcal{X}_{\mathcal{S}'} \setminus \mathcal{G}_\sigma\\\text{s.t. } [\phi_\sigma(\bm{x}'_{\mathcal{S}'})]_{\mathcal{B}'_r} =\bm{x}'_{\mathcal{B}'_r}}} \frac{1}{|\mathcal{X}'_{\mathcal{S}'}|} \right) \log \left(\sum_{\substack{\bm{x}'_{\mathcal{S}'} \in \mathcal{X}_{\mathcal{S}'}\\\text{s.t. } [\phi_\sigma(\bm{x}'_{\mathcal{S}'})]_{\mathcal{B}'_r} =\bm{x}'_{\mathcal{B}'_r}}} \frac{1}{|\mathcal{X}'_{\mathcal{S}'}|} \right)\\
    &\leq - \sum_{\bm{x}'_{\mathcal{B}'_r}} \left(\sum_{\substack{\bm{x}'_{\mathcal{S}'} \in \mathcal{G}_\sigma\\\text{s.t. } [\phi_\sigma(\bm{x}'_{\mathcal{S}'})]_{\mathcal{B}'_r} =\bm{x}'_{\mathcal{B}'_r}}} \frac{1}{|\mathcal{X}'_{\mathcal{S}'}|} \right)   \log \left(\sum_{\substack{\bm{x}'_{\mathcal{S}'} \in \mathcal{G}_\sigma \\\text{s.t. } [\phi_\sigma(\bm{x}'_{\mathcal{S}'})]_{\mathcal{B}'_r} =\bm{x}'_{\mathcal{B}'_r}}} \frac{1}{|\mathcal{X}'_{\mathcal{S}'}|} \right)\nonumber \\ &\quad - \frac{1}{|\mathcal{X}'_{\mathcal{S}'}|} \left( \sum_{\bm{x}'_{\mathcal{B}'_r}} \sum_{\substack{\bm{x}'_{\mathcal{S}'} \in \mathcal{X}_{\mathcal{S}'} \setminus \mathcal{G}_\sigma\\\text{s.t. } [\phi_\sigma(\bm{x}'_{\mathcal{S}'})]_{\mathcal{B}'_r} =\bm{x}'_{\mathcal{B}'_r}}}   \log \left( \frac{1}{ |\mathcal{X}_{\mathcal{S}'}|} \right)  \right) \label{eq:reduce-set}\\
    &\leq - \frac{1}{1 - \epsilon'}  \sum_{\bm{x}'_{\mathcal{B}'_r}} \left(\sum_{\substack{\bm{x}'_{\mathcal{S}'} \in \mathcal{G}_\sigma\\\text{s.t. } [\phi_\sigma(\bm{x}'_{\mathcal{S}'})]_{\mathcal{B}'_r} =\bm{x}'_{\mathcal{B}'_r}}} \frac{1}{|\mathcal{X}'_{\mathcal{S}'}|} \right)\nonumber \\ &\quad\quad\quad  \log \left(\sum_{\substack{\bm{x}'_{\mathcal{S}'} \in \mathcal{G}_\sigma \\\text{s.t. } [\phi_\sigma(\bm{x}'_{\mathcal{S}'})]_{\mathcal{B}'_r} =\bm{x}'_{\mathcal{B}'_r}}} \frac{1}{|\mathcal{X}'_{\mathcal{S}'}|} \right) + \epsilon' \log |\mathcal{X}_{\mathcal{S}'}|
    \label{eq:difference-2}
  \end{align}
\end{subequations}
\fi
where \eqref{eq:difference-2} follows from $\displaystyle \sum_{\bm{x}'_{\mathcal{B}'_r}} \sum_{\substack{\bm{x}'_{\mathcal{S}'} \in \mathcal{X}_{\mathcal{S}'} \setminus \mathcal{G}_\sigma\\\text{s.t. } [\phi_\sigma(\bm{x}'_{\mathcal{S}'})]_{\mathcal{B}'_r} =\bm{x}'_{\mathcal{B}'_r}}} 1 = |\mathcal{G}_\sigma^\text{c}| = \epsilon' |\mathcal{X}_{\mathcal{S}'}|$.

Combining \eqref{eq:difference-1} and \eqref{eq:difference-2}, we have Proposition~\ref{lemma:hb}. \hfill$\blacksquare$

\subsection{Proof of Proposition~\ref{lemma:hb-2}}
Define the following:
\begin{align}
  \mathcal{W}_{\hat{\bm{x}}_{{\hatcal{A}_r}}} & \equalbydef \{ \hat{\bm{x}}_{\mathcal{S}'} \in \mathcal{X}_{\mathcal{S}'}: [\hat{\bm{x}}_{\mathcal{S}'}]_{\hatcal{A}_r} = \hat{\bm{x}}_{{\hatcal{A}_r}}   \},\\
  \mathcal{W}^\text{G}_{\hat{\bm{x}}_{\hatcal{A}_r},\sigma} &\equalbydef \mathcal{W}_{\hat{\bm{x}}_{{\hatcal{A}_r}}} \cap \mathcal{G}_\sigma,\\
  \mathcal{W}^\text{B}_{\hat{\bm{x}}_{\hatcal{A}_r},\sigma} &\equalbydef \mathcal{W}_{\hat{\bm{x}}_{{\hatcal{A}_r}}} \setminus \mathcal{W}^\text{G}_{\hat{\bm{x}}_{\hatcal{A}_r},\sigma} = \mathcal{W}_{\hat{\bm{x}}_{{\hatcal{A}_r}}} \cap \mathcal{G}_\sigma^\text{c}.
\end{align}
It follows that
\begin{align}
  |\mathcal{W}_{\hat{\bm{x}}_{{\hatcal{A}_r}}}| &= |\mathcal{X}_{\mathcal{S}' \setminus \hatcal{A}_r}| = \frac{|\mathcal{X}_{\mathcal{S}'}|}{|\mathcal{X}_{\hatcal{A}_r}|},\\
  |\mathcal{W}^\text{B}_{\hat{\bm{x}}_{\hatcal{A}_r},\sigma}| &\leq |\mathcal{X}_{\mathcal{S}'} \setminus \mathcal{G}_\sigma| = \epsilon' |\mathcal{X}_{\mathcal{S}'}| = |\mathcal{G}^\text{c}_\sigma|,\\
  |\mathcal{W}^\text{G}_{\hat{\bm{x}}_{\hatcal{A}_r},\sigma}| &\geq \left( \frac{1}{|\mathcal{X}_{\hatcal{A}_r}|} - \epsilon' \right) |\mathcal{X}_{\mathcal{S}'}|.                                                             
\end{align}

Using Bayes's rule,
\begin{subequations}
  \begin{align}
  \sum_c p(a,b,c) &= p(a,b), \\
  \sum_c p(a,c) p(b|a,c) &= p(a) p (b|a), \\
  \frac{1}{p(a)} \sum_c p(a,c) p(b|a,c) &= p(b|a).
  \end{align}
\end{subequations}

Now,
\begin{subequations}
  \begin{align}
    p(\hat{\bm{x}}_{\hatcal{A}_r}|1,\sigma) &= \sum_{\hat{\bm{x}}_{\mathcal{S}'} \in \mathcal{X}_{\mathcal{S}'}} p(\hat{\bm{x}}_{\mathcal{S}'},\hat{\bm{x}}_{\hatcal{A}_r}|1,\sigma)\\
                                         &= \sum_{\hat{\bm{x}}_{\mathcal{S}'} \in \mathcal{W}^\text{G}_{\hat{\bm{x}}_{\hatcal{A}_r},\sigma}} p(\hat{\bm{x}}_{\mathcal{S}'},\hat{\bm{x}}_{\hatcal{A}_r}|1,\sigma) \label{eq:reduce-1}\\
                                         &= \sum_{\hat{\bm{x}}_{\mathcal{S}'} \in \mathcal{W}^\text{G}_{\hat{\bm{x}}_{\hatcal{A}_r},\sigma}} p(\hat{\bm{x}}_{\mathcal{S}'}|1,\sigma) \label{eq:reduce-2}\\
                                         &= \sum_{\hat{\bm{x}}_{\mathcal{S}'} \in \mathcal{W}^\text{G}_{\hat{\bm{x}}_{\hatcal{A}_r},\sigma}} \frac{1}{|\mathcal{G}_\sigma|} \label{eq:reduce-3}\\
    &= \frac{|\mathcal{W}^\text{G}_{\hat{\bm{x}}_{\hatcal{A}_r},\sigma}|}{|\mathcal{G}_\sigma|}.
  \end{align}
\end{subequations}
Here, \\
\eqref{eq:reduce-1} is derived because given that $\hat{d}=1$ and $\hat{x}_\text{b}=\sigma$, we must have $\hat{\bm{x}}_{\mathcal{S}'} \in \mathcal{G}_\sigma$, and if $\hat{\bm{x}}_{\hatcal{A}_r} = \bm{a}$, then $\hat{\bm{x}}_{\mathcal{S}'} \in \mathcal{W}_{\bm{a}}$;\\
\eqref{eq:reduce-2} follows from $p(\hat{\bm{x}}_{\hatcal{A}_r}|\hat{\bm{x}}_{\mathcal{S}'},1,\sigma)=1$ if $\hat{\bm{x}}_{\mathcal{S}'} \in \mathcal{W}^\text{G}_{\hat{\bm{x}}_{\hatcal{A}_r},\sigma}$;\\
\eqref{eq:reduce-3} follows from \eqref{eq:uniform-001}.

Also,
\begin{subequations}
  \begin{align}
    &p(\hat{\bm{x}}_{\hatcal{B}_r} | \hat{\bm{x}}_{\hatcal{A}_r}, 1, \sigma)\nonumber \\ &= \frac{1}{p(\hat{\bm{x}}_{\hatcal{A}_r}|1,\sigma)} \sum_{\hat{\bm{x}}_{\mathcal{S}'} \in \mathcal{X}_{\mathcal{S}'}} p(\hat{\bm{x}}_{\mathcal{S}'},\hat{\bm{x}}_{\hatcal{A}_r}| 1, \sigma) p(\hat{\bm{x}}_{\hatcal{B}_r} | \hat{\bm{x}}_{\mathcal{S}'},\hat{\bm{x}}_{\hatcal{A}_r}, 1, \sigma)\\
                                                                                    &= \frac{|\mathcal{G}_\sigma|}{|\mathcal{W}^\text{G}_{\hat{\bm{x}}_{\hatcal{A}_r},\sigma}|} \sum_{\hat{\bm{x}}_{\mathcal{S}'} \in \mathcal{W}^\text{G}_{\hat{\bm{x}}_{\hatcal{A}_r},\sigma}} \frac{1}{|\mathcal{G}_\sigma|} p(\hat{\bm{x}}_{\hatcal{B}_r} | \hat{\bm{x}}_{\mathcal{S}'},\hat{\bm{x}}_{\hatcal{A}_r}, 1, \sigma) \label{eq:reduce-4}\\
    &= \frac{1}{|\mathcal{W}^\text{G}_{\hat{\bm{x}}_{\hatcal{A}_r},\sigma}|} \sum_{\hat{\bm{x}}_{\mathcal{S}'} \in \mathcal{W}^\text{G}_{\hat{\bm{x}}_{\hatcal{A}_r},\sigma}} p(\hat{\bm{x}}_{\hatcal{B}_r} | \hat{\bm{x}}_{\mathcal{S}'}, 1, \sigma),
  \end{align}
\end{subequations}
where \eqref{eq:reduce-4} follows the same arguments as \eqref{eq:reduce-1}--\eqref{eq:reduce-3}.

With this, we now calculate
\ifx\doublecolumn\undefined
% ==== put single-column equations here
\begin{subequations}
  \begin{align}
    &H( \hat{\bm{X}}_{\hatcal{B}_r} | \hat{\bm{X}}_{\hatcal{A}_r},\hat{D}=1, \hat{X}_{\text{b}} = \sigma)\nonumber\\
    &= - \sum_{\substack{\hat{\bm{x}}_{\hatcal{A}_r}:\\ \text{s.t. } \exists \hat{\bm{x}}_{\mathcal{S}'} \in \mathcal{W}^\text{G}_{\hat{\bm{x}}_{\hatcal{A}_r},\sigma} }} \sum_{\hat{\bm{x}}_{\hatcal{B}_r}} p(\hat{\bm{x}}_{\hatcal{A}_r}|1,\sigma) p(\hat{\bm{x}}_{\hatcal{B}_r} | \hat{\bm{x}}_{\hatcal{A}_r}, 1, \sigma)\nonumber \\ &\quad\quad\quad  \log p(\hat{\bm{x}}_{\hatcal{B}_r} | \hat{\bm{x}}_{\hatcal{A}_r}, 1, \sigma) \label{eq:sum-special}\\
    &= - \sum_{\substack{\hat{\bm{x}}_{\hatcal{A}_r}:\\ \text{s.t. } \exists \hat{\bm{x}}_{\mathcal{S}'} \in \mathcal{W}^\text{G}_{\hat{\bm{x}}_{\hatcal{A}_r},\sigma} }} \sum_{\hat{\bm{x}}_{\hatcal{B}_r}} \frac{|\mathcal{W}^\text{G}_{\hat{\bm{x}}_{\hatcal{A}_r},\sigma}|}{|\mathcal{G}_\sigma|}\nonumber \\ &\quad\quad\quad \left( \frac{1}{|\mathcal{W}^\text{G}_{\hat{\bm{x}}_{\hatcal{A}_r},\sigma}|} \sum_{\hat{\bm{x}}_{\mathcal{S}'} \in \mathcal{W}^\text{G}_{\hat{\bm{x}}_{\hatcal{A}_r},\sigma}} p(\hat{\bm{x}}_{\hatcal{B}_r} | \hat{\bm{x}}_{\mathcal{S}'}, 1, \sigma) \right)\nonumber \\ &\quad\quad\quad \log \left( \frac{1}{|\mathcal{W}^\text{G}_{\hat{\bm{x}}_{\hatcal{A}_r},\sigma}|} \sum_{\hat{\bm{x}}_{\mathcal{S}'} \in \mathcal{W}^\text{G}_{\hat{\bm{x}}_{\hatcal{A}_r},\sigma}} p(\hat{\bm{x}}_{\hatcal{B}_r} | \hat{\bm{x}}_{\mathcal{S}'}, 1, \sigma) \right)\\
    &= \frac{1} {(1-\epsilon')|\mathcal{X}_{\mathcal{S}'}|} \sum_{\substack{\hat{\bm{x}}_{\hatcal{A}_r}:\\ \text{s.t. } \exists \hat{\bm{x}}_{\mathcal{S}'} \in \mathcal{W}^\text{G}_{\hat{\bm{x}}_{\hatcal{A}_r},\sigma} }} \sum_{\hat{\bm{x}}_{\hatcal{B}_r}}  \sum_{\hat{\bm{x}}_{\mathcal{S}'} \in \mathcal{W}^\text{G}_{\hat{\bm{x}}_{\hatcal{A}_r},\sigma}} \nonumber \\ &\quad\quad\quad p(\hat{\bm{x}}_{\hatcal{B}_r} | \hat{\bm{x}}_{\mathcal{S}'}, 1, \sigma) \log |\mathcal{W}^\text{G}_{\hat{\bm{x}}_{\hatcal{A}_r},\sigma}| \nonumber\\
    &\quad - \frac{1} {(1-\epsilon')|\mathcal{X}_{\mathcal{S}'}|} \sum_{\substack{\hat{\bm{x}}_{\hatcal{A}_r}:\\ \text{s.t. } \exists \hat{\bm{x}}_{\mathcal{S}'} \in \mathcal{W}^\text{G}_{\hat{\bm{x}}_{\hatcal{A}_r},\sigma} }} \sum_{\hat{\bm{x}}_{\hatcal{B}_r}}  \sum_{\hat{\bm{x}}_{\mathcal{S}'} \in \mathcal{W}^\text{G}_{\hat{\bm{x}}_{\hatcal{A}_r},\sigma}} \nonumber \\ &\quad\quad\quad p(\hat{\bm{x}}_{\hatcal{B}_r} | \hat{\bm{x}}_{\mathcal{S}'}, 1, \sigma) \log \left( \sum_{\hat{\bm{x}}_{\mathcal{S}'} \in \mathcal{W}^\text{G}_{\hat{\bm{x}}_{\hatcal{A}_r},\sigma}} p(\hat{\bm{x}}_{\hatcal{B}_r} | \hat{\bm{x}}_{\mathcal{S}'}, 1, \sigma) \right). \label{eq:combine-101}
  \end{align}
\end{subequations}
\else
% ==== put double-column equations here
\begin{subequations}
  \begin{align}
    &H( \hat{\bm{X}}_{\hatcal{B}_r} | \hat{\bm{X}}_{\hatcal{A}_r},\hat{D}=1, \hat{X}_{\text{b}} = \sigma)\nonumber\\
    &= - \sum_{\substack{\hat{\bm{x}}_{\hatcal{A}_r}:\\ \text{s.t. } \exists \hat{\bm{x}}_{\mathcal{S}'} \in \mathcal{W}^\text{G}_{\hat{\bm{x}}_{\hatcal{A}_r},\sigma} }} \sum_{\hat{\bm{x}}_{\hatcal{B}_r}} p(\hat{\bm{x}}_{\hatcal{A}_r}|1,\sigma) p(\hat{\bm{x}}_{\hatcal{B}_r} | \hat{\bm{x}}_{\hatcal{A}_r}, 1, \sigma)\nonumber \\ &\quad\quad\quad  \log p(\hat{\bm{x}}_{\hatcal{B}_r} | \hat{\bm{x}}_{\hatcal{A}_r}, 1, \sigma) \label{eq:sum-special}\\
    &= - \sum_{\substack{\hat{\bm{x}}_{\hatcal{A}_r}:\\ \text{s.t. } \exists \hat{\bm{x}}_{\mathcal{S}'} \in \mathcal{W}^\text{G}_{\hat{\bm{x}}_{\hatcal{A}_r},\sigma} }} \sum_{\hat{\bm{x}}_{\hatcal{B}_r}} \frac{|\mathcal{W}^\text{G}_{\hat{\bm{x}}_{\hatcal{A}_r},\sigma}|}{|\mathcal{G}_\sigma|}\nonumber \\ &\quad\quad\quad \left( \frac{1}{|\mathcal{W}^\text{G}_{\hat{\bm{x}}_{\hatcal{A}_r},\sigma}|} \sum_{\hat{\bm{x}}_{\mathcal{S}'} \in \mathcal{W}^\text{G}_{\hat{\bm{x}}_{\hatcal{A}_r},\sigma}} p(\hat{\bm{x}}_{\hatcal{B}_r} | \hat{\bm{x}}_{\mathcal{S}'}, 1, \sigma) \right)\nonumber \\ &\quad\quad\quad \log \left( \frac{1}{|\mathcal{W}^\text{G}_{\hat{\bm{x}}_{\hatcal{A}_r},\sigma}|} \sum_{\hat{\bm{x}}_{\mathcal{S}'} \in \mathcal{W}^\text{G}_{\hat{\bm{x}}_{\hatcal{A}_r},\sigma}} p(\hat{\bm{x}}_{\hatcal{B}_r} | \hat{\bm{x}}_{\mathcal{S}'}, 1, \sigma) \right)\\
    &= \frac{1} {(1-\epsilon')|\mathcal{X}_{\mathcal{S}'}|} \sum_{\substack{\hat{\bm{x}}_{\hatcal{A}_r}:\\ \text{s.t. } \exists \hat{\bm{x}}_{\mathcal{S}'} \in \mathcal{W}^\text{G}_{\hat{\bm{x}}_{\hatcal{A}_r},\sigma} }} \sum_{\hat{\bm{x}}_{\hatcal{B}_r}}  \sum_{\hat{\bm{x}}_{\mathcal{S}'} \in \mathcal{W}^\text{G}_{\hat{\bm{x}}_{\hatcal{A}_r},\sigma}} \nonumber \\ &\quad\quad\quad p(\hat{\bm{x}}_{\hatcal{B}_r} | \hat{\bm{x}}_{\mathcal{S}'}, 1, \sigma) \log |\mathcal{W}^\text{G}_{\hat{\bm{x}}_{\hatcal{A}_r},\sigma}| \nonumber\\
    &\quad - \frac{1} {(1-\epsilon')|\mathcal{X}_{\mathcal{S}'}|} \sum_{\substack{\hat{\bm{x}}_{\hatcal{A}_r}:\\ \text{s.t. } \exists \hat{\bm{x}}_{\mathcal{S}'} \in \mathcal{W}^\text{G}_{\hat{\bm{x}}_{\hatcal{A}_r},\sigma} }} \sum_{\hat{\bm{x}}_{\hatcal{B}_r}}  \sum_{\hat{\bm{x}}_{\mathcal{S}'} \in \mathcal{W}^\text{G}_{\hat{\bm{x}}_{\hatcal{A}_r},\sigma}} \nonumber \\ &\quad\quad\quad p(\hat{\bm{x}}_{\hatcal{B}_r} | \hat{\bm{x}}_{\mathcal{S}'}, 1, \sigma) \log \left( \sum_{\hat{\bm{x}}_{\mathcal{S}'} \in \mathcal{W}^\text{G}_{\hat{\bm{x}}_{\hatcal{A}_r},\sigma}} p(\hat{\bm{x}}_{\hatcal{B}_r} | \hat{\bm{x}}_{\mathcal{S}'}, 1, \sigma) \right). \label{eq:combine-101}
  \end{align}
\end{subequations}
\fi
Note that in \eqref{eq:sum-special}, we only need to sum over $\hat{\bm{x}}_{\hatcal{A}_r}$ where where exists some $\hat{\bm{x}}_{\mathcal{S}'} \in \mathcal{W}^\text{G}_{\hat{\bm{x}}_{\hatcal{A}_r},\sigma}$, because we impose the condition that $\hat{D}=1$; the rest give $p(\hat{\bm{x}}_{\hatcal{A}_r}|1,\sigma)=0$.

To calculate $H(\bm{X}'_{\mathcal{B}'_r}|\bm{X}'_{\mathcal{A}'_r})$, we first find $p(\bm{x}'_{\mathcal{A}'_r})$ and $p(\bm{x}'_{\mathcal{B}'_r} | \bm{x}'_{\mathcal{A}'_r})$.
\begin{subequations}
  \begin{align}
    p(\bm{x}'_{\mathcal{A}'_r})
    &= \sum_{\bm{x}'_{\mathcal{S}'} \in \mathcal{X}_{\mathcal{S}}'} p(\bm{x}'_{\mathcal{S}'},\bm{x}'_{\mathcal{A}'_r})\\
    &= \sum_{ \bm{x}'_{\mathcal{S}'} \in \mathcal{W}_{\bm{x}'_{\mathcal{A}'_r}}} p(\bm{x}'_{\mathcal{S}'})\\
    &= \sum_{\bm{x}'_{\mathcal{S}'} \in \mathcal{W}_{\bm{x}'_{\mathcal{A}'_r}} \cap  \mathcal{G}_\sigma} p(\bm{x}'_{\mathcal{S}'}) + \sum_{\bm{x}'_{\mathcal{S}'} \in \mathcal{W}_{\bm{x}'_{\mathcal{A}'_r}} \cap  \mathcal{G}^\text{c}_\sigma} p(\bm{x}'_{\mathcal{S}'})\\
    &= \sum_{\bm{x}'_{\mathcal{S}'} \in \mathcal{W}^\text{G}_{\bm{x}'_{\mathcal{A}'_r},\sigma}} \frac{1}{|\mathcal{X}_{\mathcal{S}'}|} + \sum_{\bm{x}'_{\mathcal{S}'} \in \mathcal{W}^\text{B}_{\bm{x}'_{\mathcal{A}'_r},\sigma} } \frac{1}{|\mathcal{X}_{\mathcal{S}'}|}\\
    &= \frac{|\mathcal{W}^\text{G}_{\bm{x}'_{\mathcal{A}'_r},\sigma}|}{|\mathcal{X}_{\mathcal{S}'}|} + \epsilon'_{{\bm{x}'_{\mathcal{A}'_r},\sigma}},
  \end{align}
\end{subequations}
where $\epsilon'_{{\bm{x}'_{\mathcal{A}'_r},\sigma}} \equalbydef \frac{|\mathcal{W}^\text{B}_{\bm{x}'_{\mathcal{A}'_r},\sigma}|}{|\mathcal{X}_{\mathcal{S}'}|} \leq \epsilon'$. Also note that $p(\bm{x}'_{\mathcal{A}'_r}) = 1/|\mathcal{X}_{\mathcal{A}'_r}|$ as the messages  are uniformly distributed in $\mathbb{N}'$.

\ifx\doublecolumn\undefined
% ==== put single-column equations here
\begin{subequations}
  \begin{align}
    p(\bm{x}'_{\mathcal{B}'_r} | \bm{x}'_{\mathcal{A}'_r})
    &= \frac{1}{p(\bm{x}'_{\mathcal{A}'_r})} \sum_{\bm{x}'_{\mathcal{S}'} \in \mathcal{X}_{\mathcal{S}}'} p(\bm{x}'_{\mathcal{S}'},\bm{x}'_{\mathcal{A}'_r}) p(\bm{x}'_{\mathcal{B}'_r} |\bm{x}'_{\mathcal{S}'}, \bm{x}'_{\mathcal{A}'_r})\\
    &= \frac{1}{p(\bm{x}'_{\mathcal{A}'_r})} \sum_{ \bm{x}'_{\mathcal{S}'} \in \mathcal{W}_{\bm{x}'_{\mathcal{A}'_r}}} p(\bm{x}'_{\mathcal{S}'}) p(\bm{x}'_{\mathcal{B}'_r} |\bm{x}'_{\mathcal{S}'}, \bm{x}'_{\mathcal{A}'_r})\\
    &= \frac{1}{p(\bm{x}'_{\mathcal{A}'_r})} \sum_{ \bm{x}'_{\mathcal{S}'} \in \mathcal{W}_{\bm{x}'_{\mathcal{A}'_r}}} \frac{1}{|\mathcal{X}_{\mathcal{S}'}|} p(\bm{x}'_{\mathcal{B}'_r} |\bm{x}'_{\mathcal{S}'})\\
    &= \frac{1}{p(\bm{x}'_{\mathcal{A}'_r})} \frac{1}{|\mathcal{X}_{\mathcal{S}'}|} \left( \underbrace{\sum_{\bm{x}'_{\mathcal{S}'} \in \mathcal{W}^\text{G}_{\bm{x}'_{\mathcal{A}'_r},\sigma}} p(\bm{x}'_{\mathcal{B}'_r} |\bm{x}'_{\mathcal{S}'})}_{\equalbydef C} + \underbrace{\sum_{\bm{x}'_{\mathcal{S}'} \in \mathcal{W}^\text{B}_{\bm{x}'_{\mathcal{A}'_r},\sigma}} p(\bm{x}'_{\mathcal{B}'_r} |\bm{x}'_{\mathcal{S}'})}_{\equalbydef D}  \right).
    % & \leq \frac{1}{p(\bm{x}'_{\mathcal{A}'_r})} \frac{1}{|\mathcal{X}_{\mathcal{S}'}|} \sum_{\bm{x}'_{\mathcal{S}'} \in \mathcal{W}^\text{G}_{\bm{x}'_{\mathcal{A}'_r},\sigma}} p(\bm{x}'_{\mathcal{B}'_r} |\bm{x}'_{\mathcal{S}'}) + \epsilon'|\mathcal{X}_{\mathcal{A}'_r}|.
  \end{align}
\end{subequations}
\else
% ==== put double-column equations here
\begin{subequations}
  \begin{align*}
    p(\bm{x}'_{\mathcal{B}'_r} | \bm{x}'_{\mathcal{A}'_r})
    &= \frac{1}{p(\bm{x}'_{\mathcal{A}'_r})} \sum_{\bm{x}'_{\mathcal{S}'} \in \mathcal{X}_{\mathcal{S}}'} p(\bm{x}'_{\mathcal{S}'},\bm{x}'_{\mathcal{A}'_r}) p(\bm{x}'_{\mathcal{B}'_r} |\bm{x}'_{\mathcal{S}'}, \bm{x}'_{\mathcal{A}'_r})\nonumber \\
    &= \frac{1}{p(\bm{x}'_{\mathcal{A}'_r})} \sum_{ \bm{x}'_{\mathcal{S}'} \in \mathcal{W}_{\bm{x}'_{\mathcal{A}'_r}}} p(\bm{x}'_{\mathcal{S}'}) p(\bm{x}'_{\mathcal{B}'_r} |\bm{x}'_{\mathcal{S}'}, \bm{x}'_{\mathcal{A}'_r})\\
    &= \frac{1}{p(\bm{x}'_{\mathcal{A}'_r})} \sum_{ \bm{x}'_{\mathcal{S}'} \in \mathcal{W}_{\bm{x}'_{\mathcal{A}'_r}}} \frac{1}{|\mathcal{X}_{\mathcal{S}'}|} p(\bm{x}'_{\mathcal{B}'_r} |\bm{x}'_{\mathcal{S}'})\\
    &= \frac{1}{p(\bm{x}'_{\mathcal{A}'_r})} \frac{1}{|\mathcal{X}_{\mathcal{S}'}|} \Bigg( \underbrace{\sum_{\bm{x}'_{\mathcal{S}'} \in \mathcal{W}^\text{G}_{\bm{x}'_{\mathcal{A}'_r},\sigma}} p(\bm{x}'_{\mathcal{B}'_r} |\bm{x}'_{\mathcal{S}'})}_{\equalbydef C} \nonumber\\ &\quad + \underbrace{\sum_{\bm{x}'_{\mathcal{S}'} \in \mathcal{W}^\text{B}_{\bm{x}'_{\mathcal{A}'_r},\sigma}} p(\bm{x}'_{\mathcal{B}'_r} |\bm{x}'_{\mathcal{S}'})}_{\equalbydef D}  \Bigg).
    % & \leq \frac{1}{p(\bm{x}'_{\mathcal{A}'_r})} \frac{1}{|\mathcal{X}_{\mathcal{S}'}|} \sum_{\bm{x}'_{\mathcal{S}'} \in \mathcal{W}^\text{G}_{\bm{x}'_{\mathcal{A}'_r},\sigma}} p(\bm{x}'_{\mathcal{B}'_r} |\bm{x}'_{\mathcal{S}'}) + \epsilon'|\mathcal{X}_{\mathcal{A}'_r}|.
  \end{align*}
\end{subequations}
\fi

Now,
\ifx\doublecolumn\undefined
% ==== put single-column equations here
\begin{subequations}
  \begin{align}
    & H(\bm{X}'_{\mathcal{B}'_r} | \bm{X}'_{\mathcal{A}'_r}) \nonumber \\
    &= - \sum_{\bm{x}'_{\mathcal{A}'_r}} \sum_{\bm{x}'_{\mathcal{B}'_r}} p(\bm{x}'_{\mathcal{A}'_r}) p(\bm{x}'_{\mathcal{B}'_r} | \bm{x}'_{\mathcal{A}'_r}) \log p(\bm{x}'_{\mathcal{B}'_r} | \bm{x}'_{\mathcal{A}'_r})\\
    &\geq - \sum_{\substack{\bm{x}'_{\mathcal{A}'_r}:\\ \text{s.t. } \exists \bm{x}'_{\mathcal{S}'} \in \mathcal{W}^\text{G}_{\bm{x}'_{\mathcal{A}'_r},\sigma} }} \sum_{\bm{x}'_{\mathcal{B}'_r}} p(\bm{x}'_{\mathcal{A}'_r}) p(\bm{x}'_{\mathcal{B}'_r} | \bm{x}'_{\mathcal{A}'_r}) \log p(\bm{x}'_{\mathcal{B}'_r} | \bm{x}'_{\mathcal{A}'_r}) \label{eq:b-given-a-1}\\
    &= - \sum_{\substack{\bm{x}'_{\mathcal{A}'_r}:\\ \text{s.t. } \exists \bm{x}'_{\mathcal{S}'} \in \mathcal{W}^\text{G}_{\bm{x}'_{\mathcal{A}'_r},\sigma} }} \sum_{\bm{x}'_{\mathcal{B}'_r}}  \frac{1}{|\mathcal{X}_{\mathcal{S}'}|} ( C + D)  \log p(\bm{x}'_{\mathcal{B}'_r} | \bm{x}'_{\mathcal{A}'_r})\\
    &\geq - \frac{1}{|\mathcal{X}_{\mathcal{S}'}|} \sum_{\substack{\bm{x}'_{\mathcal{A}'_r}:\\ \text{s.t. } \exists \bm{x}'_{\mathcal{S}'} \in \mathcal{W}^\text{G}_{\bm{x}'_{\mathcal{A}'_r},\sigma} }} \sum_{\bm{x}'_{\mathcal{B}'_r}} C \log p(\bm{x}'_{\mathcal{B}'_r} | \bm{x}'_{\mathcal{A}'_r})\\
    &= - \frac{1}{|\mathcal{X}_{\mathcal{S}'}|} \sum_{\substack{\bm{x}'_{\mathcal{A}'_r}:\\ \text{s.t. } \exists \bm{x}'_{\mathcal{S}'} \in \mathcal{W}^\text{G}_{\bm{x}'_{\mathcal{A}'_r},\sigma} }} \sum_{\bm{x}'_{\mathcal{B}'_r}} C \log \frac{C+D}{p(\bm{x}'_{\mathcal{A}'_r})|\mathcal{X}_{\mathcal{S}'}|}\\
    &= \frac{1}{|\mathcal{X}_{\mathcal{S}'}|} \sum_{\substack{\bm{x}'_{\mathcal{A}'_r}:\\ \text{s.t. } \exists \bm{x}'_{\mathcal{S}'} \in \mathcal{W}^\text{G}_{\bm{x}'_{\mathcal{A}'_r},\sigma} }} \sum_{\bm{x}'_{\mathcal{B}'_r}} C \log \left( |\mathcal{W}^\text{G}_{\bm{x}'_{\mathcal{A}'_r},\sigma}| + \epsilon'_{{\bm{x}'_{\mathcal{A}'_r},\sigma}} |\mathcal{X}_{\mathcal{S}'}| \right) - \frac{1}{|\mathcal{X}_{\mathcal{S}'}|} \sum_{\substack{\bm{x}'_{\mathcal{A}'_r}:\\ \text{s.t. } \exists \bm{x}'_{\mathcal{S}'} \in \mathcal{W}^\text{G}_{\bm{x}'_{\mathcal{A}'_r},\sigma} }} \sum_{\bm{x}'_{\mathcal{B}'_r}} C \log (C+D)\\
    &\geq \frac{1}{|\mathcal{X}_{\mathcal{S}'}|} \sum_{\substack{\bm{x}'_{\mathcal{A}'_r}:\\ \text{s.t. } \exists \bm{x}'_{\mathcal{S}'} \in \mathcal{W}^\text{G}_{\bm{x}'_{\mathcal{A}'_r},\sigma} }} \sum_{\bm{x}'_{\mathcal{B}'_r}} C \log |\mathcal{W}^\text{G}_{\bm{x}'_{\mathcal{A}'_r},\sigma}| - \frac{1}{|\mathcal{X}_{\mathcal{S}'}|} \sum_{\substack{\bm{x}'_{\mathcal{A}'_r}:\\ \text{s.t. } \exists \bm{x}'_{\mathcal{S}'} \in \mathcal{W}^\text{G}_{\bm{x}'_{\mathcal{A}'_r},\sigma} }} \sum_{\bm{x}'_{\mathcal{B}'_r}} C \log (C+D)\\
    &= \frac{1}{|\mathcal{X}_{\mathcal{S}'}|} \sum_{\substack{\bm{x}'_{\mathcal{A}'_r}:\\ \text{s.t. } \exists \bm{x}'_{\mathcal{S}'} \in \mathcal{W}^\text{G}_{\bm{x}'_{\mathcal{A}'_r},\sigma} }} \sum_{\bm{x}'_{\mathcal{B}'_r}} C \log |\mathcal{W}^\text{G}_{\bm{x}'_{\mathcal{A}'_r},\sigma}| - \frac{1}{|\mathcal{X}_{\mathcal{S}'}|} \sum_{\substack{\bm{x}'_{\mathcal{A}'_r}:\\ \text{s.t. } \exists \bm{x}'_{\mathcal{S}'} \in \mathcal{W}^\text{G}_{\bm{x}'_{\mathcal{A}'_r},\sigma} }} \sum_{\bm{x}'_{\mathcal{B}'_r}} C \left( \log C + \log( 1 + D/C) \right)\\
    &= \frac{1}{|\mathcal{X}_{\mathcal{S}'}|} \sum_{\substack{\bm{x}'_{\mathcal{A}'_r}:\\ \text{s.t. } \exists \bm{x}'_{\mathcal{S}'} \in \mathcal{W}^\text{G}_{\bm{x}'_{\mathcal{A}'_r},\sigma} }} \sum_{\bm{x}'_{\mathcal{B}'_r}} C \log |\mathcal{W}^\text{G}_{\bm{x}'_{\mathcal{A}'_r},\sigma}| - \frac{1}{|\mathcal{X}_{\mathcal{S}'}|} \sum_{\substack{\bm{x}'_{\mathcal{A}'_r}:\\ \text{s.t. } \exists \bm{x}'_{\mathcal{S}'} \in \mathcal{W}^\text{G}_{\bm{x}'_{\mathcal{A}'_r},\sigma} }} \sum_{\bm{x}'_{\mathcal{B}'_r}} C \log C \nonumber \\
    &\quad - \frac{1}{|\mathcal{X}_{\mathcal{S}'}|} \sum_{\substack{\bm{x}'_{\mathcal{A}'_r}:\\ \text{s.t. } \exists \bm{x}'_{\mathcal{S}'} \in \mathcal{W}^\text{G}_{\bm{x}'_{\mathcal{A}'_r},\sigma} }} \sum_{\bm{x}'_{\mathcal{B}'_r}} C \log (1 + D/C) \\
    &\geq \frac{1}{|\mathcal{X}_{\mathcal{S}'}|} \sum_{\substack{\bm{x}'_{\mathcal{A}'_r}:\\ \text{s.t. } \exists \bm{x}'_{\mathcal{S}'} \in \mathcal{W}^\text{G}_{\bm{x}'_{\mathcal{A}'_r},\sigma} }} \sum_{\bm{x}'_{\mathcal{B}'_r}} C \log |\mathcal{W}^\text{G}_{\bm{x}'_{\mathcal{A}'_r},\sigma}| - \frac{1}{|\mathcal{X}_{\mathcal{S}'}|} \sum_{\substack{\bm{x}'_{\mathcal{A}'_r}:\\ \text{s.t. } \exists \bm{x}'_{\mathcal{S}'} \in \mathcal{W}^\text{G}_{\bm{x}'_{\mathcal{A}'_r},\sigma} }} \sum_{\bm{x}'_{\mathcal{B}'_r}} C \log C \nonumber \\
    &\quad - \frac{1}{|\mathcal{X}_{\mathcal{S}'}|} \sum_{\substack{\bm{x}'_{\mathcal{A}'_r}:\\ \text{s.t. } \exists \bm{x}'_{\mathcal{S}'} \in \mathcal{W}^\text{G}_{\bm{x}'_{\mathcal{A}'_r},\sigma} }} \sum_{\bm{x}'_{\mathcal{B}'_r}} C (\log e) \frac{D}{C}  \label{eq:b-given-a-2} \\
    &= \frac{1}{|\mathcal{X}_{\mathcal{S}'}|} \sum_{\substack{\bm{x}'_{\mathcal{A}'_r}:\\ \text{s.t. } \exists \bm{x}'_{\mathcal{S}'} \in \mathcal{W}^\text{G}_{\bm{x}'_{\mathcal{A}'_r},\sigma} }} \sum_{\bm{x}'_{\mathcal{B}'_r}} \sum_{\bm{x}'_{\mathcal{S}'} \in \mathcal{W}^\text{G}_{\bm{x}'_{\mathcal{A}'_r},\sigma}} p(\bm{x}'_{\mathcal{B}'_r} |\bm{x}'_{\mathcal{S}'})  \log |\mathcal{W}^\text{G}_{\bm{x}'_{\mathcal{A}'_r},\sigma}|\nonumber \\
    &\quad - \frac{1}{|\mathcal{X}_{\mathcal{S}'}|} \sum_{\substack{\bm{x}'_{\mathcal{A}'_r}:\\ \text{s.t. } \exists \bm{x}'_{\mathcal{S}'} \in \mathcal{W}^\text{G}_{\bm{x}'_{\mathcal{A}'_r},\sigma} }} \sum_{\bm{x}'_{\mathcal{B}'_r}} \sum_{\bm{x}'_{\mathcal{S}'} \in \mathcal{W}^\text{G}_{\bm{x}'_{\mathcal{A}'_r},\sigma}} p(\bm{x}'_{\mathcal{B}'_r} |\bm{x}'_{\mathcal{S}'}) \log \sum_{\bm{x}'_{\mathcal{S}'} \in \mathcal{W}^\text{G}_{\bm{x}'_{\mathcal{A}'_r},\sigma}} p(\bm{x}'_{\mathcal{B}'_r} |\bm{x}'_{\mathcal{S}'}) \nonumber \\
    &\quad - \log e  \underbrace{\frac{1}{|\mathcal{X}_{\mathcal{S}'}|} \sum_{\substack{\bm{x}'_{\mathcal{A}'_r}:\\ \text{s.t. } \exists \bm{x}'_{\mathcal{S}'} \in \mathcal{W}^\text{G}_{\bm{x}'_{\mathcal{A}'_r},\sigma} }} \sum_{\bm{x}'_{\mathcal{B}'_r}} D}_{\equalbydef E}, \label{eq:combine-102}
  \end{align}
\end{subequations}
\else
% ==== put double-column equations here
\begin{subequations}
  \begin{align}
    & H(\bm{X}'_{\mathcal{B}'_r} | \bm{X}'_{\mathcal{A}'_r}) \nonumber \\
    &= - \sum_{\bm{x}'_{\mathcal{A}'_r}} \sum_{\bm{x}'_{\mathcal{B}'_r}} p(\bm{x}'_{\mathcal{A}'_r}) p(\bm{x}'_{\mathcal{B}'_r} | \bm{x}'_{\mathcal{A}'_r}) \log p(\bm{x}'_{\mathcal{B}'_r} | \bm{x}'_{\mathcal{A}'_r})\\
    &\geq - \sum_{\substack{\bm{x}'_{\mathcal{A}'_r}:\\ \text{s.t. } \exists \bm{x}'_{\mathcal{S}'} \in \mathcal{W}^\text{G}_{\bm{x}'_{\mathcal{A}'_r},\sigma} }} \sum_{\bm{x}'_{\mathcal{B}'_r}} p(\bm{x}'_{\mathcal{A}'_r}) p(\bm{x}'_{\mathcal{B}'_r} | \bm{x}'_{\mathcal{A}'_r}) \log p(\bm{x}'_{\mathcal{B}'_r} | \bm{x}'_{\mathcal{A}'_r}) \label{eq:b-given-a-1}\\
    &= - \sum_{\substack{\bm{x}'_{\mathcal{A}'_r}:\\ \text{s.t. } \exists \bm{x}'_{\mathcal{S}'} \in \mathcal{W}^\text{G}_{\bm{x}'_{\mathcal{A}'_r},\sigma} }} \sum_{\bm{x}'_{\mathcal{B}'_r}}  \frac{1}{|\mathcal{X}_{\mathcal{S}'}|} ( C + D)  \log p(\bm{x}'_{\mathcal{B}'_r} | \bm{x}'_{\mathcal{A}'_r})\\
    &\geq - \frac{1}{|\mathcal{X}_{\mathcal{S}'}|} \sum_{\substack{\bm{x}'_{\mathcal{A}'_r}:\\ \text{s.t. } \exists \bm{x}'_{\mathcal{S}'} \in \mathcal{W}^\text{G}_{\bm{x}'_{\mathcal{A}'_r},\sigma} }} \sum_{\bm{x}'_{\mathcal{B}'_r}} C \log p(\bm{x}'_{\mathcal{B}'_r} | \bm{x}'_{\mathcal{A}'_r})\\
    &= - \frac{1}{|\mathcal{X}_{\mathcal{S}'}|} \sum_{\substack{\bm{x}'_{\mathcal{A}'_r}:\\ \text{s.t. } \exists \bm{x}'_{\mathcal{S}'} \in \mathcal{W}^\text{G}_{\bm{x}'_{\mathcal{A}'_r},\sigma} }} \sum_{\bm{x}'_{\mathcal{B}'_r}} C \log \frac{C+D}{p(\bm{x}'_{\mathcal{A}'_r})|\mathcal{X}_{\mathcal{S}'}|}\\
    &= \frac{1}{|\mathcal{X}_{\mathcal{S}'}|} \sum_{\substack{\bm{x}'_{\mathcal{A}'_r}:\\ \text{s.t. } \exists \bm{x}'_{\mathcal{S}'} \in \mathcal{W}^\text{G}_{\bm{x}'_{\mathcal{A}'_r},\sigma} }} \sum_{\bm{x}'_{\mathcal{B}'_r}} C \log \left( |\mathcal{W}^\text{G}_{\bm{x}'_{\mathcal{A}'_r},\sigma}| + \epsilon'_{{\bm{x}'_{\mathcal{A}'_r},\sigma}} |\mathcal{X}_{\mathcal{S}'}| \right) \nonumber\\ &\quad - \frac{1}{|\mathcal{X}_{\mathcal{S}'}|} \sum_{\substack{\bm{x}'_{\mathcal{A}'_r}:\\ \text{s.t. } \exists \bm{x}'_{\mathcal{S}'} \in \mathcal{W}^\text{G}_{\bm{x}'_{\mathcal{A}'_r},\sigma} }} \sum_{\bm{x}'_{\mathcal{B}'_r}} C \log (C+D)\\
    &\geq \frac{1}{|\mathcal{X}_{\mathcal{S}'}|} \sum_{\substack{\bm{x}'_{\mathcal{A}'_r}:\\ \text{s.t. } \exists \bm{x}'_{\mathcal{S}'} \in \mathcal{W}^\text{G}_{\bm{x}'_{\mathcal{A}'_r},\sigma} }} \sum_{\bm{x}'_{\mathcal{B}'_r}} C \log |\mathcal{W}^\text{G}_{\bm{x}'_{\mathcal{A}'_r},\sigma}| \nonumber\\ &\quad - \frac{1}{|\mathcal{X}_{\mathcal{S}'}|} \sum_{\substack{\bm{x}'_{\mathcal{A}'_r}:\\ \text{s.t. } \exists \bm{x}'_{\mathcal{S}'} \in \mathcal{W}^\text{G}_{\bm{x}'_{\mathcal{A}'_r},\sigma} }} \sum_{\bm{x}'_{\mathcal{B}'_r}} C \log (C+D)\\
    % &= \frac{1}{|\mathcal{X}_{\mathcal{S}'}|} \sum_{\substack{\bm{x}'_{\mathcal{A}'_r}:\\ \text{s.t. } \exists \bm{x}'_{\mathcal{S}'} \in \mathcal{W}^\text{G}_{\bm{x}'_{\mathcal{A}'_r},\sigma} }} \sum_{\bm{x}'_{\mathcal{B}'_r}} C \log |\mathcal{W}^\text{G}_{\bm{x}'_{\mathcal{A}'_r},\sigma}| \nonumber\\ &\quad - \frac{1}{|\mathcal{X}_{\mathcal{S}'}|} \sum_{\substack{\bm{x}'_{\mathcal{A}'_r}:\\ \text{s.t. } \exists \bm{x}'_{\mathcal{S}'} \in \mathcal{W}^\text{G}_{\bm{x}'_{\mathcal{A}'_r},\sigma} }} \sum_{\bm{x}'_{\mathcal{B}'_r}} C \left( \log C + \log( 1 + D/C) \right)\\
    &= \frac{1}{|\mathcal{X}_{\mathcal{S}'}|} \sum_{\substack{\bm{x}'_{\mathcal{A}'_r}:\\ \text{s.t. } \exists \bm{x}'_{\mathcal{S}'} \in \mathcal{W}^\text{G}_{\bm{x}'_{\mathcal{A}'_r},\sigma} }} \sum_{\bm{x}'_{\mathcal{B}'_r}} C \log |\mathcal{W}^\text{G}_{\bm{x}'_{\mathcal{A}'_r},\sigma}| \nonumber\\ &\quad - \frac{1}{|\mathcal{X}_{\mathcal{S}'}|} \sum_{\substack{\bm{x}'_{\mathcal{A}'_r}:\\ \text{s.t. } \exists \bm{x}'_{\mathcal{S}'} \in \mathcal{W}^\text{G}_{\bm{x}'_{\mathcal{A}'_r},\sigma} }} \sum_{\bm{x}'_{\mathcal{B}'_r}} C \log C \nonumber \\
    &\quad - \frac{1}{|\mathcal{X}_{\mathcal{S}'}|} \sum_{\substack{\bm{x}'_{\mathcal{A}'_r}:\\ \text{s.t. } \exists \bm{x}'_{\mathcal{S}'} \in \mathcal{W}^\text{G}_{\bm{x}'_{\mathcal{A}'_r},\sigma} }} \sum_{\bm{x}'_{\mathcal{B}'_r}} C \log (1 + D/C) \\
    &\geq \frac{1}{|\mathcal{X}_{\mathcal{S}'}|} \sum_{\substack{\bm{x}'_{\mathcal{A}'_r}:\\ \text{s.t. } \exists \bm{x}'_{\mathcal{S}'} \in \mathcal{W}^\text{G}_{\bm{x}'_{\mathcal{A}'_r},\sigma} }} \sum_{\bm{x}'_{\mathcal{B}'_r}} C \log |\mathcal{W}^\text{G}_{\bm{x}'_{\mathcal{A}'_r},\sigma}| \nonumber\\ &\quad - \frac{1}{|\mathcal{X}_{\mathcal{S}'}|} \sum_{\substack{\bm{x}'_{\mathcal{A}'_r}:\\ \text{s.t. } \exists \bm{x}'_{\mathcal{S}'} \in \mathcal{W}^\text{G}_{\bm{x}'_{\mathcal{A}'_r},\sigma} }} \sum_{\bm{x}'_{\mathcal{B}'_r}} C \log C \nonumber \\
    &\quad - \frac{1}{|\mathcal{X}_{\mathcal{S}'}|} \sum_{\substack{\bm{x}'_{\mathcal{A}'_r}:\\ \text{s.t. } \exists \bm{x}'_{\mathcal{S}'} \in \mathcal{W}^\text{G}_{\bm{x}'_{\mathcal{A}'_r},\sigma} }} \sum_{\bm{x}'_{\mathcal{B}'_r}} C (\log e) \frac{D}{C}  \label{eq:b-given-a-2} \\
    &= \frac{1}{|\mathcal{X}_{\mathcal{S}'}|} \sum_{\substack{\bm{x}'_{\mathcal{A}'_r}:\\ \text{s.t. } \exists \bm{x}'_{\mathcal{S}'} \in \mathcal{W}^\text{G}_{\bm{x}'_{\mathcal{A}'_r},\sigma} }} \sum_{\bm{x}'_{\mathcal{B}'_r}} \sum_{\bm{x}'_{\mathcal{S}'} \in \mathcal{W}^\text{G}_{\bm{x}'_{\mathcal{A}'_r},\sigma}} p(\bm{x}'_{\mathcal{B}'_r} |\bm{x}'_{\mathcal{S}'})  \nonumber\\ &\quad\quad\quad  \log |\mathcal{W}^\text{G}_{\bm{x}'_{\mathcal{A}'_r},\sigma}|\nonumber \\
    &\quad - \frac{1}{|\mathcal{X}_{\mathcal{S}'}|} \sum_{\substack{\bm{x}'_{\mathcal{A}'_r}:\\ \text{s.t. } \exists \bm{x}'_{\mathcal{S}'} \in \mathcal{W}^\text{G}_{\bm{x}'_{\mathcal{A}'_r},\sigma} }} \sum_{\bm{x}'_{\mathcal{B}'_r}} \sum_{\bm{x}'_{\mathcal{S}'} \in \mathcal{W}^\text{G}_{\bm{x}'_{\mathcal{A}'_r},\sigma}} p(\bm{x}'_{\mathcal{B}'_r} |\bm{x}'_{\mathcal{S}'}) \nonumber\\ &\quad\quad\quad \log \sum_{\bm{x}'_{\mathcal{S}'} \in \mathcal{W}^\text{G}_{\bm{x}'_{\mathcal{A}'_r},\sigma}} p(\bm{x}'_{\mathcal{B}'_r} |\bm{x}'_{\mathcal{S}'}) \nonumber \\
    &\quad - \log e  \underbrace{\frac{1}{|\mathcal{X}_{\mathcal{S}'}|} \sum_{\substack{\bm{x}'_{\mathcal{A}'_r}:\\ \text{s.t. } \exists \bm{x}'_{\mathcal{S}'} \in \mathcal{W}^\text{G}_{\bm{x}'_{\mathcal{A}'_r},\sigma} }} \sum_{\bm{x}'_{\mathcal{B}'_r}} D}_{\equalbydef E}, \label{eq:combine-102}
  \end{align}
\end{subequations}
\fi
where\\
\eqref{eq:b-given-a-1} is obtained as we take a subset over which the first summation is evaluated;\\
\eqref{eq:b-given-a-2} follow from $x \geq \ln (1 + x)$ for all $x \geq 0$;

Now,
\
\begin{subequations}
  \begin{align}
    E &= \frac{1}{|\mathcal{X}_{\mathcal{S}'}|} \sum_{\substack{\bm{x}'_{\mathcal{A}'_r}:\\ \text{s.t. } \exists \bm{x}'_{\mathcal{S}'} \in \mathcal{W}^\text{G}_{\bm{x}'_{\mathcal{A}'_r},\sigma} }} \sum_{\bm{x}'_{\mathcal{B}'_r}} \sum_{\bm{x}'_{\mathcal{S}'} \in \mathcal{W}_{\hat{\bm{x}}_{{\hatcal{A}_r}}} \cap \mathcal{G}^\text{c}_\sigma} p(\bm{x}'_{\mathcal{B}'_r} |\bm{x}'_{\mathcal{S}'})\\
      &\leq \frac{1}{|\mathcal{X}_{\mathcal{S}'}|} \sum_{\bm{x}'_{\mathcal{A}'_r}} \sum_{\bm{x}'_{\mathcal{B}'_r}} \sum_{\bm{x}'_{\mathcal{S}'} \in \mathcal{W}_{\hat{\bm{x}}_{{\hatcal{A}_r}}} \cap \mathcal{G}^\text{c}_\sigma} p(\bm{x}'_{\mathcal{B}'_r} |\bm{x}'_{\mathcal{S}'})\\
      & = \frac{1}{|\mathcal{X}_{\mathcal{S}'}|} \sum_{\bm{x}'_{\mathcal{B}'_r}} \sum_{\bm{x}'_{\mathcal{S}'} \in  \mathcal{G}^\text{c}_\sigma} p(\bm{x}'_{\mathcal{B}'_r} |\bm{x}'_{\mathcal{S}'}) \label{eq:non-overlap}\\
      & = \frac{1}{|\mathcal{X}_{\mathcal{S}'}|} |\mathcal{G}^\text{c}_\sigma| \label{eq:deterministic-010}\\
    &= \epsilon', \label{eq:combine-103}
  \end{align}
\end{subequations}
where\\ \eqref{eq:non-overlap} follows as $\mathcal{W}_{\bm{a}}$ and $\mathcal{W}_{\bm{b}}$ do not overlap for $\bm{a} \neq \bm{b}$; \eqref{eq:deterministic-010} is obtained by noting that $p(\bm{x}'_{\mathcal{B}'_r} |\bm{x}'_{\mathcal{S}'})$ is a deterministic function of $\bm{x}'_{\mathcal{S}'}$.

So,  combining \eqref{eq:combine-000}, \eqref{eq:combine-101} and \eqref{eq:combine-102}, and \eqref{eq:combine-103}, we have
\ifx\doublecolumn\undefined
% ==== put single-column equations here
\begin{subequations}
  \begin{align}
    &H( \hat{\bm{X}}_{\hatcal{B}_r} | \hat{\bm{X}}_{\hatcal{A}_r},\hat{D}=1, \hat{X}_{\text{b}} = \sigma) - H(\bm{X}'_{\mathcal{B}'_r} | \bm{X}'_{\mathcal{A}'_r}) \nonumber \\
    &\leq \epsilon' \log e +  \frac{\epsilon'}{(1-\epsilon')|\mathcal{X}_{\mathcal{S}'}|} \sum_{\substack{\bm{x}'_{\mathcal{A}'_r}:\\ \text{s.t. } \exists \bm{x}'_{\mathcal{S}'} \in \mathcal{W}^\text{G}_{\bm{x}'_{\mathcal{A}'_r},\sigma} }} \sum_{\bm{x}'_{\mathcal{B}'_r}} \sum_{\bm{x}'_{\mathcal{S}'} \in \mathcal{W}^\text{G}_{\bm{x}'_{\mathcal{A}'_r},\sigma}}\nonumber\\ & \quad\quad p(\bm{x}'_{\mathcal{B}'_r} |\bm{x}'_{\mathcal{S}'})  \log |\mathcal{W}^\text{G}_{\bm{x}'_{\mathcal{A}'_r},\sigma}| \nonumber\\
    &\quad- \frac{\epsilon'} {(1-\epsilon')|\mathcal{X}_{\mathcal{S}'}|}  \sum_{\substack{\bm{x}'_{\mathcal{A}'_r}:\\ \text{s.t. } \exists \bm{x}'_{\mathcal{S}'} \in \mathcal{W}^\text{G}_{\bm{x}'_{\mathcal{A}'_r},\sigma} }} \sum_{\bm{x}'_{\mathcal{B}'_r}} \sum_{\bm{x}'_{\mathcal{S}'} \in \mathcal{W}^\text{G}_{\bm{x}'_{\mathcal{A}'_r},\sigma}} p(\bm{x}'_{\mathcal{B}'_r} |\bm{x}'_{\mathcal{S}'})\nonumber\\ & \quad\quad \log \sum_{\bm{x}'_{\mathcal{S}'} \in \mathcal{W}^\text{G}_{\bm{x}'_{\mathcal{A}'_r},\sigma}} p(\bm{x}'_{\mathcal{B}'_r} |\bm{x}'_{\mathcal{S}'}))\\
    &\leq \epsilon' \log e + \frac{\epsilon'}{1-\epsilon'} \left(  E \log e + H(\bm{X}'_{\mathcal{B}'_r} | \bm{X}'_{\mathcal{A}'_r}) \right) \label{eq:combine-103}\\
    &\leq \frac{\epsilon'}{1-\epsilon'} \left(  (1-\epsilon')\log e + \epsilon' \log e + H(\bm{X}'_{\mathcal{E}'})  \right)\\
    & \leq \frac{\epsilon'}{1-\epsilon'} \left(  \log e + \hat{n}  \right),
  \end{align}
\end{subequations}\else
% ==== put double-column equations here
\begin{subequations}
  \begin{align}
    &H( \hat{\bm{X}}_{\hatcal{B}_r} | \hat{\bm{X}}_{\hatcal{A}_r},\hat{D}=1, \hat{X}_{\text{b}} = \sigma) - H(\bm{X}'_{\mathcal{B}'_r} | \bm{X}'_{\mathcal{A}'_r}) \nonumber \\
    &\leq \epsilon' \log e +  \frac{\epsilon'}{(1-\epsilon')|\mathcal{X}_{\mathcal{S}'}|} \sum_{\substack{\bm{x}'_{\mathcal{A}'_r}:\\ \text{s.t. } \exists \bm{x}'_{\mathcal{S}'} \in \mathcal{W}^\text{G}_{\bm{x}'_{\mathcal{A}'_r},\sigma} }} \sum_{\bm{x}'_{\mathcal{B}'_r}} \sum_{\bm{x}'_{\mathcal{S}'} \in \mathcal{W}^\text{G}_{\bm{x}'_{\mathcal{A}'_r},\sigma}}\nonumber\\ & \quad\quad p(\bm{x}'_{\mathcal{B}'_r} |\bm{x}'_{\mathcal{S}'})  \log |\mathcal{W}^\text{G}_{\bm{x}'_{\mathcal{A}'_r},\sigma}| \nonumber\\
    &\quad- \frac{\epsilon'} {(1-\epsilon')|\mathcal{X}_{\mathcal{S}'}|}  \sum_{\substack{\bm{x}'_{\mathcal{A}'_r}:\\ \text{s.t. } \exists \bm{x}'_{\mathcal{S}'} \in \mathcal{W}^\text{G}_{\bm{x}'_{\mathcal{A}'_r},\sigma} }} \sum_{\bm{x}'_{\mathcal{B}'_r}} \sum_{\bm{x}'_{\mathcal{S}'} \in \mathcal{W}^\text{G}_{\bm{x}'_{\mathcal{A}'_r},\sigma}} p(\bm{x}'_{\mathcal{B}'_r} |\bm{x}'_{\mathcal{S}'})\nonumber\\ & \quad\quad \log \sum_{\bm{x}'_{\mathcal{S}'} \in \mathcal{W}^\text{G}_{\bm{x}'_{\mathcal{A}'_r},\sigma}} p(\bm{x}'_{\mathcal{B}'_r} |\bm{x}'_{\mathcal{S}'}))\\
    &\leq \epsilon' \log e + \frac{\epsilon'}{1-\epsilon'} \left(  E \log e + H(\bm{X}'_{\mathcal{B}'_r} | \bm{X}'_{\mathcal{A}'_r}) \right) \label{eq:combine-103}\\
    &\leq \frac{\epsilon'}{1-\epsilon'} \left(  (1-\epsilon')\log e + \epsilon' \log e + H(\bm{X}'_{\mathcal{E}'})  \right)\\
    & \leq \frac{\epsilon'}{1-\epsilon'} \left(  \log e + \hat{n}  \right),
  \end{align}
\end{subequations}
\fi
where \eqref{eq:combine-103} follows from multiplying \eqref{eq:combine-102} with $\epsilon'/(1-\epsilon')$. \hfill $\blacksquare$
%\end{IEEEproof}

\end{document}